\documentclass[a4paper,reprint,aps,unsortedaddress]{revtex4}

\usepackage[utf8]{inputenc}
\usepackage[english]{babel}

\usepackage[hidelinks]{hyperref}

\usepackage{graphicx}
\usepackage{subcaption}
\usepackage{parskip}
\usepackage{appendix}
\usepackage{float}
\usepackage{booktabs}
\usepackage{makecell}
\usepackage{multirow}

\usepackage[figurename=Fig.]{caption}
\usepackage[tablename=Tab.]{caption}
\usepackage[font=small]{caption}

\usepackage{amsmath}
\usepackage{amssymb}
\usepackage{amsfonts}
\usepackage{mathtools}
\usepackage{upgreek}
\usepackage{xcolor}
\usepackage[normalem]{ulem}
\usepackage{cancel}
\usepackage{physics}

\newcommand{\ee}{\end{equation}}
\newcommand{\eea}{\end{eqnarray}}
\newcommand{\be}{\begin{equation}}
\newcommand{\bea}{\begin{eqnarray}}

\begin{document}

\title{Circular orbits and particle collisions\\
in the space-time of boson stars and their frozen states}

\date{\today}

\author{Maria Chivers}
 \email{maria.chivers.20@ucl.ac.uk}
\affiliation{Department of Mathematics, University College London, Gower Street, London, WC1E 6BT, UK}

\author{Betti Hartmann}
 \email{b.hartmann@ucl.ac.uk}
\affiliation{Department of Mathematics, University College London, Gower Street, London, WC1E 6BT, UK}

\author{Katherine Horton}
 \email{katherine.horton.22@ucl.ac.uk}
\affiliation{Department of Mathematics, University College London, Gower Street, London, WC1E 6BT, UK}

\author{Yves Brihaye}
 \email{yves.brihaye@umons.ac.uk}
\affiliation{Physique de l'Univers, Universit\'e de
Mons, 7000 Mons, Belgium}

\begin{abstract}
In this paper, we discuss the dynamics of test particles in the space-time of spherically symmetric, charged boson stars and their frozen states. These latter objects consist of a de Sitter core and a black hole exterior with a shell of finite thickness interpolating between the two. We discuss charged and uncharged massive particles, find the ISCOs and determine whether collisions of massive particles close to the core of (frozen) boson stars can lead to large center of mass energies. We also discuss massless test particles and find that photons rings as well as light points are possible in the space-time of frozen boson stars. We compare our results with those in the corresponding Reissner-Nordstr\"om space-time and point out differences. 

\end{abstract}

\maketitle

\section{Introduction}
Ever since Penrose's and Hawking's singularity theorems \cite{penrose1965,penrosehawking1970} which predict the formation of a space-time singularity in gravitational collapse (formation of a black hole), there has been huge effort in the scientific community to understand how quantum physics can be incorporated into a geometrical description of space-time. Common belief is that at the Planck scale quantum effects will resolve the singularity which otherwise would violate the unitary evolution of quantum states. 
Up to now, a consistent theory of quantum gravity is, however, missing.  Alternatives to black holes -- black hole mimickers --
are consequently discussed extensively (see e.g. \cite{cardosopani2019, Bambi:2025wjx}). 
Since these objects need to be ultracompact to account for observations that are typically attributed to black holes, the gravitational
effects and composition of these objects needs to be studied carefully, in particular space-time curvature close to their cores needs to be so strong that they resemble the region close to the event horizon of a black hole. 
Boson stars \cite{Kaup:1968zz, Schunck:1999pm, Friedberg:1986tp, Jetzer:1991jr, Schunck:2003kk, Liebling:2012fv} are globally regular, soliton-like objects
and are made off a complex scalar field. They can be ultracompact in certain parameter regimes and constitute a form of macroscopic Bose-Einstein condensate being supported by field dynamics rather than having a vacuum interior. Regular black holes, on the other hand, possess an event horizon, but no space-time singularity. In the Bardeen model \cite{bardeen, Ayon-Beato:1998hmi, Ayon-Beato:2000mjt}, the source of the regular black hole is a magnetic monopole within non-linear electrodynamics. Adding scalar fields to regular black hole models allows the construction of a third class of black hole mimickers, so-called {\it frozen stars} \cite{frozenbardeen1,frozenbardeen2}.
In these solutions the event horizon is replaced by a thin shell that interpolates between the singularity free interior and the black hole exterior. It goes back to the gravastar idea of Mazur and Mottola \cite{Mazur:2001fv} that interpolates a de Sitter interior with a Schwarzschild exterior across an infinitly thin shell. It was later shown that these gravastars -- if composed of a fluid -- need  anisotropic stresses \cite{Cattoen:2005he}, very similar to boson stars which, when interpreted as fluids, possess radial pressure unequal to the tangential pressure. Interestingly, a solution akin to gravastars also exists when replacing the vacuum interior of black holes with a maximally entropic fluid described by a state that is highly quantum \cite{brustein1, brustein2}, which can be described classically \cite{Brustein:2021lnr, Brustein:2023cvf}.

The U(1) symmetry present in the boson star model can be gauged, making the resulting soliton-like objects electrically charged boson stars \cite{Brihaye:2014gua,Lopez:2023phk}. It has been pointed out that these boson stars can have a frozen star limit \cite{Brihaye:2025dlq} such that the interior of the star resides in the false vacuum of the scalar field potential leading to a de Sitter state, while
the exterior corresponds to the electro-vaccum solution, in this case the Reissner-Nordstr\"om black hole solution.

The question then remains whether charged boson stars and their frozen states can be distinguished from charged black holes in any way.
One of the best ways to test the geometry around ultracompact objects is to understand how test particles behave when moving in their space-time. In this paper, we will discuss circular motion of massive and massless test particles as well as the collision of test particles.

The innermost stable circular orbit (ISCO) is defined as the smallest marginally stable circular orbit on which a massive test particle can orbit a compact object. ISCOs are most commeonly studied around black holes. In these cases, any orbits lying within the ISCO are unstable, and hence small perturbations will cause particles to fall into the black hole. Around Reissner-Nordstr\"om (RN) black holes, two circular orbits occur for each value of the angular momentum above a minimum $L\geq2\sqrt{2}$, occurring in unstable-stable pairs \cite{Pradhan:2011}. In \cite{Chivers:2026}, black holes with scalar hair were found in some cases to have multiple pairs of circular orbits, again occurring in unstable-stable pairs. The motion of charged particles around both black holes and horizonless space-times has also been studied, such as in \cite{Isomura:2023} and \cite{Rayimbaev:2020}.
Around RN black holes, stable light rings occur on the horizon for extremal RN black holes, but not for non-extremal black holes. An additional unstable lightlike orbit occurs outside the horizon in both cases \cite{Pradhan:2011}. For regular space-times, sufficiently compact objects can also exhibit light rings. Examples can be found in \cite{Guo:2021}, which finds cases with stable-unstable pairs of light rings, and \cite{Berry:2020}, which finds again pairs of photon spheres around regular black holes with asymptotically Minkowski cores, certain cases of which have two stable orbits.

When considering the collision of test particles the resulting center-of-mass energy $E_{\text{\tiny{C.M.}}}$ during the collision matters. Previous studies, such as \cite{Olivares:2020} have examined particle motion in boson-star space-times, focusing on massive particles in non-rotating configurations. In particular, these studies show that, for certain boson-star solutions, stable circular orbits can extend all the way to the center of the star. This is particularly relevant to our analysis of $E_{\text{\tiny{C.M.}}}$ in the limit $r\rightarrow 0$, since at the center of a regular boson star there is no singularity. In the black hole case, arbitrarily large $E_{\text{\tiny{C.M.}}}$ is typically associated with the horizon and the presence of a critical particle, as in the Ba\~nados-Silk-West (BSW) mechanism \cite{Banados:2009}. A regular boson star, however, possesses no horizon, and its metric functions remain finite and regular at the center. Consequently, the behaviour of the center-of-mass energy near \(r=0\) must be investigated using regular small-\(r\) expansions of the metric functions, rather than the near-horizon expansions commonly employed in black hole analyses. This provides a different setting from the standard BSW scenario and allows us to determine whether large or divergent $E_{\text{\tiny{C.M.}}}$ can arise purely from the behaviour of particle trajectories in regular interior of a boson star.

\section{The space-time}
A spherically symmetric, electrically charged boson star is described by the following model:
\begin{equation}
\label{eq:action}
{\cal S}= \int {\rm d}^4 x \sqrt{-g} \left(\frac{1}{16 \pi G} R - D^{\mu} \Psi^{\dagger}  D_{\mu} \Psi - U(\Phi) - \frac{1}{4} F^{\mu \nu} F_{\mu \nu}\right)
\end{equation}
with scalar field potential of the form
\be 
\label{action}
U(\Phi) = \mu^2 \eta^2 \left[1 - \exp\left(-\frac{\vert\Psi\vert^2}{\eta^2}\right)\right]  \ .
\ee
$R$ is the Ricci scalar, $\Psi$ is a complex-valued scalar field with potential $U(|\Psi|)$
and $F_{\mu\nu}=\partial_{\mu} A_{\nu} - \partial_{\nu} A_{\mu}$ is the field strength tensor of a U(1) gauge field $A_{\mu}$. The scalar field is minimally coupled to this gauge field through the covariant derivative $D_{\mu} = \partial_{\mu} + i g A_{\mu}$.
$G$ is Newton's constant and $g$ the gauge coupling. Variation of the action (\ref{eq:action}) with respect to the matter fields and the metric result in a set of differential equations. Using the following Ansatz for the metric and matter fields
\begin{eqnarray}
& & {\rm d}s^2 = -(\sigma(r))^2 N(r) {\rm d}t^2 + \frac{1}{N(r)} {\rm d}r^2 + r^2\left({\rm d}\theta^2 + \sin^2 \theta {\rm d}\varphi^2 \right) \ \ , \ \  N(r) = 1 - \frac{2m(r)}{r}
\nonumber\\
& & 
A_{\mu} {\rm d} x^{\mu} = V(r) {\rm d} t  \ \ , \ \ 
\Psi=\psi(r) \exp(-i\omega t) \
\label{eq:ansatz}
\end{eqnarray} 
leads to a set of four coupled, non-linear ordinary differential equations that can only be solved numerically (for more details, see e.g. \cite{Brihaye:2025dlq}) subject to appropriate boundary conditions which assure regularity at the origin and finite energy of the solutions. These read~:
\begin{equation}
\label{eq:bcs}
N(0)=1 \ \ ,  \ \  \psi'(0)=0 \ \ , \ \  V'(0)=0 \ \ , \ \
\sigma(r\rightarrow \infty) \rightarrow 1  \ \ , \ \ \psi(r\rightarrow \infty) \rightarrow 0 \ \ , \ \ V(r\rightarrow \infty) \rightarrow V_{\infty} \ , 
\end{equation}
with $V_{\infty}$ a real-valued constant. In the following, we will use dimensionless coordinates and functions and rescale
$r\rightarrow r/\mu$, $\omega \rightarrow \mu \omega$, $\psi \rightarrow \eta \psi$, $m(r)\rightarrow m(r)/\mu$. We will also use the abbreviation $\alpha=4\pi G$ in the following. 

The physical parameters of the solutions are the (dimensionless) mass $M$, the (dimensionless) electric charge $Q$ and the
(dimensionless) Noether charge $Q_N$.  These can be read off from the metric and matter field functions at infinity
\begin{equation}
\label{eq:infty}
N(r \gg 1)=1-\frac{2M}{r} + \frac{\alpha Q^2}{r^2}+ ..... \ \ , \ \ 
V(r)=V_{\infty}  - \frac{Q}{r} + ....
\end{equation}
and the following integral of the $t$-component of the locally conserved Noether current~:
\begin{equation}
\label{eq:noether}
Q_N=\int\limits_0^{\infty} {\rm d} r \ \frac{2r^2 q V\psi^2}{N\sigma}  \ .
\end{equation}
This globally conserved quantity can be interpreted as the number of scalar bosons and 
hence $qQ_N\equiv Q$. 

For $\psi\equiv 0$ and disregarding the boundary conditions at $r=0$, the unique solution of the equations is the Reissner-Nordstr\"om (RN) solution~:
\begin{equation}
N= 1- \frac{2M}{r} + \frac{\alpha Q^2}{r^2} \ \  , \ \ \sigma\equiv 1 \ \ , \ \
V(r)=V_{\infty} - \frac{Q}{r} \ 
\end{equation}
with event horizon at $r_{+}=M(1+\sqrt{1-2\alpha Q^2/M^2})$. The extremal RN solution has
$r_+=M=\sqrt{\alpha}Q$. Note that this solution is not globally regular,  which is why the boundary conditions at $r = 0$ cannot hold.

Finally, in order to make the connection to the gravastar solution of Mazur-Mottola \cite{Mazur:2001fv}, we give the components of the energy density $\rho=-T^t_t$, radial pressure $p_r=T^r_r$ and transverse pressure $p_{t}=T_{\theta}^{\theta}=T^{\varphi}_{\varphi}$ of the solutions. These read~:
\begin{align}
\label{eq:density_pressure}
\rho = \frac{V'^2}{2\sigma^2}  + N {{\psi}}'^2 + \frac{(\omega - q V)^2 {{\psi}}^2}{N \sigma^2} +  U({\psi})  \ , \\
p_r =  - \frac{V'^2}{2\sigma^2}  +  N {{\psi}}'^2 + \frac{(\omega - q V)^2 {{\psi}}^2}{N \sigma^2} - U({{\psi}}) \ , \\
p_{t}=\frac{V'^2}{2\sigma^2}  - N {{\psi}}'^2 +\frac{(\omega - q V)^2 {{\psi}}^2}{N \sigma^2} - U({{\psi}})   \ .
\end{align}
Clearly, $\rho \geq p_r$ and
$\rho \geq p_{t}$, i.e. the model fulfills the causality bound as well as the weak energy condition. 

\section{Test particles}
We consider the motion and collisions of massive and massless test particles in the spherically symmetric space-time described by the line element (\ref{eq:ansatz}). The two cases can be treated simulatenously by introducing the parameter
$\delta$ with $\delta=-1$ for massive particles and $\delta=0$ for massless particles, respectively.
For $\delta=-1$, the particles can also carry electric charge \(q\), whereas in the massless case we consider neutral particles, \(q=0\), only.
\newline The equations describing the dynamics of the particles can be derived from the Hamilton-Jacobi equation \cite{Grunau:2010gd}~:
\begin{equation}
\label{eq:S}
\frac{\partial S}{\partial\tau}=\frac{1}{2}g^{\mu\nu}\left(\frac{\partial S}{\partial x^{\mu}}-qA_{\mu}\right)\left(\frac{\partial S}{\partial x^{\nu}}-qA_{\nu}\right) \ .
\end{equation}
A separable solution is given by
\begin{equation} 
\label{eq:S_solution}
S=\frac{1}{2}\delta\tau-Et+L_z\varphi+S_r(r)+S_{\theta}(\theta) \ ,
\end{equation}
where \(E\) and \(L_z\) denote the conserved energy and the component of angular momentum along the \(z\)-axis, respectively. Here, \(\tau\) is an affine parameter, while \(S_r\) and \(S_{\theta}\) depend only on \(r\) and \(\theta\) respectively.  Substitution of (\ref{eq:S_solution}) into the Hamilton-Jacobi equation gives the equations of motion:
\begin{eqnarray}
\label{eq:geodesic_components}
\dot{t}=\frac{E+qV}{N\sigma^2} \ \ \ &,& \ \ \ \dot{\varphi}=\frac{L_z}{r^2\sin^2\theta} \nonumber \\
\dot{r}^2=\frac{(E+qV)^2}{\sigma^2}+N\bigg(\delta-\frac{C}{r^2}\bigg) \ \ &,& \ \ \dot{\theta}=\frac{1}{r^2}\sqrt{C-\frac{L_z^2}{\sin^2\theta}} \ ,
\end{eqnarray}
where \(C\) is the separation constant defined as $C=(r^2 \dot{\theta})^2+L_z^2/\sin^2\theta$. We restrict the motion to the equatorial plane, \(\theta=\pi/2\) and therefore \(L_z=L\) and \(C=L^2\). For neutral particles, this restriction follows from the spherical symmetry of the space-time, whereas for charged particles it is adapted as part of our setup. Using the normalization condition for the four-velocity, $\dot{x}_{\mu}\dot{x}^{\mu}=\delta$, we then obtain~:
\begin{equation}
E= \sigma\sqrt{ \dot{r}^2 + \frac{NL^2}{r^2\sin^2\theta} - N\delta} - q V \ .
\end{equation}
The equation governing the radial motion can therefore be written in the unified form as 
\begin{equation}\label{eq:rdot}
\dot{r}^2+ V_{\rm eff}(r)=0 \ \ , \ \ V_{\rm eff}=-\frac{(E+qV)^2}{\sigma^2}-N\bigg(\delta-\frac{L^2}{r^2}\bigg)  \  ,
\end{equation}
with the effective potential $V_{\rm eff}$. Using the boundary conditions (\ref{eq:bcs}), we find that $V_{\rm eff}(r\rightarrow 0)\rightarrow +\infty$ for $L\neq 0$, while $V_{\rm eff}(r\rightarrow \infty)\rightarrow -(E+qV_{\infty})^2-\delta$.

\subsection{Circular orbits}
Circular orbits with radius $r=r_c$ correspond to stationary points of the effective potential, i.e. 
\begin{equation}
\label{eq:circular_condition}
V_{\rm eff}(r_c) = 0 \ , \quad V'_{\rm eff}(r_c) = 0  \ ,
\end{equation}
where the prime denotes the derivative with respect to $r$. 
Unstable circular orbits additionally fulfill $V''_{\rm eff}(r_c) < 0$, while stable circular orbits have $V''_{\rm eff}(r_c) > 0$.
Inserting the effective potential in (\ref{eq:rdot}) into (\ref{eq:circular_condition}), we can express the angular momentum 
of the particle on the circular orbit in terms of the metric functions $N$, $\sigma$ and the gauge field function $V$ and their respective derivatives \cite{Chivers:2026}. For massive test particles, the radius of the circular orbit is a function of the angular momentum $L$. The smallest possible value of $r_c$ allowing a massive test particle to move on a stable circular orbit corresponds to the Innermost stable circular orbit (ISCO) and fulfills the additional condition $V''_{\rm eff}(r_c) = 0$.  
Massless test particles, on the other hand, can only move on circular orbits, so-called photon rings, for a very specific value of $L$.  

\subsection{Particle collisions}

We consider particles originating at infinity with finite conserved energy and angular momentum. In the asymptotic limit, the particles have vanishing velocity, while their conserved angular momenta remain finite. Hence, for a massive particle the conserved energy is \(E=1-qV_{\infty}\). With the chosen gauge, this energy differs from unity whenever the particle carries a non-zero charge \(q\neq0\). For massless (and uncharged) particles, the conserved energy is \(E=1\). 
The possibility of achieving arbitrarily large center-of-mass energies in particle collisions is intimately connected to the existence of circular orbits, as pointed out in \cite{Harada:2014vka}. Consequently, the analysis of particle collisions reduces to determining the critical angular momentum \(L\) and the corresponding critical radius \(r_c\). The critical angular momentum is therefore determined by analysing the effective potential and imposing the conditions for circular motion, which read
\[V_{\rm eff}(r_c) = 0 \ , \quad V'_{\rm eff}(r_c) = 0 \ .\]
Since only unstable circular orbits allow the particle to reverse its radial motion, we additionally require $V''_{\rm eff}(r_c) < 0$.

\subsubsection{Massive particles}
The center-of-mass energy is one of the key quantities of interest in the study of particle collisions. For massive test participles, \(E_{\text{\tiny{C.M.}}}\) can be determined using the formalism developed in \cite{Bardeen:1972}. Restricting to two particle of equal rest mass \(m_0\), the general expression for the center-of-mass energy takes the form
\begin{equation}
\label{ECM}\frac{E^2_{\text{\tiny{C.M.}}}}{2m_0^2}=1-g_{\mu\nu}u^{\mu}_{(1)}u^{\nu}_{(2)} \ .
\end{equation}
The four-velocity vectors associated with two particles in the space-time characterised by the line element (\ref{eq:ansatz}) can be written as
\begin{equation}
\label{4static}
u^{\mu}_{(i)}=\bigg(\frac{\tilde{E}_{i}}{N\sigma^2},-\sqrt{\frac{\tilde{E}^2_i}{\sigma^2}-N\bigg(1+\frac{L^2_i}{r^2}\bigg)},0,\frac{L_i}{r^2}\bigg) \ , \quad i=1,2 \  ,
\end{equation}
where $\tilde{E}_i=E_i + q_i V(r)$. Substituting (\ref{4static}) into (\ref{ECM}), we find the center-of-mass energy to be~:
\begin{equation}
\label{CMEstatic}
\frac{E^2_{\text{\tiny{C.M.}}}}{2m_0^2}=1+\frac{\tilde{E}_1 \tilde{E}_2}{N\sigma^2}-\frac{L_1L_2}{r^2}-\frac{1}{N}\sqrt{\frac{\tilde{E}_1^2}{\sigma^2}-N\bigg(1+\frac{L^2_1}{r^2}\bigg)}\sqrt{\frac{\tilde{E}^2_2}{\sigma^2}-N\bigg(1+\frac{L^2_2}{r^2}\bigg)} \ .
\end{equation}
In order to obtain arbitrarily large center-of-mass energies, \(E_{\text{\tiny{C.M.}}}\) must be evaluated in a limiting regime rather than directly at the collision point.

\subsubsection{Massless Particles}
The center-of-mass energy can be defined for massless particles since it depends only on the particles' 4-momenta which are determined  from the null geodesic equations. Consider two particles moving in curved space-time with 4-momenta: \(p^{\mu}_{1}\) and \(p^{\mu}_2\) where the total 4-momentum is
\[P^{\mu}=p^{\mu}_1+p^{\mu}_2 \ .\]
The center-of-mass energy is defined as
\begin{equation}E^2_{\text{\tiny{C.M.}}}=-P_{\mu}P^{\mu} \ .
\end{equation}
For massless particle we have \(p^{\mu}p_{\mu}=0\), therefore 
\begin{equation}\label{ECMmassless}
\frac{E^2_{\text{\tiny{C.M.}}}}{2}=-g_{\mu\nu}p^{\mu}_{1}p^{\mu}_2 \ .
\end{equation}
The four-momenta associated with two particles on the equatorial plane can be written as
\begin{equation}
\label{4static}
p^{\mu}_{(i)}=\bigg(\frac{E_{i}}{N\sigma^2},\epsilon_i\sqrt{\frac{E^2_i}{\sigma^2}-\frac{NL^2_i}{r^2}},0,\frac{L_i}{r^2}\bigg) \ , \ i=1,2 \ ,
\end{equation}
where \(\epsilon_i=\pm1\) determines the radial direction of motion. Substitution of (29) into (26) gives  
\begin{equation}
\label{CMEstaticMassless}
\frac{E^2_{\text{\tiny{C.M.}}}}{2}=\frac{E_1 E_2}{N\sigma^2}-\frac{L_1L_2}{r^2}-\frac{\epsilon_1\epsilon_2}{N}\sqrt{\frac{E_1^2}{\sigma^2}-\frac{NL^2_1}{r^2}}\sqrt{\frac{E^2_2}{\sigma^2}-\frac{NL^2_2}{r^2}} \ ,
\end{equation}
where the sign of the radial square-root term depends on the relative directions of motion of the two particles. For particles moving in the same radial direction, \(\epsilon_1\epsilon_2=1\), and the square-root contribution enters with a negative sign. For particles moving in opposite radial directions, \(\epsilon_1\epsilon_2=-1\), and the contribution enters with a positive sign.

\section{Particle motion in boson star space-times}
In the following, we will study particle motion in the space-time of boson star solutions of the model (\ref{eq:action}). The solutions have to be constructed numerically and have been discussed in detail in \cite{Brihaye:2025dlq}. In the following, we will fix
$\alpha=0.012$ and $g=0.08$ and vary the electric charge $Q$. We believe that this choice of parameters represents the general behaviour quite well. After having obtained the numerical solutions we have interpolated the data
for the metric functions to discuss circular orbits and particle collisions. The interpolation was done using cubic splines in Python and the interpolation routine in MATHEMATICA, respectively. 

\subsection{Circular orbits}
We find it often useful to define an energy-independent potential $\tilde{V}_\text{eff}$. Using (\ref{eq:geodesic_components}) we find the following 
\begin{equation}
    \dot{r}^2=\frac{1}{\sigma^2}\left(E^2-\tilde{V}_\text{eff}\right) \ \ , \ \  \tilde{V}_\text{eff}=\sigma^2N\left(\frac{L^2}{r^2}-\delta\right)  \ .
\end{equation}
Circular orbits then occur at stationary points of $\tilde{V}_\text{eff}(r)$, with energy equal to $\sqrt{\tilde{V}_\text{eff}}$.

\subsubsection{Massive uncharged test particles}\label{sec:orbits_boson_massive_uncharged}
\begin{figure}[!h]
	\centering
	\includegraphics[width=0.5\linewidth]{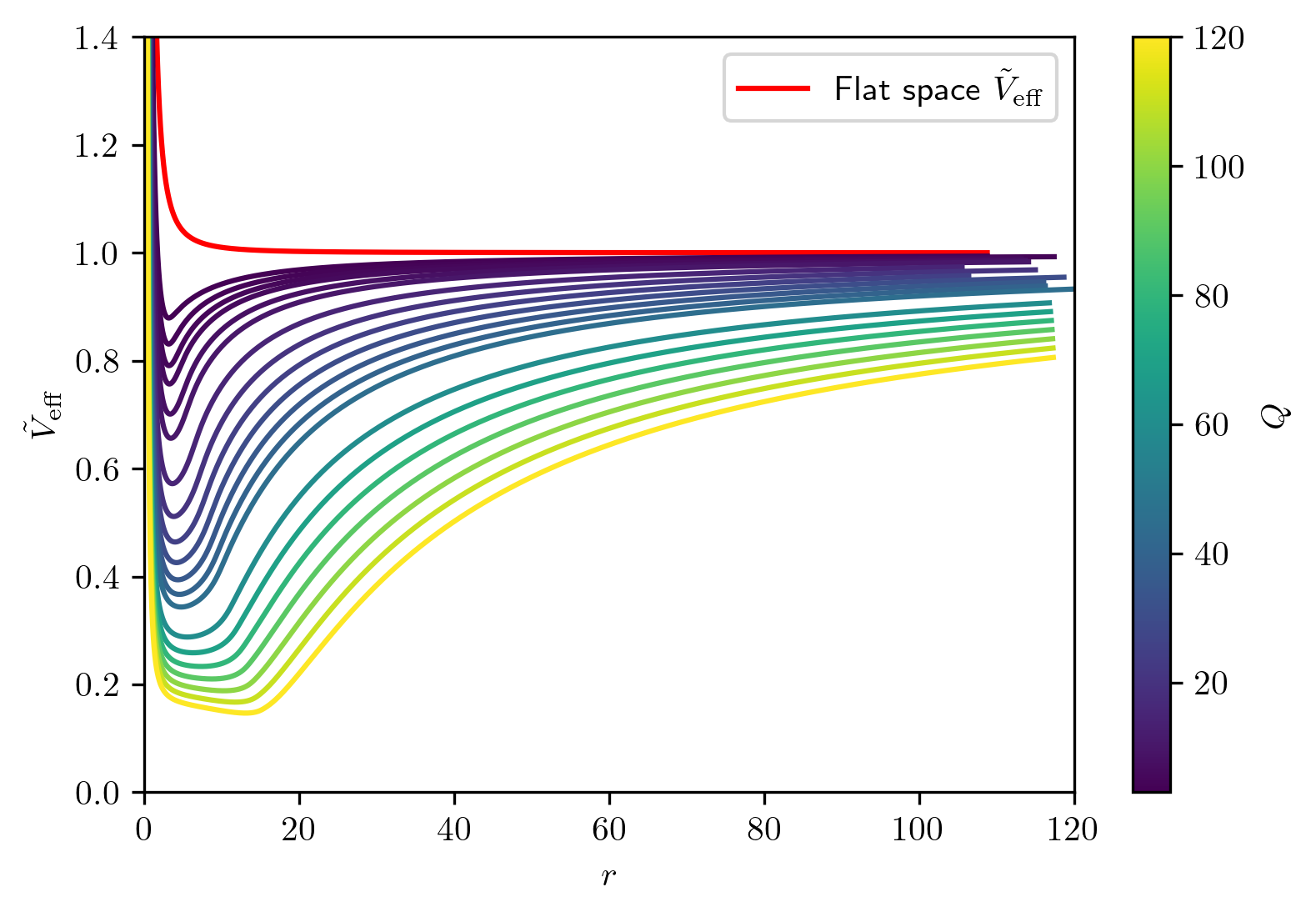}
	\caption{We show the effective potential $\tilde{V}_\text{eff}$ for a massive particle with $L=1$ in the space-time of boson stars with varying $Q$. For comparison, we also show the flat space-time limit (red).}
	\label{fig:pot_massive_L1}
\end{figure}

In Fig.~\ref{fig:pot_massive_L1} we show $\tilde{V}_\text{eff}$ for massive particles with conserved angular momentum $L=1$ in these boson star space-times, as well as $\tilde{V}_\text{eff}$ for flat space-time (in red.)
We see that as $Q\to3$ and $r\to r_\infty$, $\tilde{V}_\text{eff}$ approaches that for flat space-time. However, for small $r$, while the flat space-time potential approaches $\infty$ monotonically from below (as the potential is given by $\frac{L^2}{r^2}+1$ in flat space-time), the potentials for the boson star space-times have a minimum, representing a stable orbit. Further, as $Q$ increases, this minimum broadens and shifts to larger $r$.

\begin{figure}[!h]
	\centering
	\begin{subfigure}{0.4\textwidth}
		\includegraphics[width=\textwidth]{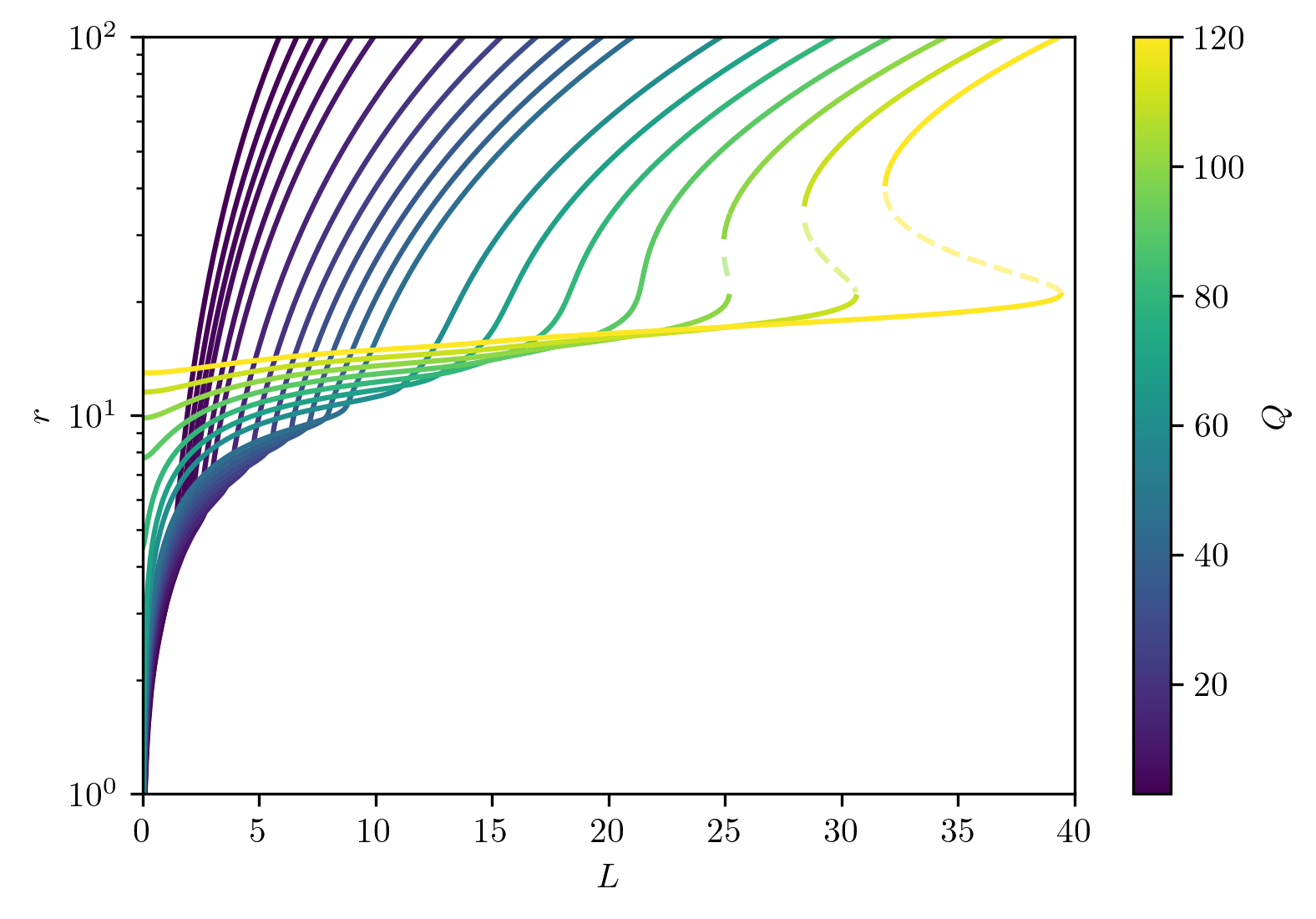}
	\end{subfigure}
    \begin{subfigure}{0.4\textwidth}
		\includegraphics[width=\textwidth]{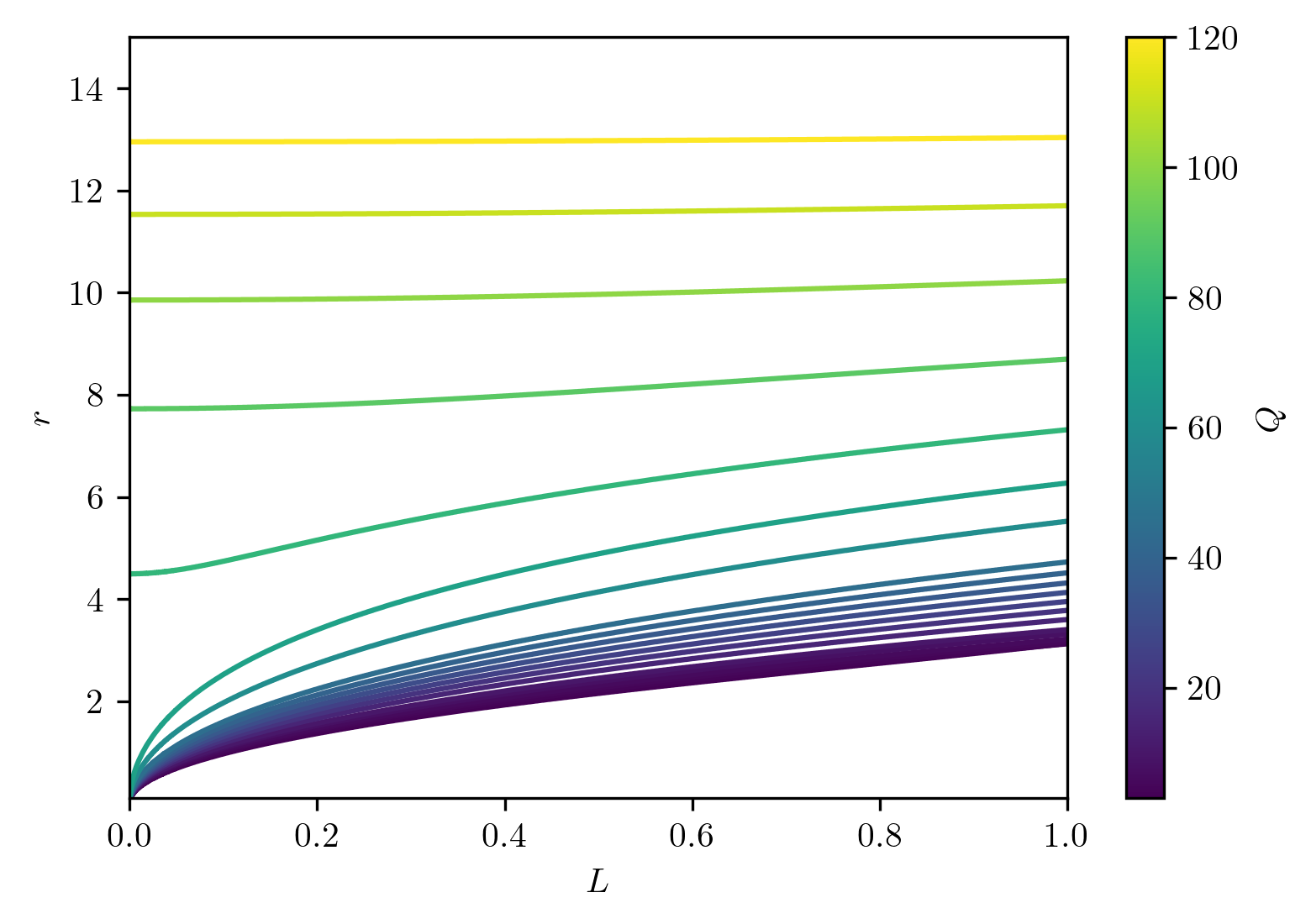}
	\end{subfigure}
    \caption{We show the value of $r$ of the stable circular orbit against the angular momentum $L$ for solutions with varying $Q$. The right-hand side shows a zoom of the small $L$ region.}
    \label{fig:st_pts_Q_range}
\end{figure}

Fig.~\ref{fig:st_pts_Q_range} shows the radius of the circular orbits in dependence of $L$ for a variety of values of $Q$. Unlike for black hole solutions, where stationary points occur in maximum-minimum pairs (which can be demonstrate by considering the gradient of the effective potential), there is usually only one stable orbit for each $L$, with no unstable orbits. For small $Q$, the position of that stable orbit tends to $r=0$ as $L\to 0$, while for larger $Q$ it tends to a finite non-zero value. Further, for $Q\geq100$, an additional unstable-stable pair of orbits forms for certain $L$.
For $Q<80$, we can have circular orbits with radii arbitrarily small when $L\rightarrow 0$. The limit $r`rigntarrow 0$ is, of course, a point and not an orbit. This exists always due to the spherical symmetry of the space-time. For $Q\geq 80$, we find that as $L\to 0$, $r$ tends to a finite non-zero value $r_\text{ISCO}$, representing a static orbit with $\dot{r}=\dot{\varphi}=0$.
Table \ref{tab:Q_shell_ISCO} shows the radius of this static orbit, $r_\text{static}$, as well as its energy $E_\text{static}$, for $Q\geq80$.
\begin{center}
	\begin{tabular}{||c | c c||} 
		\hline
		$Q$ &  $r_\text{static}$ & $E_\text{static}$ \\ [0.5ex] 
		\hline\hline
		80 & 4.496 & 0.477 \\ 
		\hline
		90 & 7.726 & 0.455  \\
		\hline
		100 & 9.856 & 0.432 \\
		\hline
		110 & 11.532 & 0.407 \\
		\hline
		120 & 12.953 & 0.382 \\ [1ex] 
		\hline
	\end{tabular}
	\captionof{table}{The values of the radius, $r_\text{static}$, and energy, $E_\text{static}$, of the static orbit ($L=0$) for $Q\geq80$.}\label{tab:Q_shell_ISCO}
\end{center}

\begin{figure}[!h]
	\centering
	\begin{subfigure}{0.4\textwidth}
		\includegraphics[width=\textwidth]{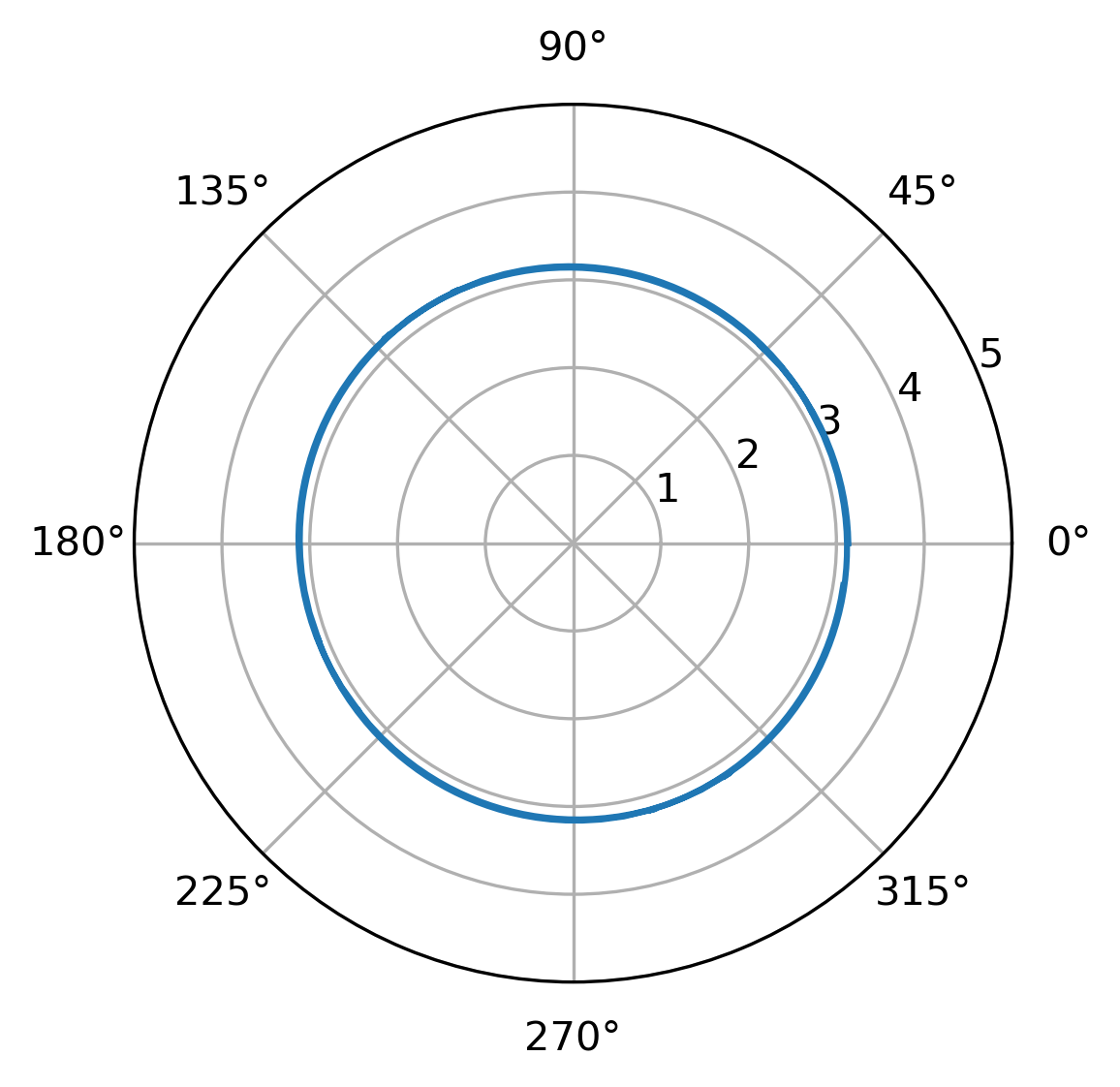}
		\caption{$L=1, E=0.9376$}
		\label{fig:orb3_l1_1}
	\end{subfigure}
	\begin{subfigure}{0.4\textwidth}
		\includegraphics[width=\textwidth]{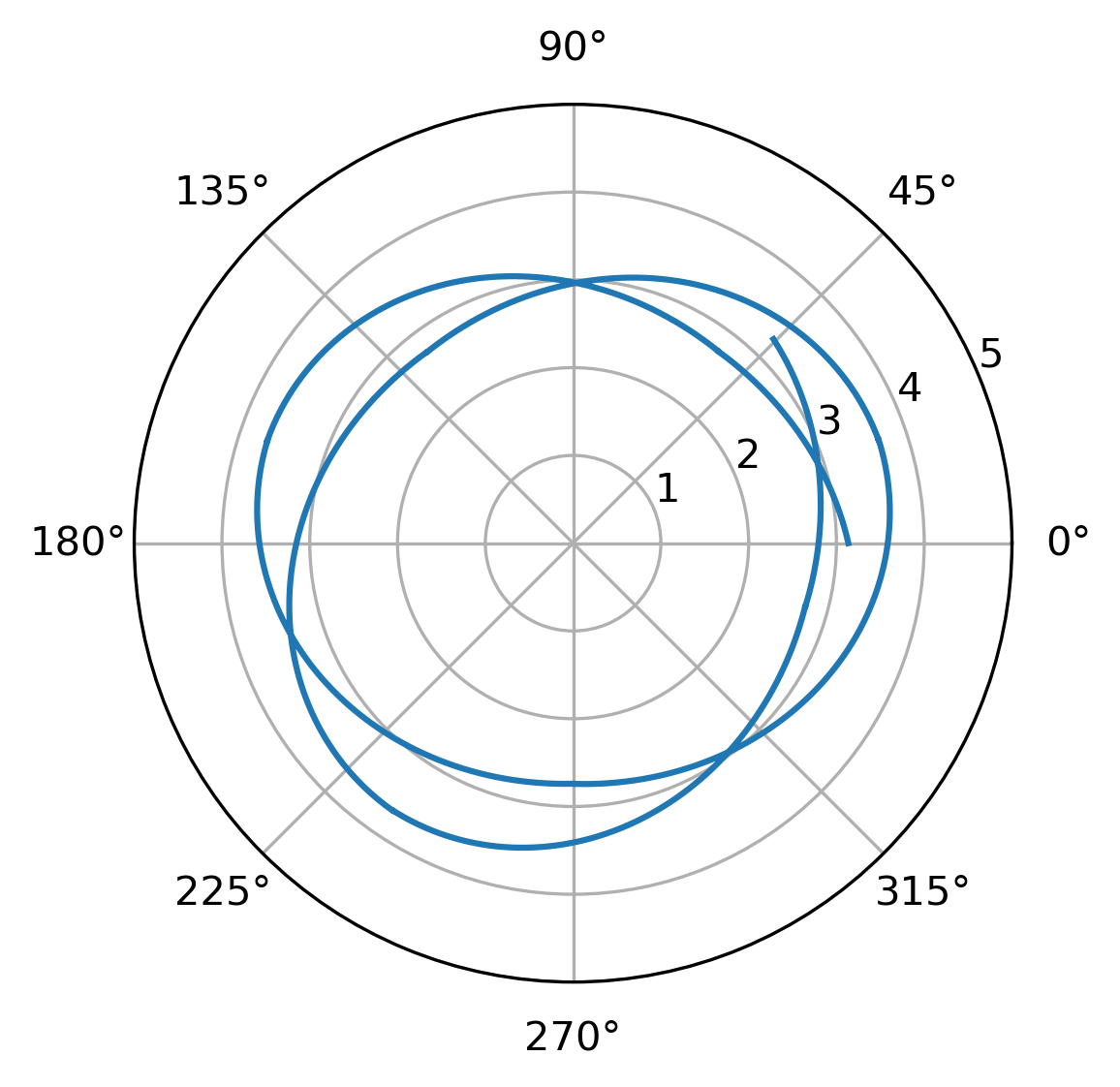}
		\caption{$L=1,E=0.94$}
		\label{fig:orb3_l1_2}
	\end{subfigure}
	\caption{Orbits for $Q=3$, $L=1$}
    \label{fig:orb3_l1}
\end{figure}

Fig.~\ref{fig:orb3_l1} shows two orbits for $Q=3$ with $L=1$, both starting at the location of the stable orbit. 
In Fig.~\ref{fig:orb3_l1_1} the particle has energy very near $\sqrt{\tilde{V}_\text{eff}}$, $E=0.9376$, giving a circular orbit, while in
Fig.~\ref{fig:orb3_l1_2} it has slightly greater energy, $E=0.94$, causing an elliptical orbit with significant procession. 

\begin{figure}[!h]
	\centering
	\includegraphics[width=0.4\textwidth]{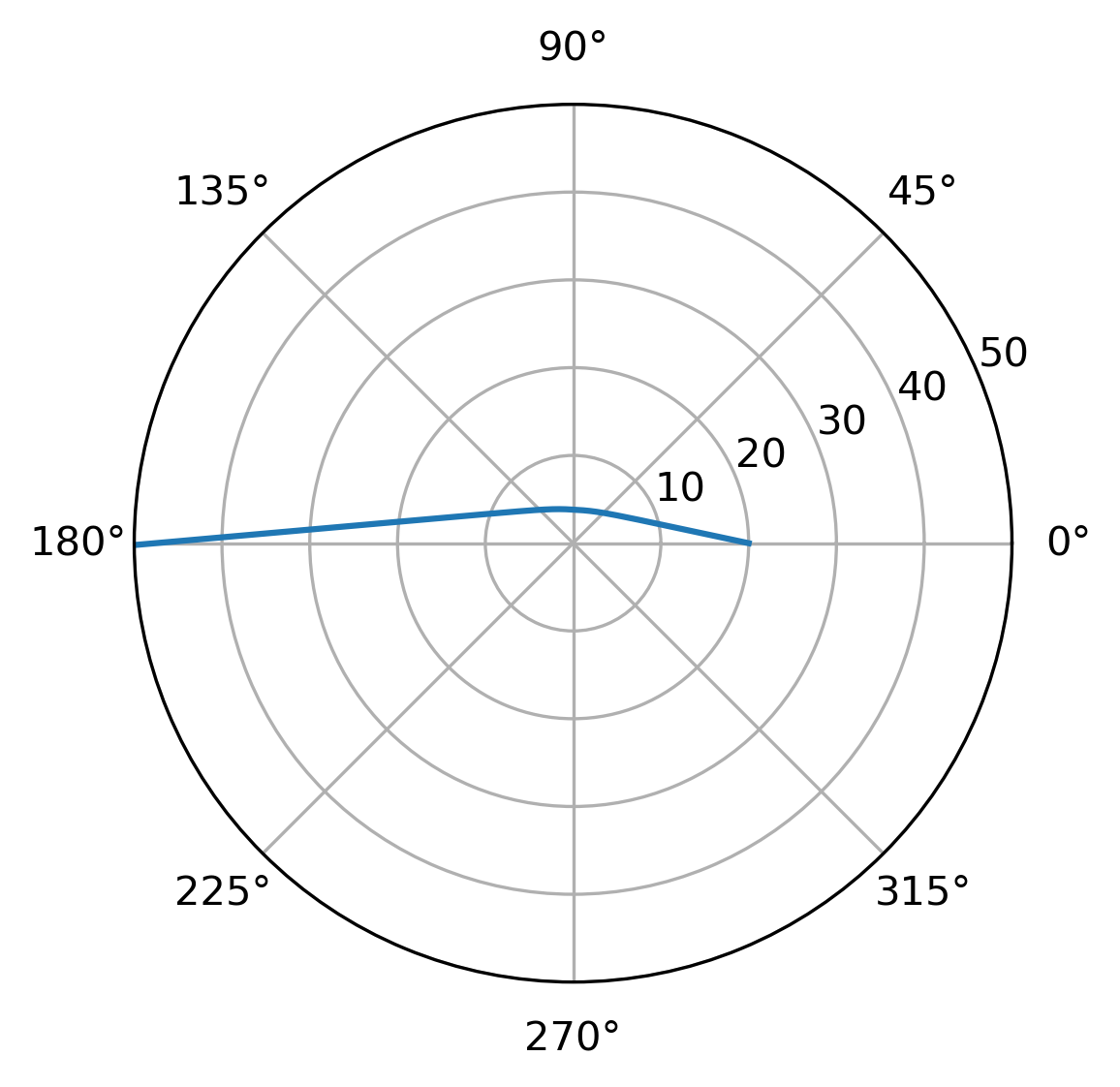}
    \caption{Orbit for $Q=3$, $L=50$, $E=12.0$ }
    \label{fig:orb3_l50}
\end{figure}

Let us now check what happens for larger angular momentum. In Fig.~\ref{fig:orb3_l50} we show an orbit for $Q=3$ with $L=50$, demonstrating deflection due to the boson star and no possibility for a circular orbit.

\begin{figure}[!h]
	\centering
	\begin{subfigure}{0.4\textwidth}
		\includegraphics[width=\textwidth]{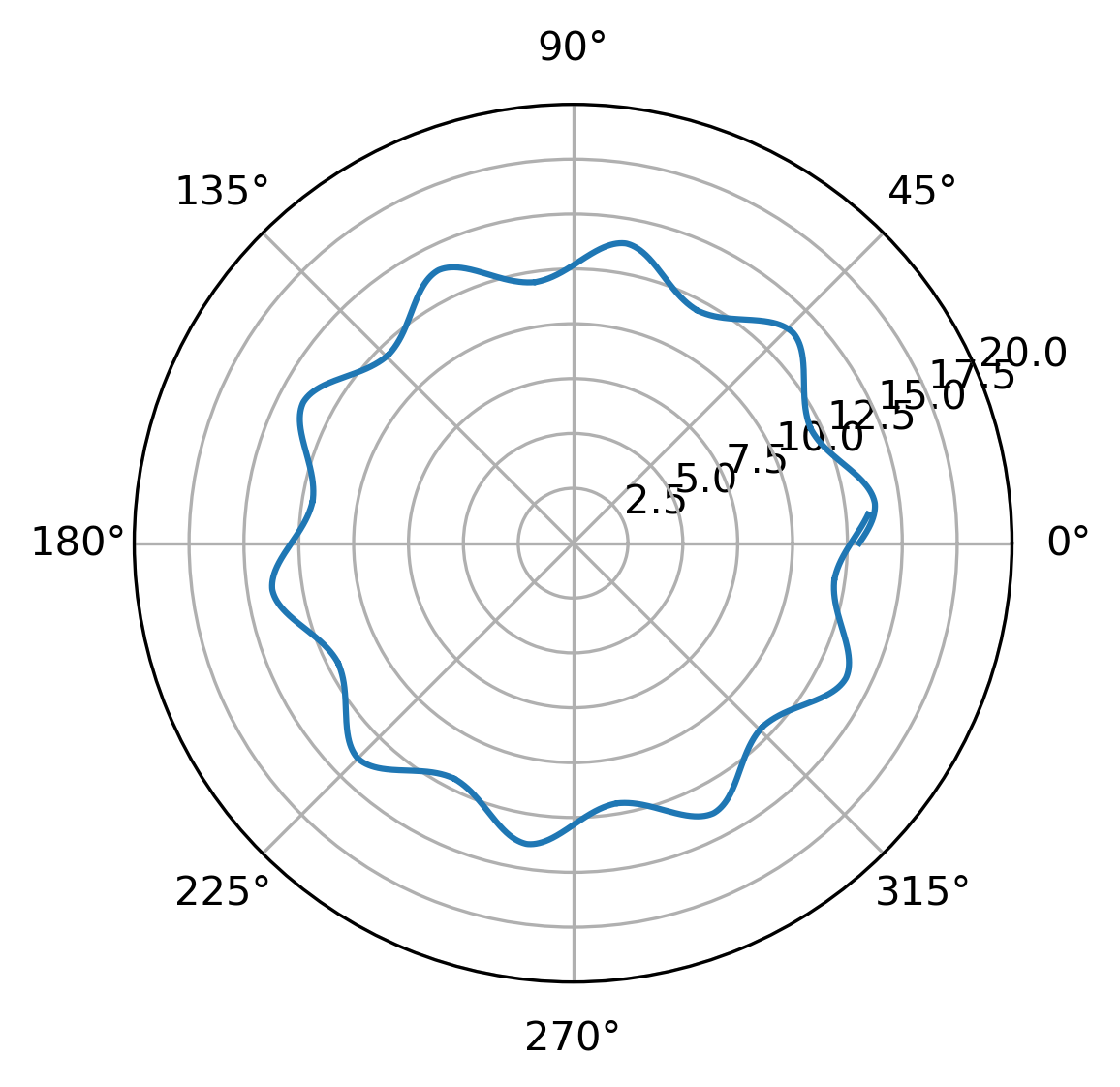}
		\caption{$L=1, E=0.384$}
		\label{fig:orb120_l1_1}
	\end{subfigure}
	\begin{subfigure}{0.4\textwidth}
		\includegraphics[width=\textwidth]{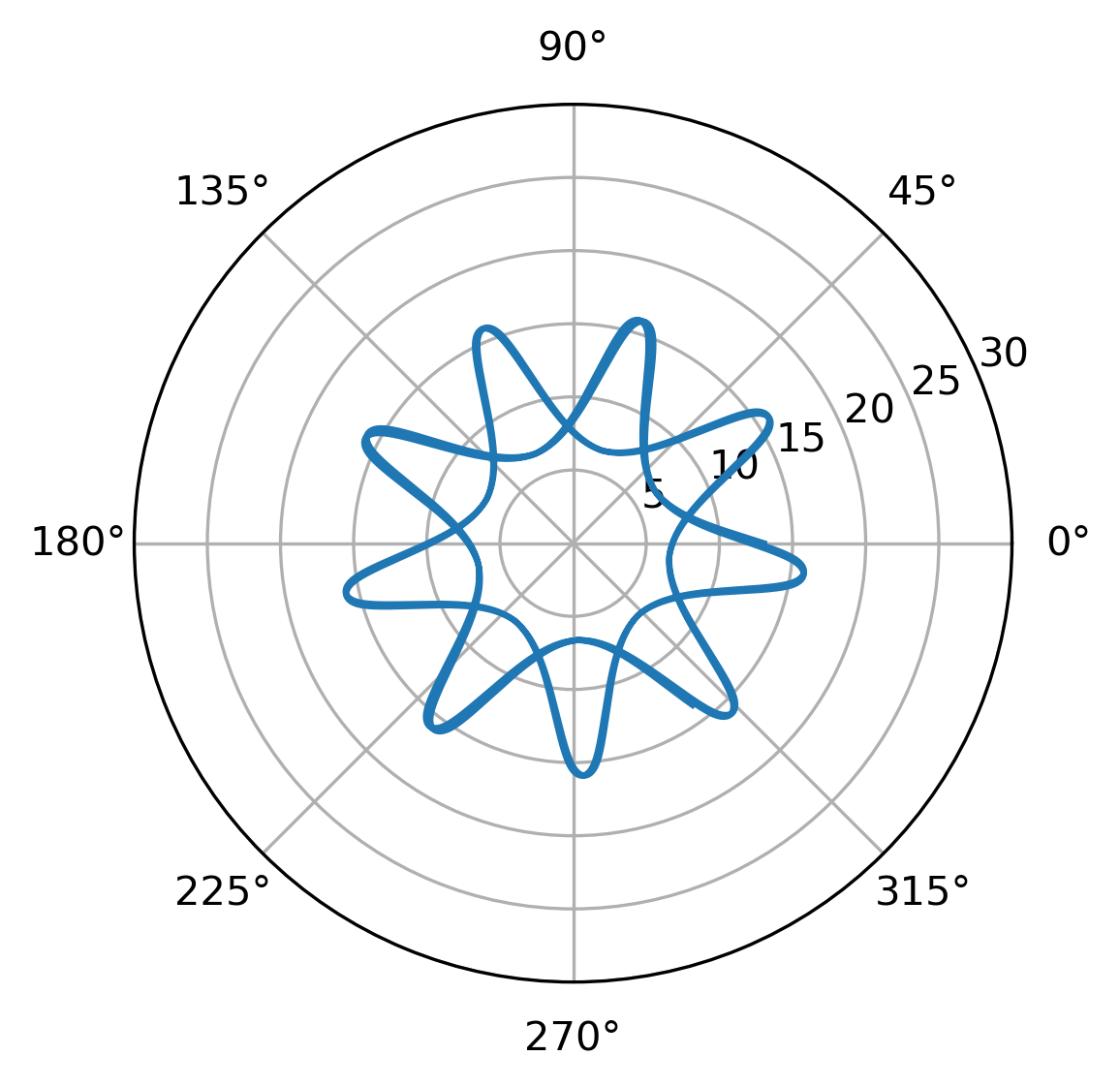}
		\caption{$L=1,E=0.4$}
		\label{fig:orb120_l1_2}
	\end{subfigure}
	\caption{Orbits for $Q=120$, $L=1$}
    \label{fig:orb120_l1}
\end{figure}

Considering now boson stars with much larger charge, $Q=120$, we find new features. This is shown in Fig.~\ref{fig:orb120_l1} for two orbits with $L=1$, both starting at the location of the stable circular orbit at $r=13.037$. Both of these display a new behaviour in comparison to that around typical massive bodies. The broadening of the minimum shown in Fig.~\ref{fig:pot_massive_L1} allows the formation of a new class of motion, in which the test particle oscillates around the circular orbit with a period less than that of a revolution around the star. For $E=0.384$, i.e. only slightly greater than the energy of the circular orbit, we find small oscillations, while increasing the energy to $E=0.4$ leads to large oscillations.

\begin{figure}[!h]
	\centering
	\begin{subfigure}{0.4\textwidth}
		\includegraphics[width=\textwidth]{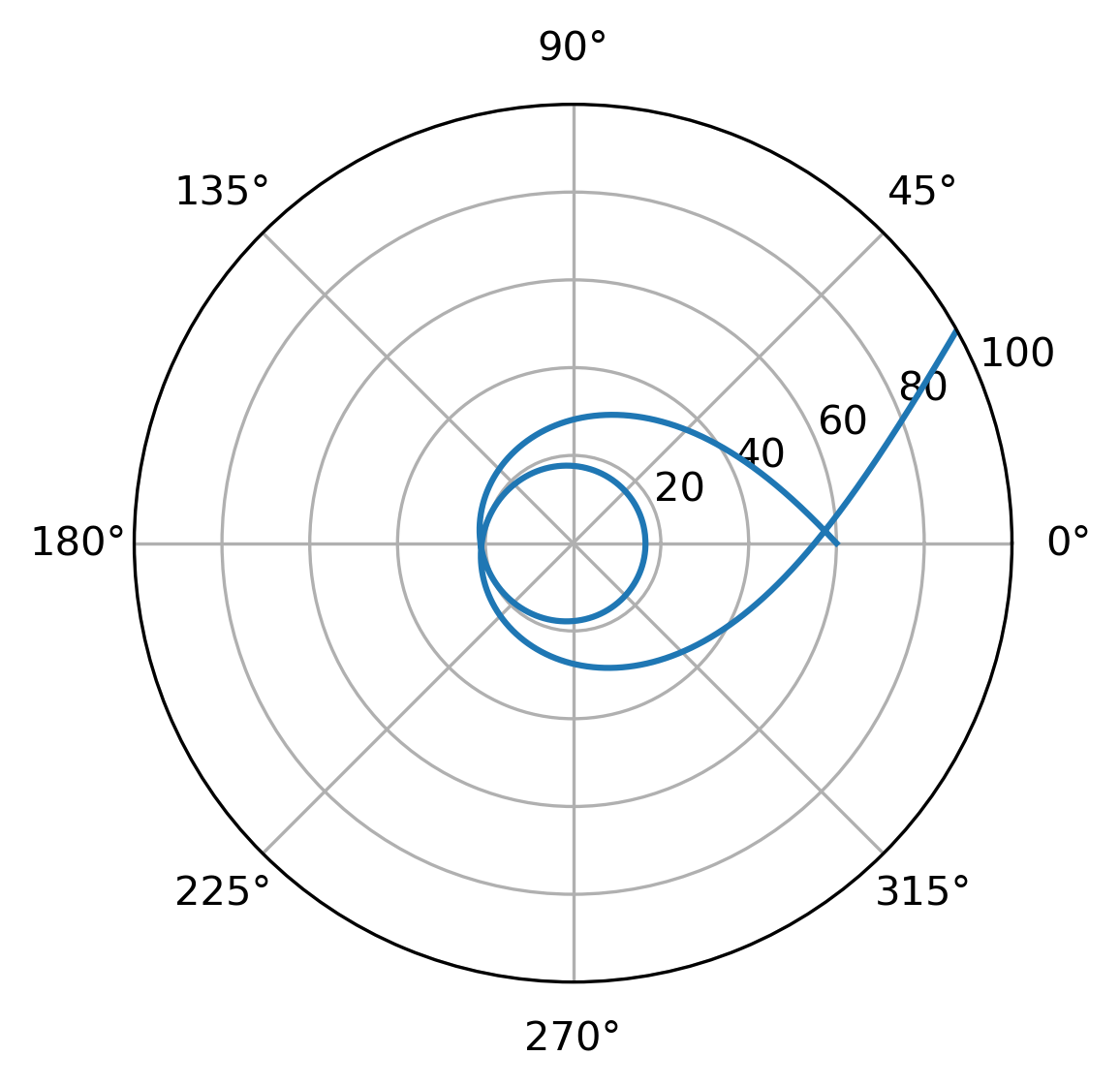}
		\caption{$L=1, E=0.384$}
		\label{fig:orb120_l50_1}
	\end{subfigure}
	\begin{subfigure}{0.4\textwidth}
		\includegraphics[width=\textwidth]{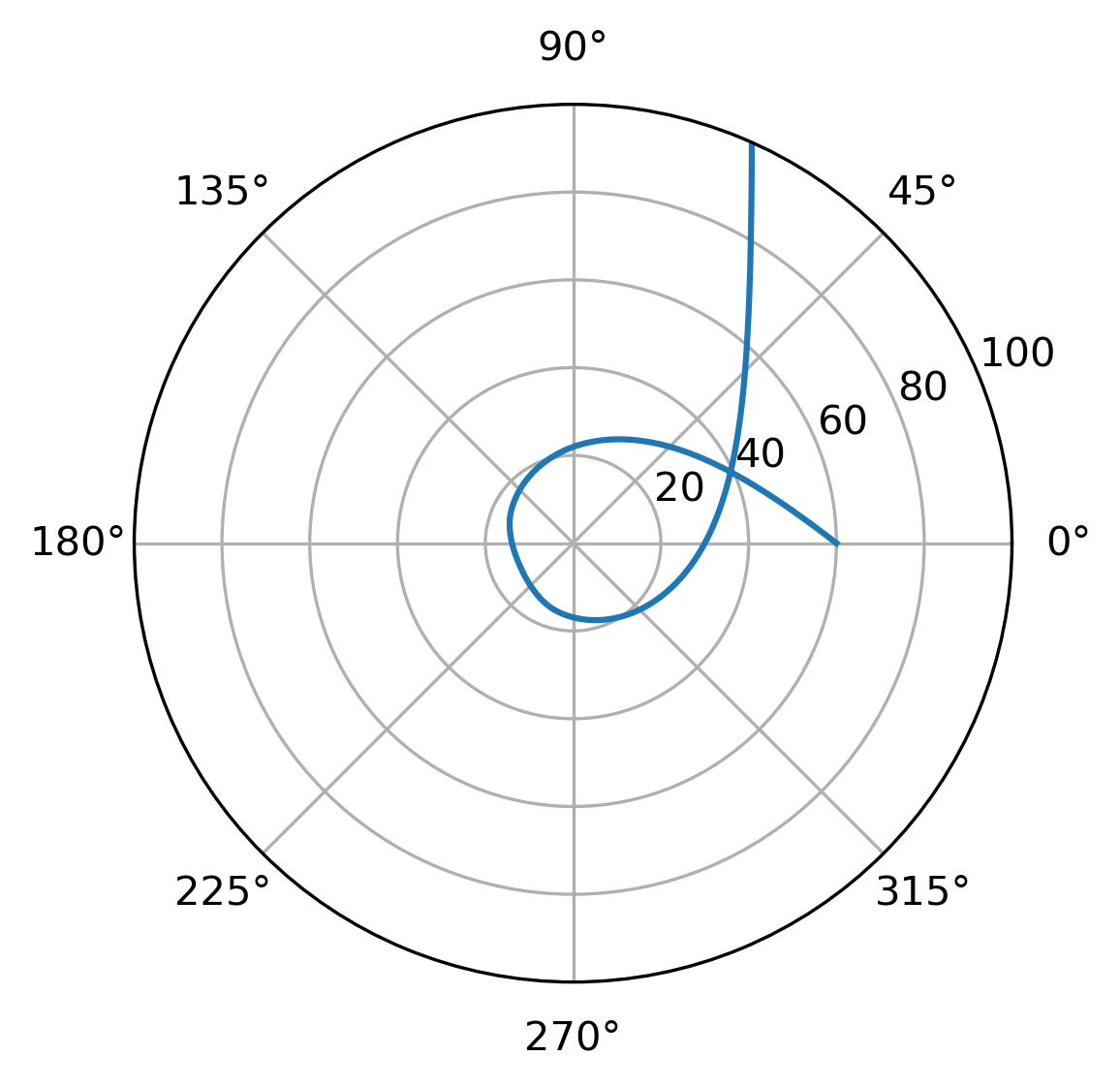}
		\caption{$L=1,E=0.4$}
		\label{fig:orb120_l50_2}
	\end{subfigure}
	\caption{Orbits for $Q=120$, $L=50$}
    \label{fig:orb120_l50}
\end{figure}

In this same space-time and for $L=50$, the larger charge $Q$ does still not allow circular motion, see Fig.~\ref{fig:orb120_l50}. However, as a particle approaches the origin we see a new behaviour - a single near-circular revolution can occur at $r\approx 20$, while motion within the region $r<20$ approaches a straight line. 

\subsubsection{Massive charged test particles}
\begin{figure}[!h]
	\centering
	\begin{subfigure}{0.4\textwidth}
		\includegraphics[width=\textwidth]{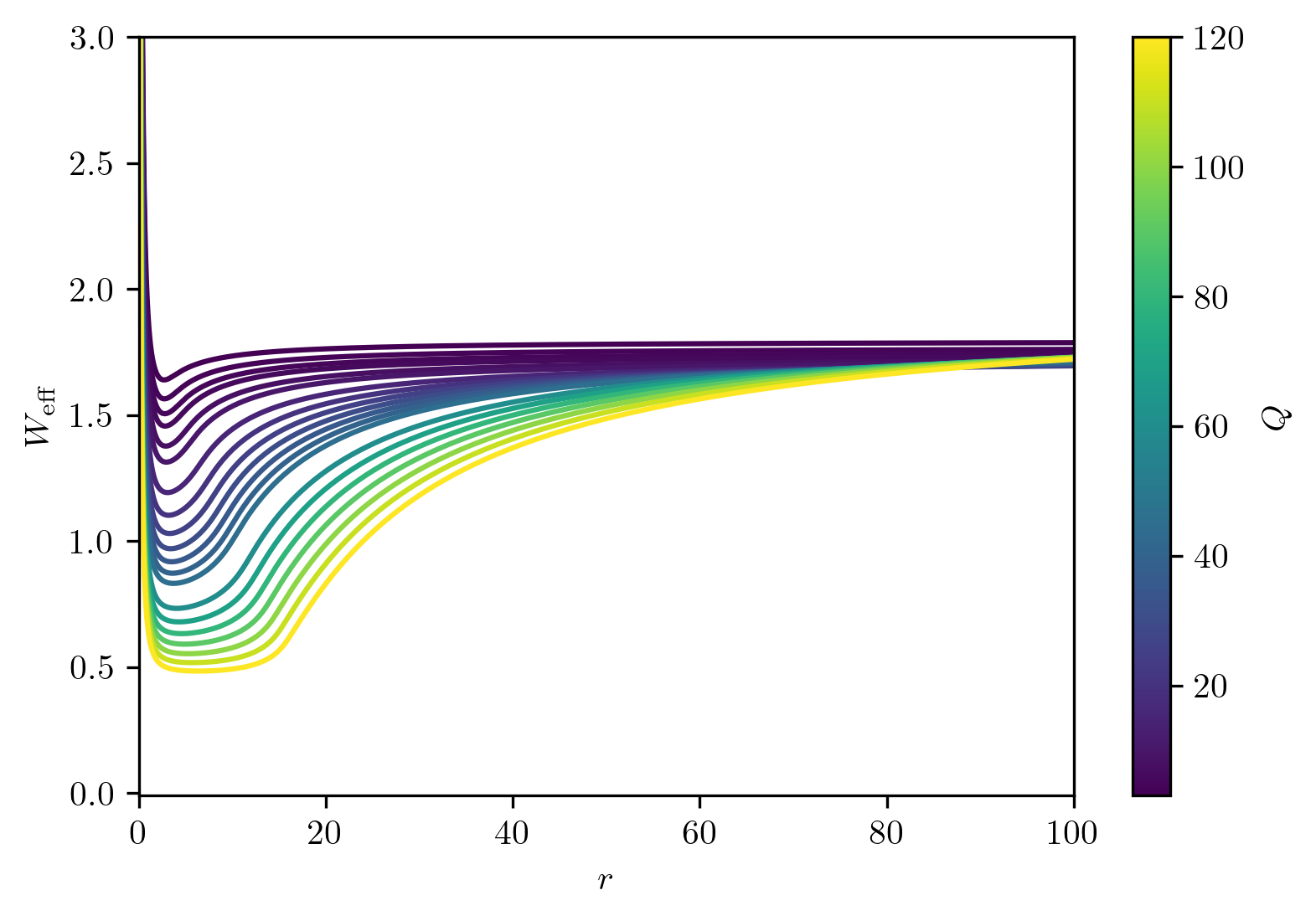}
        \caption{$L=1$}
	\end{subfigure}
    \begin{subfigure}{0.4\textwidth}
		\includegraphics[width=\textwidth]{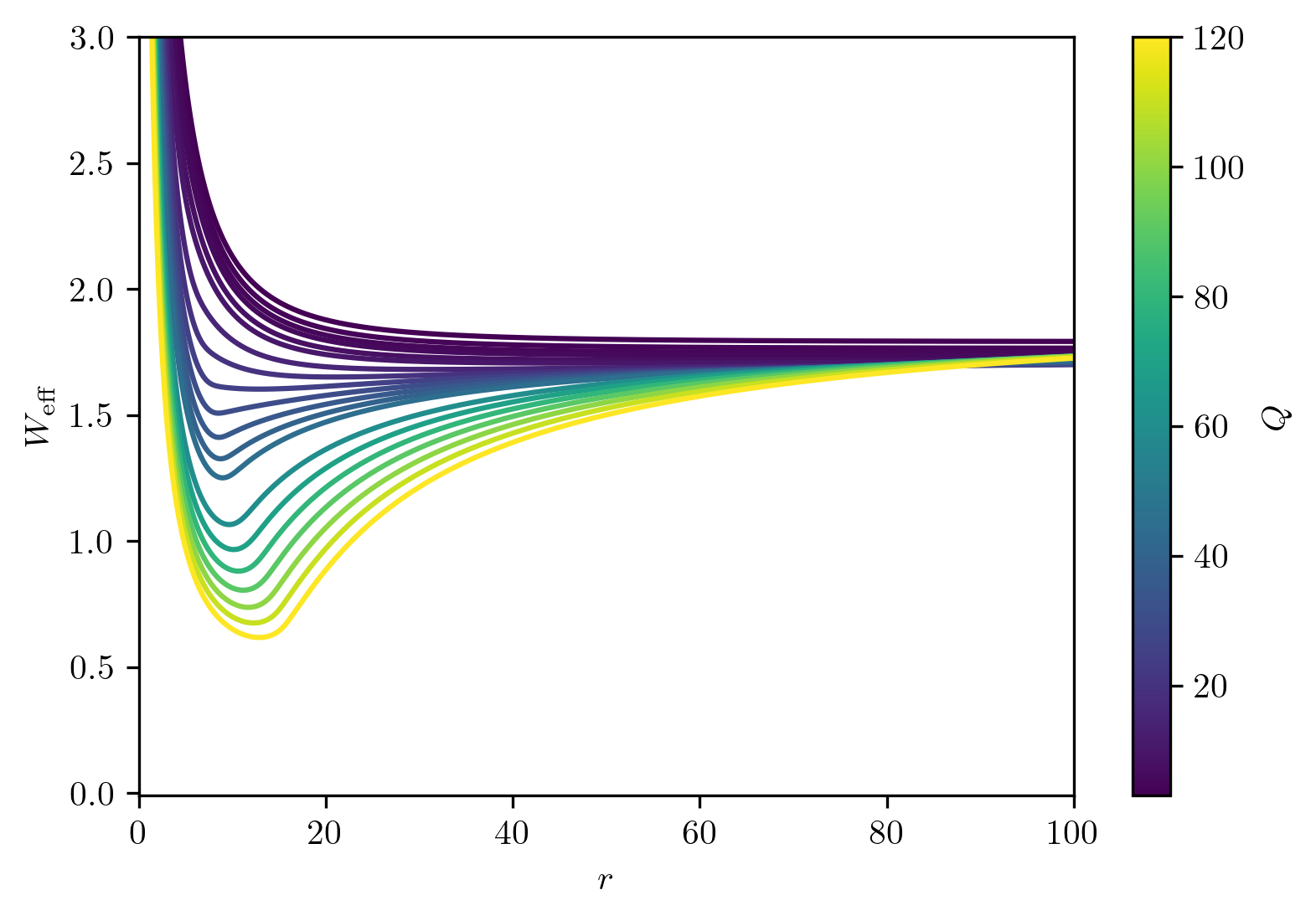}
        \caption{$L=10$}
	\end{subfigure}
    \caption{We show the modified effective potential $W_\text{eff}$ for test particles with charge $q=-0.1$ and angular momentum $L=1$ (left) and $L=10$ (right).}
    \label{fig:W_ch-0.1}
\end{figure}

From (\ref{eq:rdot}) we find the following condition for charged massive particles on circular orbits~:
\begin{equation}
	E+qV = \sqrt{\tilde{V}_\text{eff}} \ \ \Rightarrow \\
	\frac{d}{dr}\left(\sqrt{\tilde{V}_\text{eff}}-qV\right)=0 \ .
\end{equation}
We therefore define a modified effective potential 
\begin{equation} \label{eq:W_def}
	W_\text{eff}=\sqrt{\tilde{V}_\text{eff}}-qV
\end{equation}
so that circular orbits for charged particles occur at stationary points of $W_\text{eff}$.
Fig. \ref{fig:W_ch-0.1} shows $W_\text{eff}(r)$ for test particles with negative charge $q=-0.1$, for two values of the angular momentum $L$. In both cases we see that particles with non-zero angular momentum cannot approach the origin, as would be expected. However, while for $L=1$ we see that all solutions allow a global minimum below $r=20$, representing a stable orbit, for $L=10$ this only holds true for solutions with sufficiently large $Q$. Instead, for solutions with small $Q$, the particles will always escape to infinity. 

\begin{figure}[!h]
	\centering
	\begin{subfigure}{0.4\textwidth}
		\includegraphics[width=\textwidth]{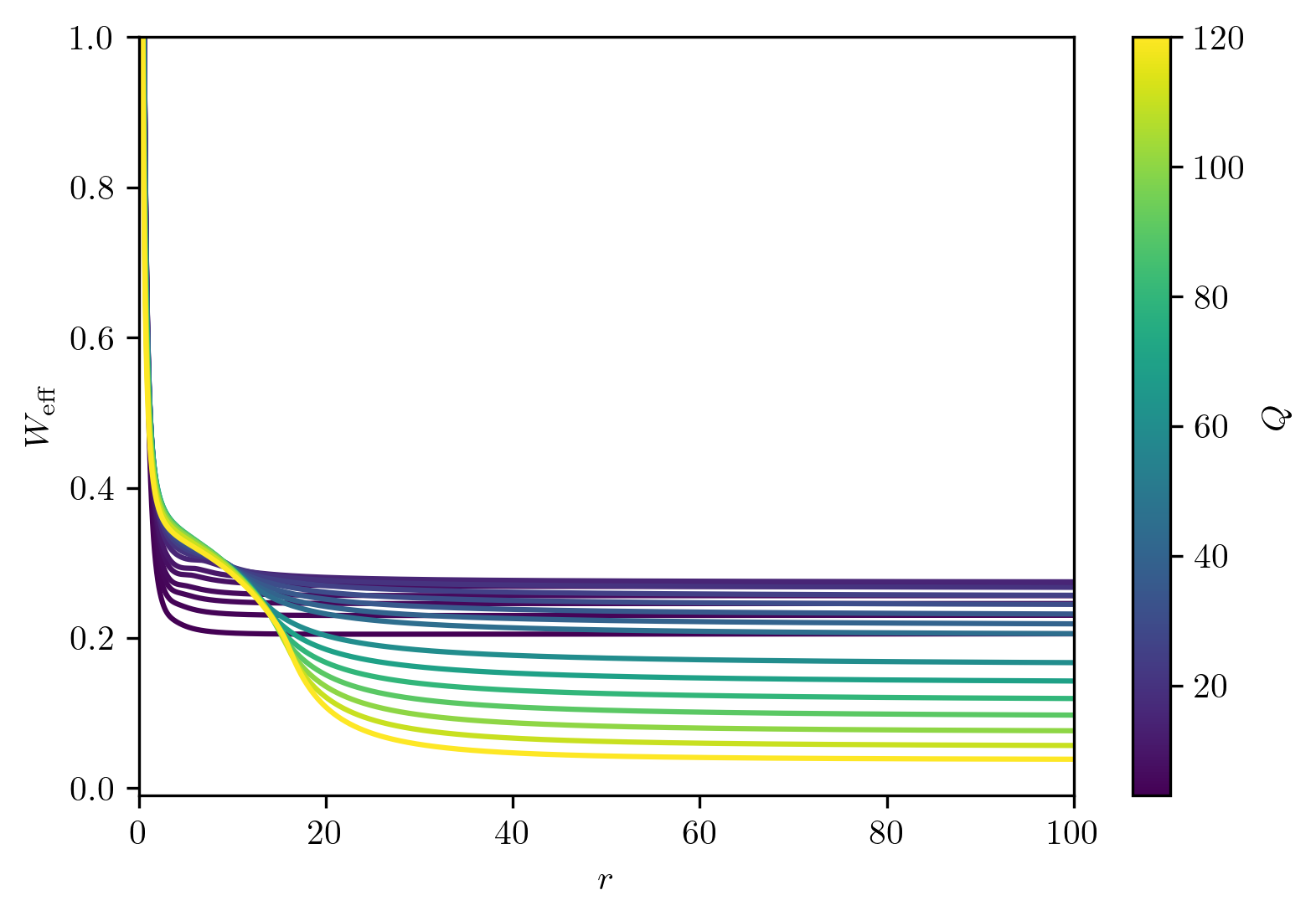}
        \caption{$L=1$}
	\end{subfigure}
    \begin{subfigure}{0.4\textwidth}
		\includegraphics[width=\textwidth]{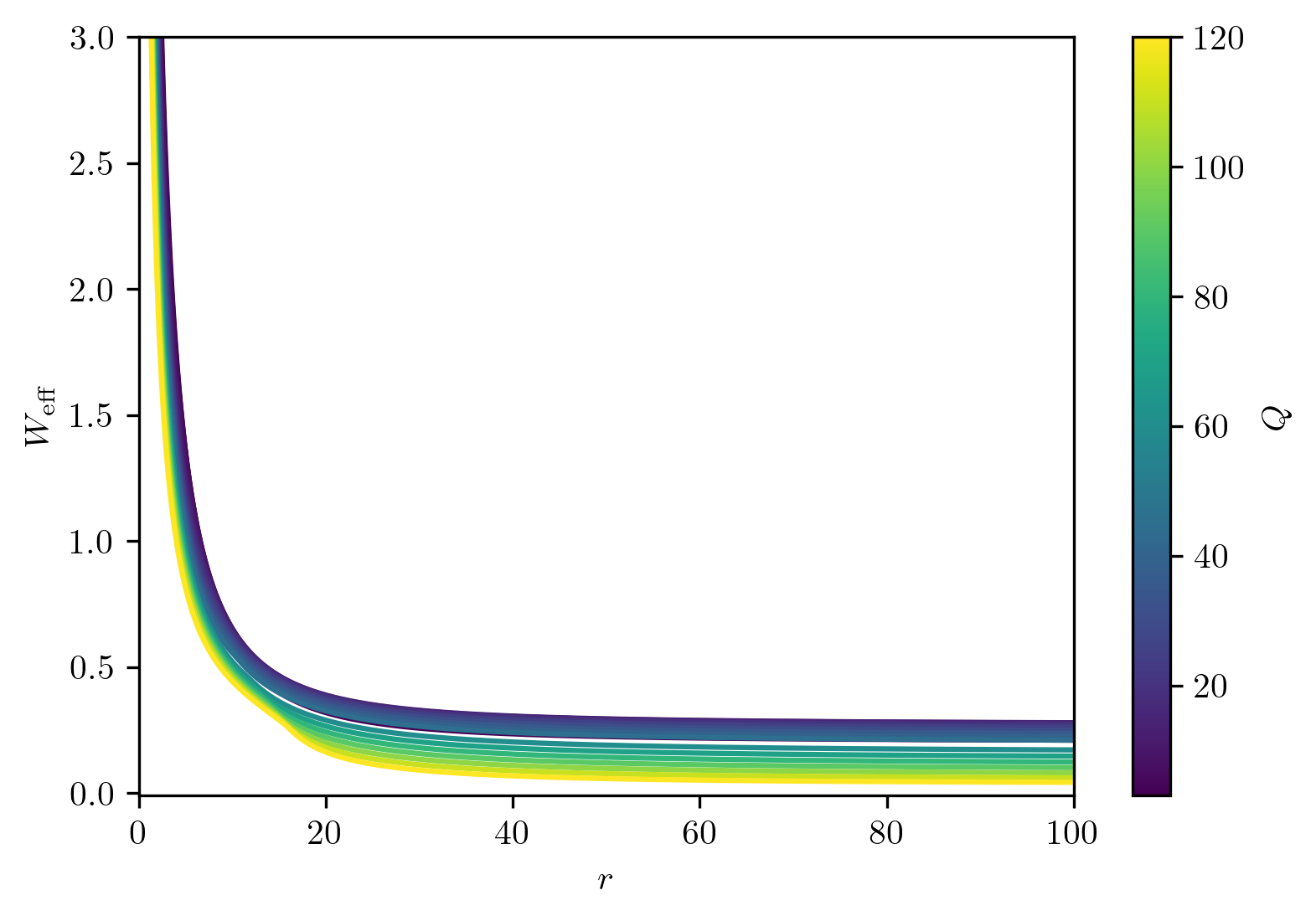}
        \caption{$L=10$}
	\end{subfigure}
    \caption{We show the modified effective potential $W_\text{eff}$ for test particles with charge $q=+0.1$ and angular momentum $L=1$ (left) and $L=10$ (right).}
    \label{fig:W_ch+0.1}
\end{figure}

Fig. \ref{fig:W_ch+0.1} shows $W_\text{eff}(r)$ for test particles with positive charge $q=0.1$, again for $L=1$ and $L=10.$ Here, the repulsive electromagnetic force between the boson star and the particle significantly restricts the number of solutions which permit circular orbits, with no orbits for $L=10$ and only solutions of low $Q$ ($Q\leq 6$) permitting circular orbits for $L=1$.

\begin{figure}[!h]
	\centering
	\begin{subfigure}{0.4\textwidth}
		\includegraphics[width=\textwidth]{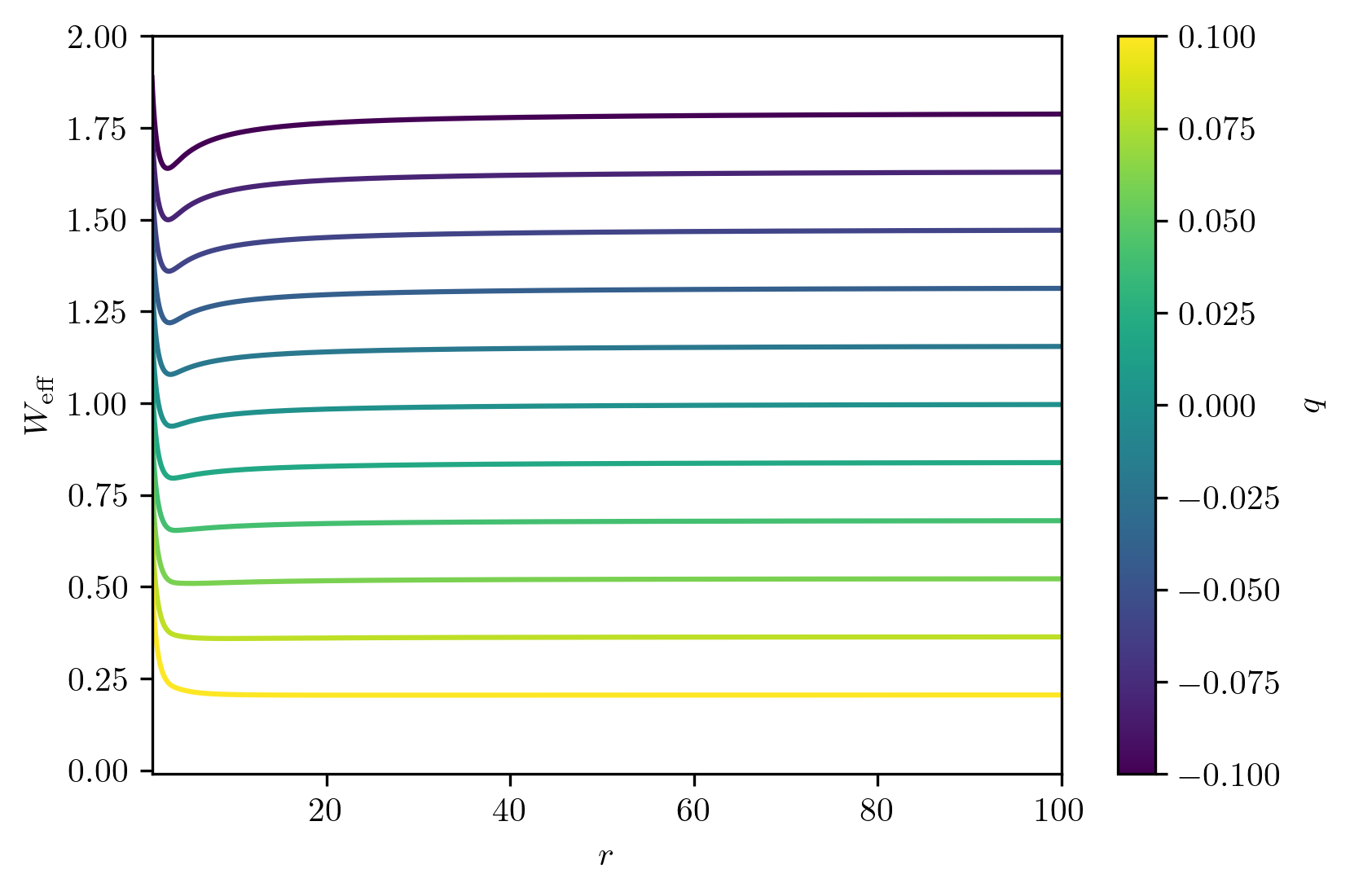}
        \caption{$L=1$}
	\end{subfigure}
    \begin{subfigure}{0.4\textwidth}
		\includegraphics[width=\textwidth]{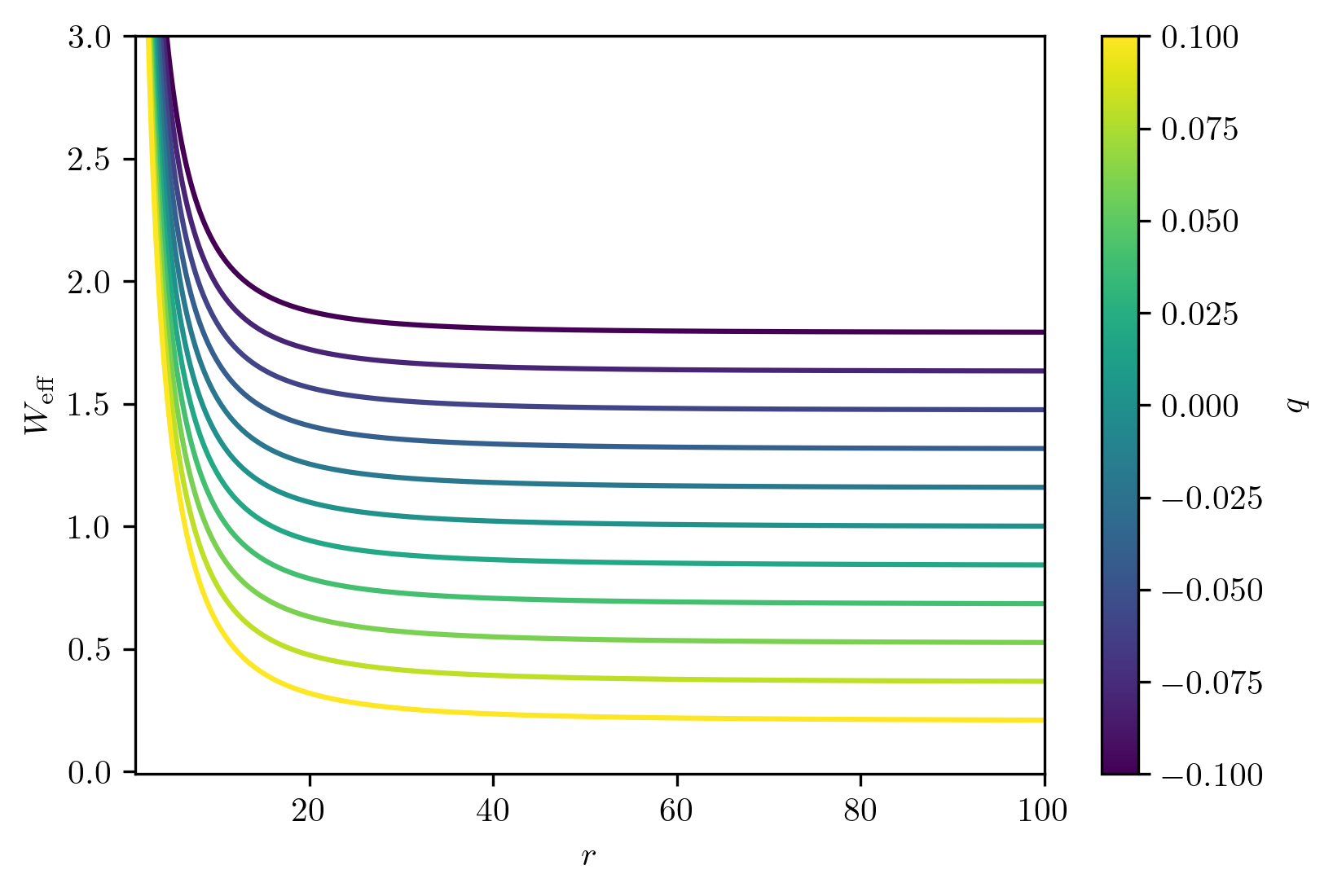}
        \caption{$L=10$}
	\end{subfigure}
    \caption{We show the modified effective potential $W_\text{eff}$ for test particles in the space-time with $Q=3$ and with angular momentum $L=1$ (left) and $L=10$ (right).}
    \label{fig:W_Q3}
\end{figure}

Fixing the charge of the boson star to $Q=3$, we investigate particles with charge $q$ in the range $[-0.1,+0.1]$. Our results are shown in Fig.~\ref{fig:W_Q3} where we give the corresponding $W_\text{eff}(r)$.

\begin{figure}[!h]
	\centering
	\begin{subfigure}{0.4\textwidth}
		\includegraphics[width=\textwidth]{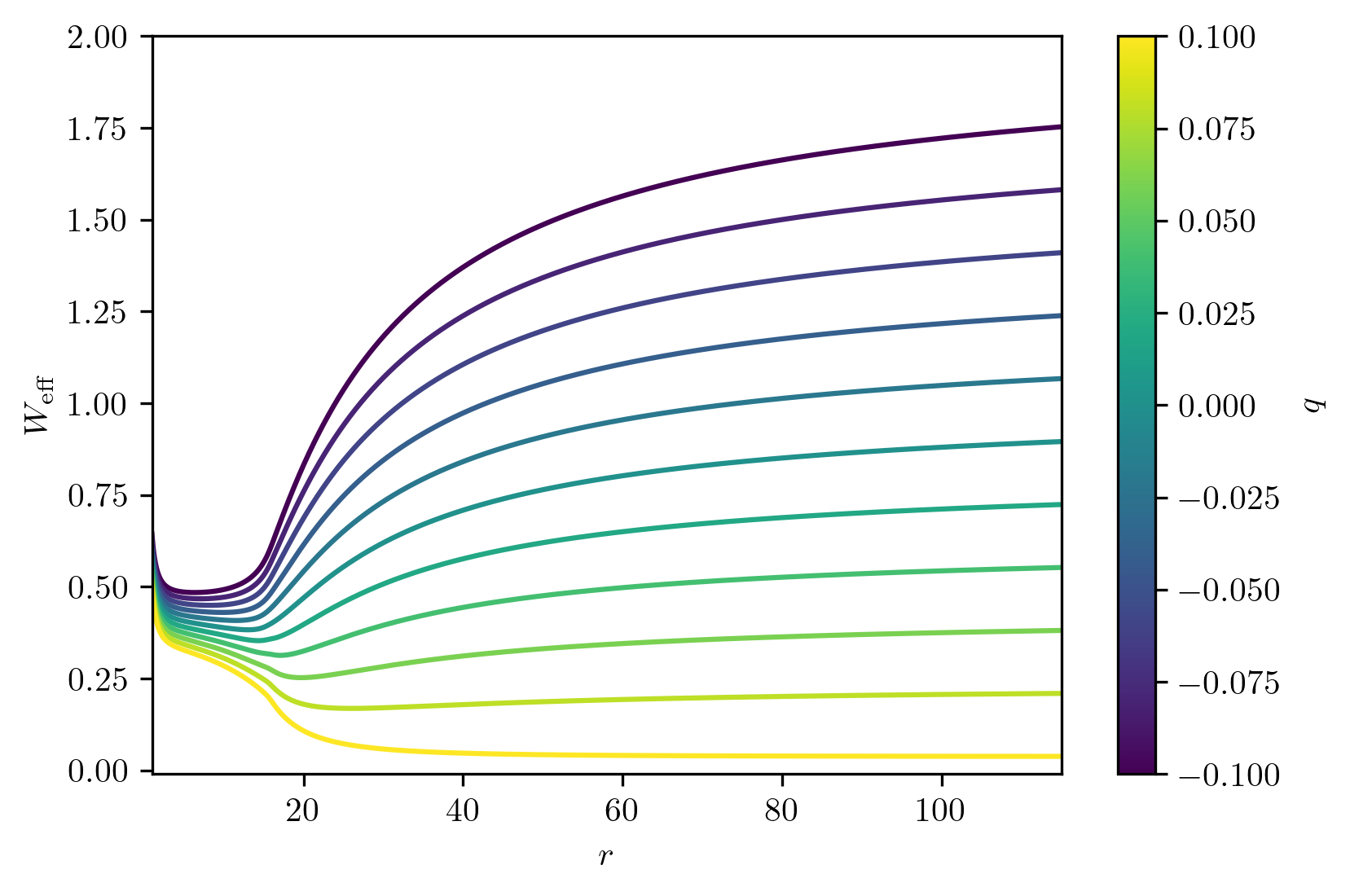}
        \caption{$L=1$}
	\end{subfigure}
    \begin{subfigure}{0.4\textwidth}
		\includegraphics[width=\textwidth]{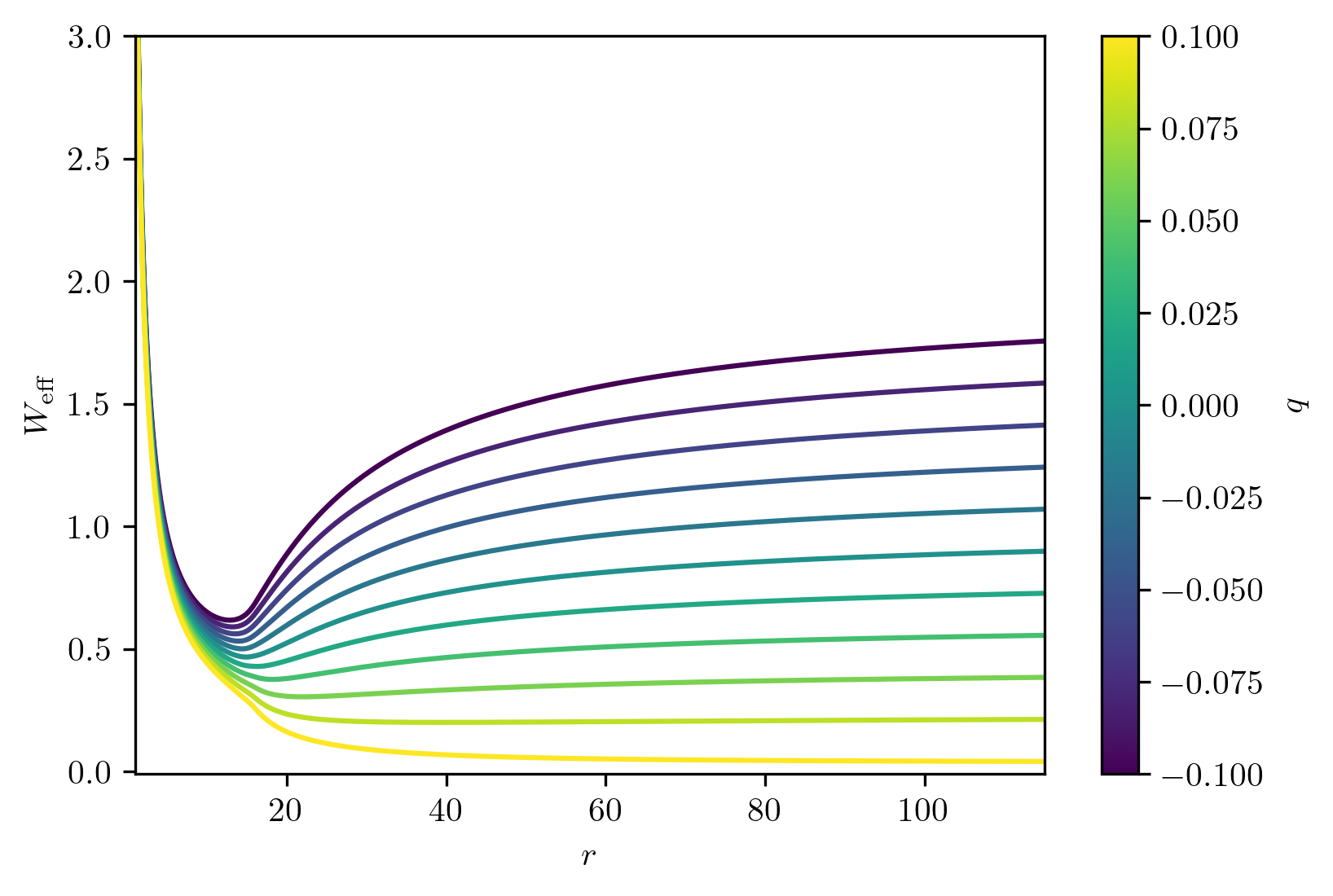}
        \caption{$L=10$}
	\end{subfigure}
    \caption{We show the modified effective potential $W_\text{eff}$ for test particles in the space-time with $Q=120$ and with angular momentum $L=1$ (left) and $L=10$ (right).}
    \label{fig:W_Q120}
\end{figure}

The corresponding case for $Q=120$ is given in Fig.~\ref{fig:W_Q120}. The broad, shallow nature of the stationary points of the effective potential for $L=1$ permits the oscillations around the stationary point shown in Fig.~\ref{fig:orb120_l1}, while increasing $L$ we find the minimum to narrow, see the effective potential for $L=10$.
In both cases, we find that there are no stationary points for $q=0.1$ and the potential is monotonically decreasing, indicating that particles will escape to infinity and circular motion is not possible.

\begin{figure}[!h]
	\centering
	\begin{subfigure}{0.4\textwidth}
		\includegraphics[width=\textwidth]{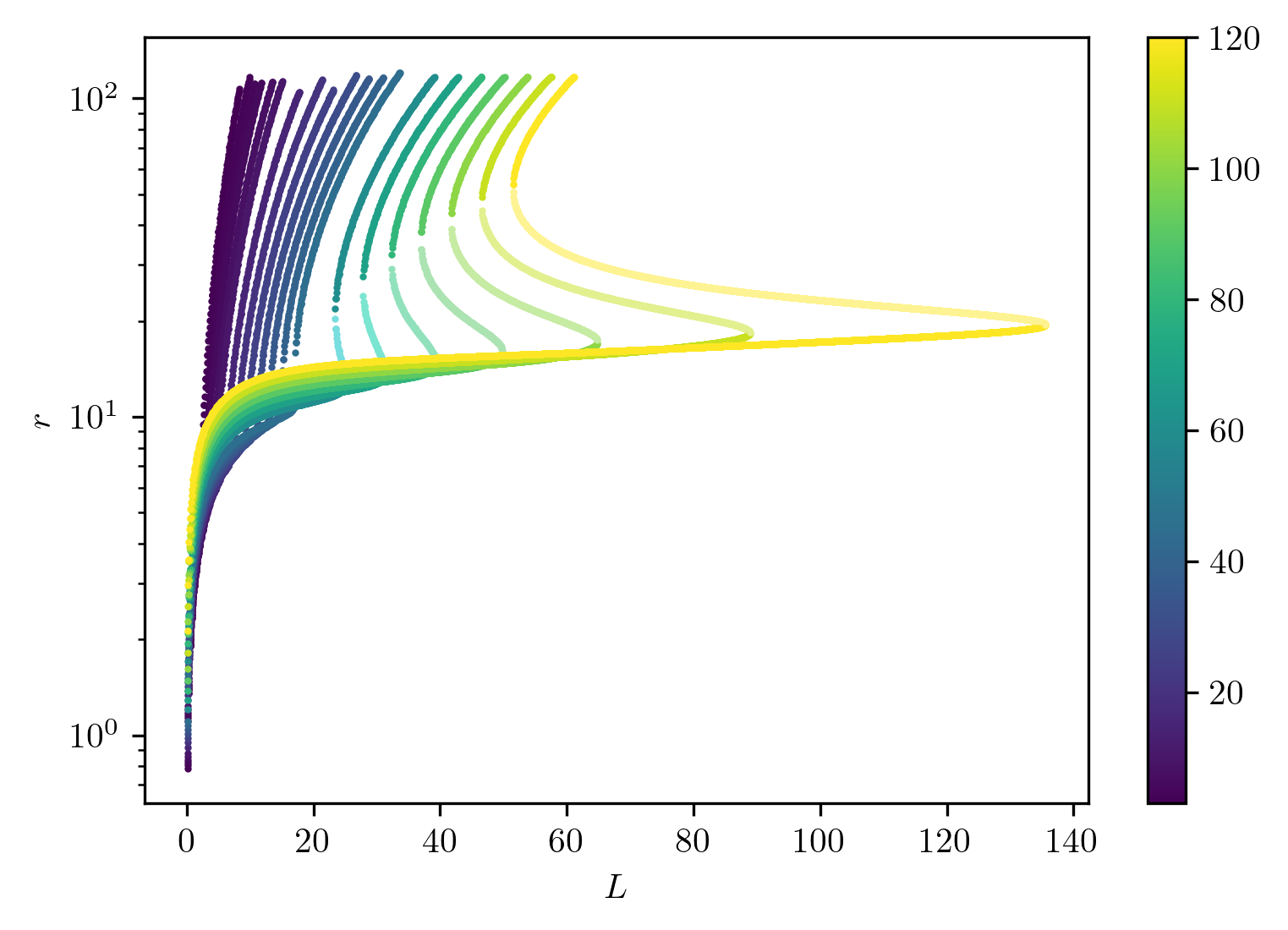}
        \caption{$q=-0.1$}
	\end{subfigure}
    \begin{subfigure}{0.4\textwidth}
		\includegraphics[width=\textwidth]{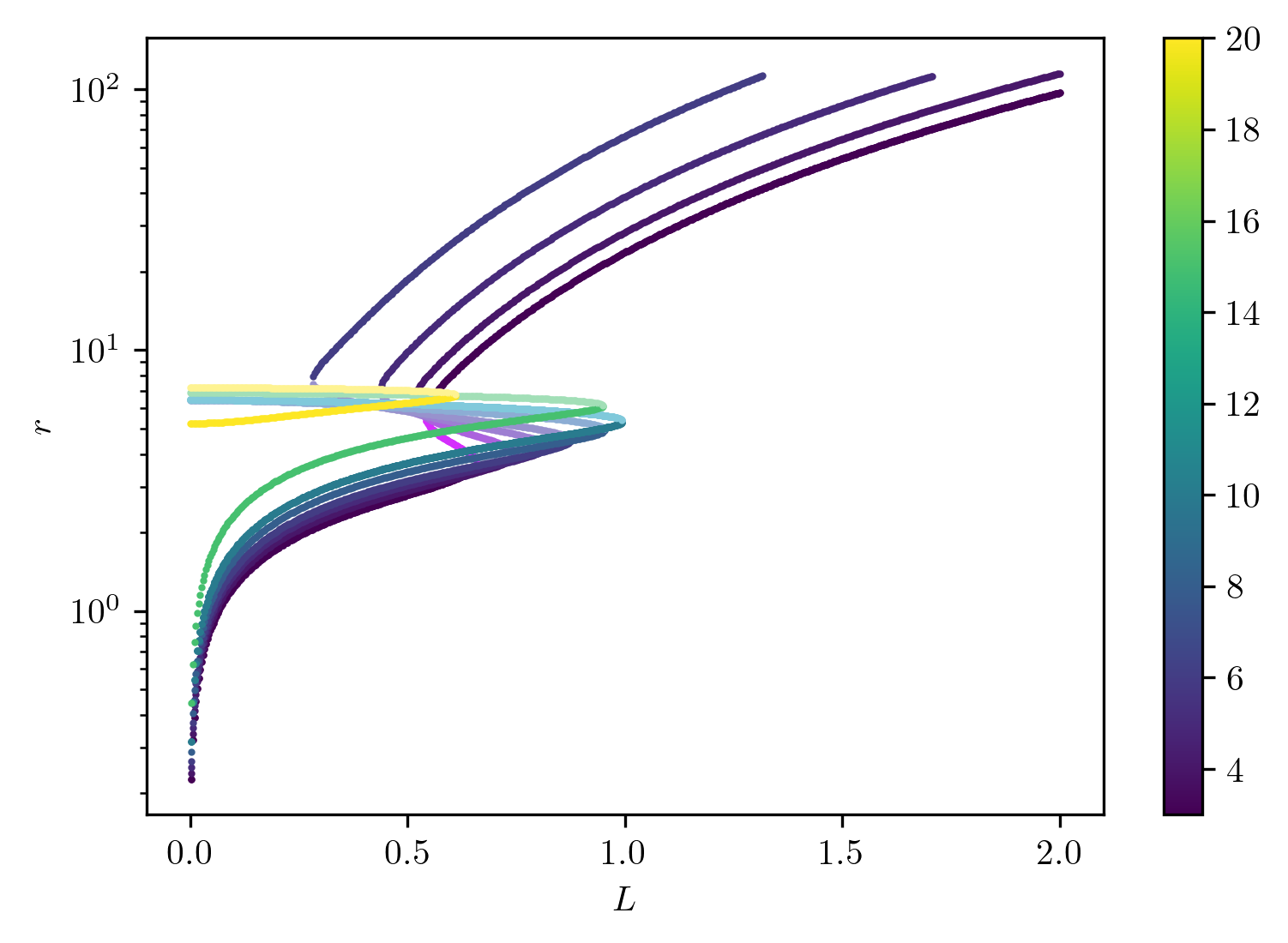}
        \caption{$q=+0.1$}
	\end{subfigure}
    \caption{We show the radius $r$ of circular orbits against the angular momentum $L$ for charged boson star space-times with varying $Q$, for particles with charge $q=\pm0.1$. Note the different scales for $Q$; for the $q=+0.1$ case and $Q>20$ no circular orbits exist. Compare this with Fig.~\ref{fig:st_pts_Q_range} where $q=0$.}
    \label{fig:st_pts_Q_cg}
\end{figure}

We will now discuss the dependence of the radius of the circular orbits on the charges.  Fig.~\ref{fig:st_pts_Q_cg} shows the radius $r$ of circular orbits in function of the angular momentum $L$ for solutions with varying $Q$, for particles with charge $q=\pm0.1$. Note the different scales for $Q$. For the $q=+0.1$ case, circular orbits are not possible when $Q>20$. Compare this with 
Fig.~\ref{fig:st_pts_Q_range} where $q=0$.

\begin{figure}[!h]
	\centering
	\begin{subfigure}{0.4\textwidth}
		\includegraphics[width=\textwidth]{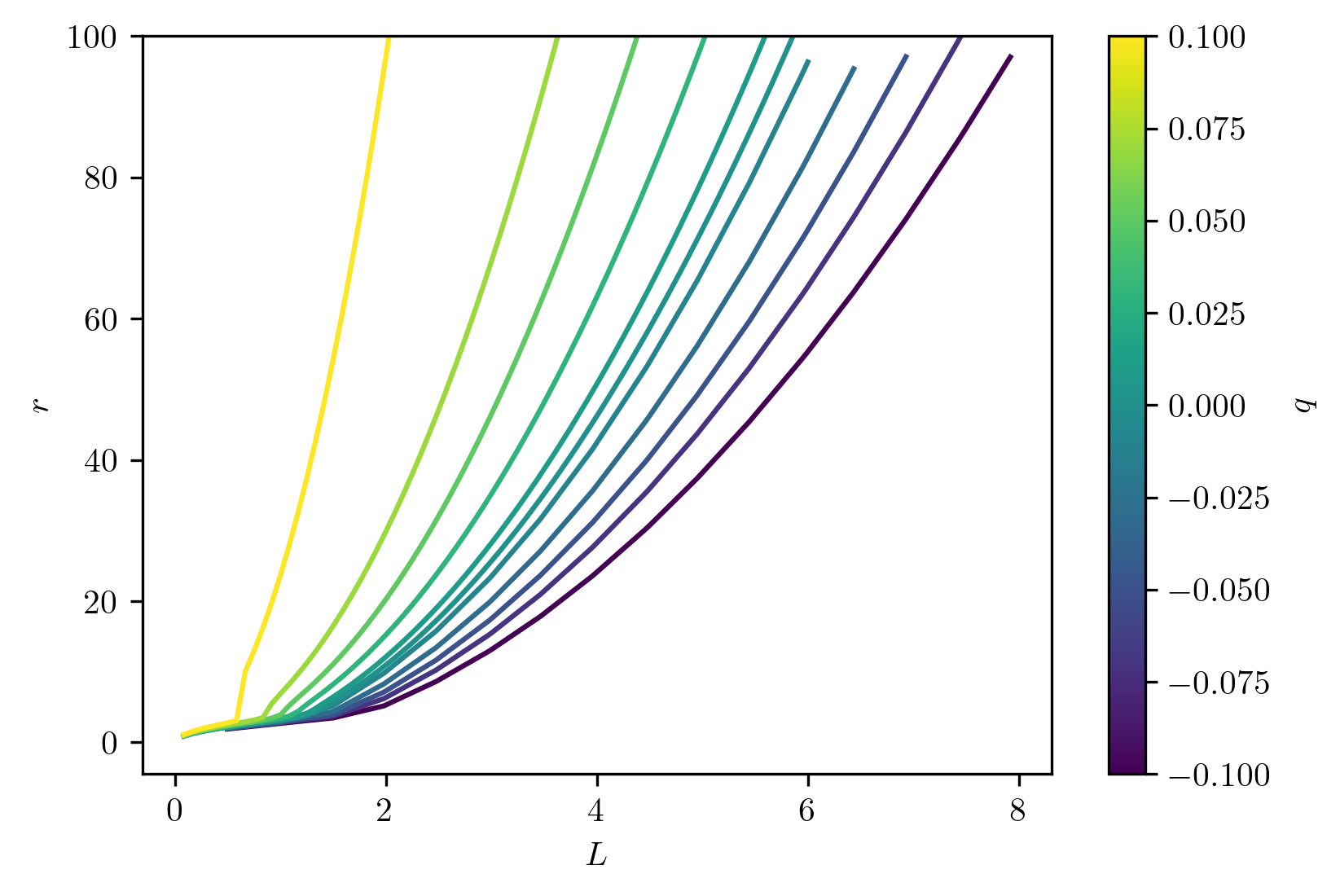}
        \caption{$Q=3$}
	\end{subfigure}
    \begin{subfigure}{0.4\textwidth}
		\includegraphics[width=\textwidth]{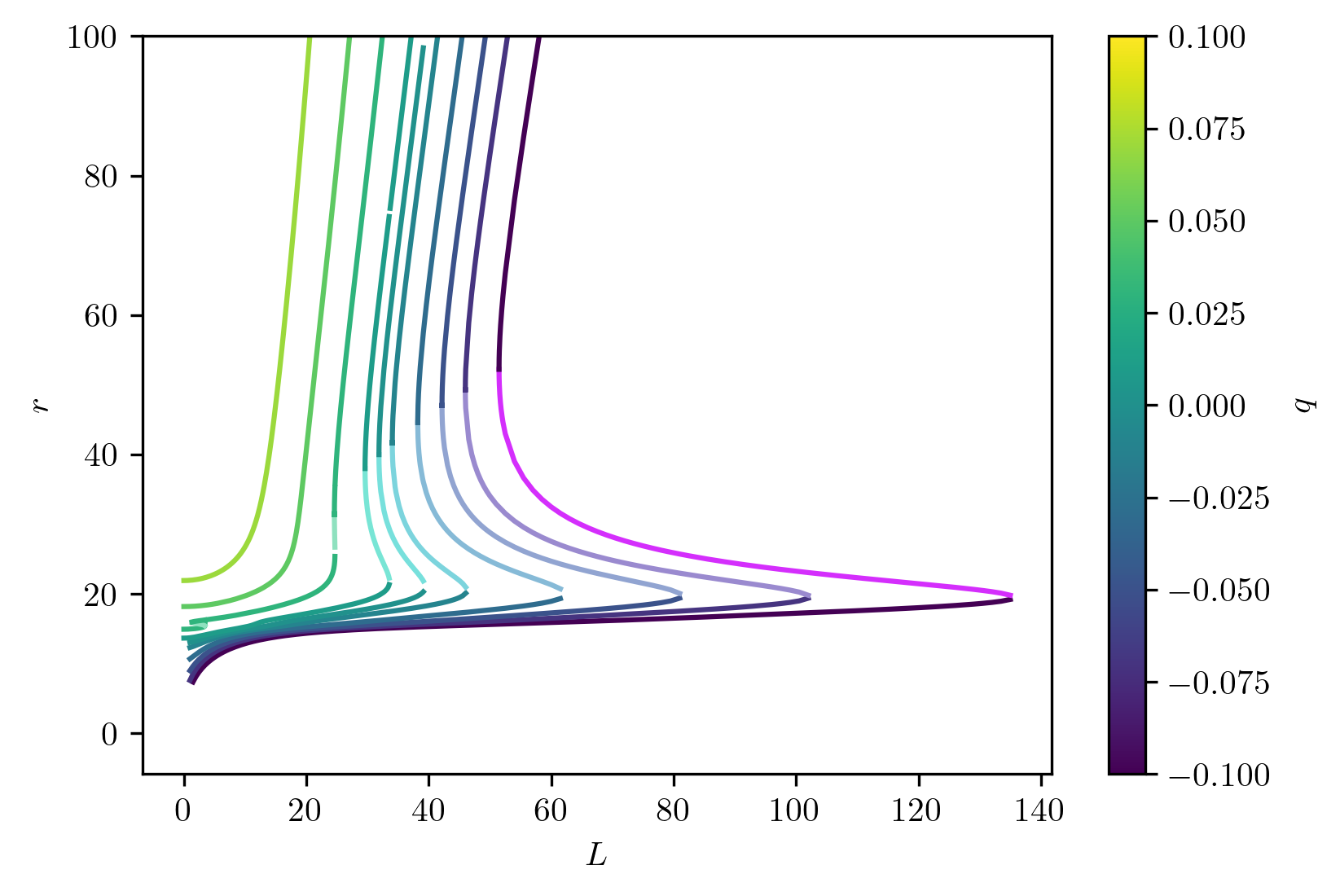}
        \caption{$Q=120$}
	\end{subfigure}
    \caption{We show the radius $r$ of circular orbits against the angular momentum $L$ for charged boson star space-times with $Q=3$ and $Q=120$, for particles with varying charge $q$}
    \label{fig:st_pts_Q_cg_overq}
\end{figure}

Fig.~\ref{fig:st_pts_Q_cg_overq} shows the radius of the circular orbits $r$ against the angular momentum $L$ in the space-time of charged boson stars with $Q=3$ and $Q=120$, respectively, for particles with varying charge $q$. Note that for $Q=120$, $q=0.1$ no circular orbits exist which matches the results in Fig.~\ref{fig:st_pts_Q_cg}.

\begin{figure}[!h]
	\centering
	\begin{subfigure}{0.4\textwidth}
		\includegraphics[width=\textwidth]{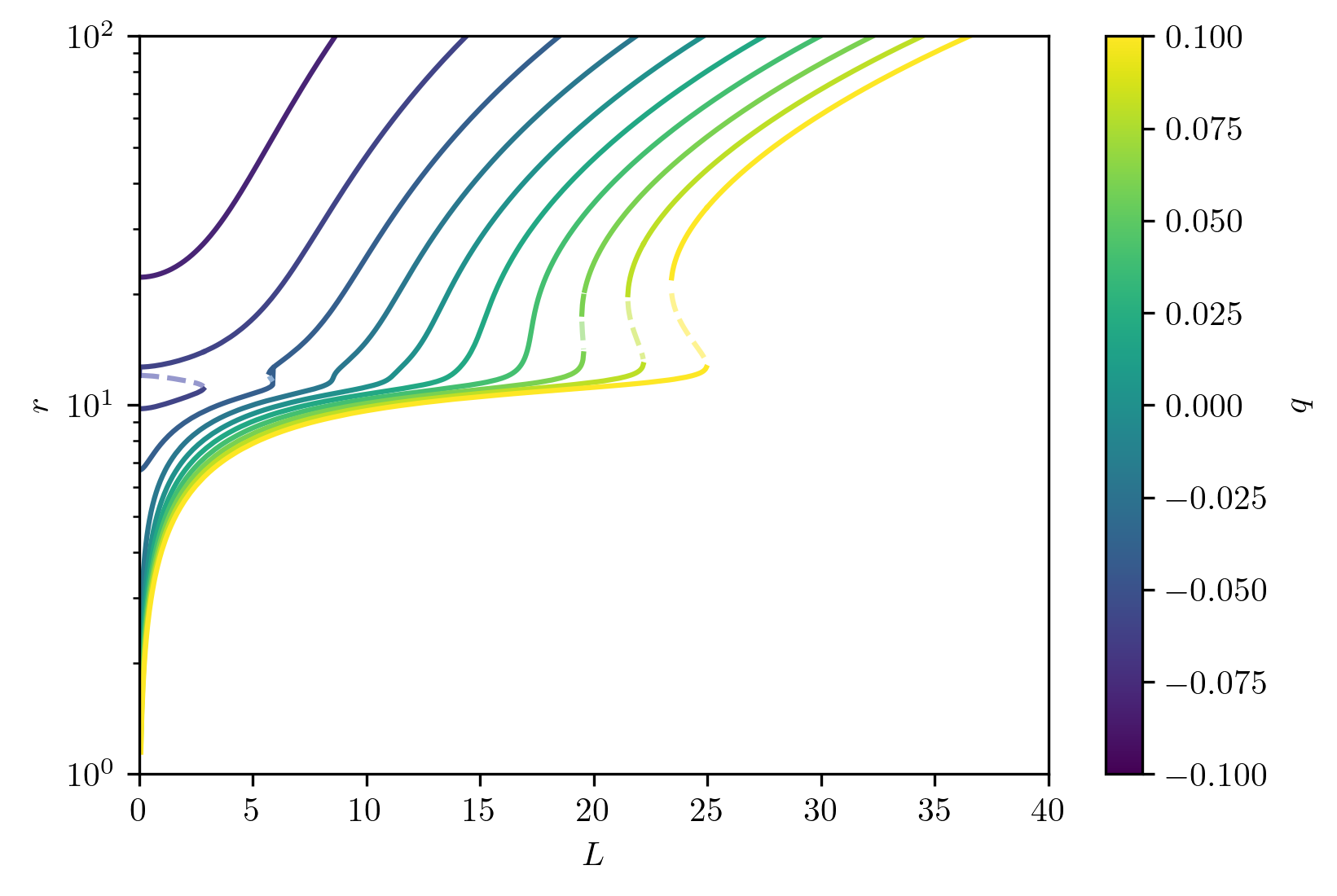}
	\end{subfigure}
    \begin{subfigure}{0.4\textwidth}
		\includegraphics[width=\textwidth]{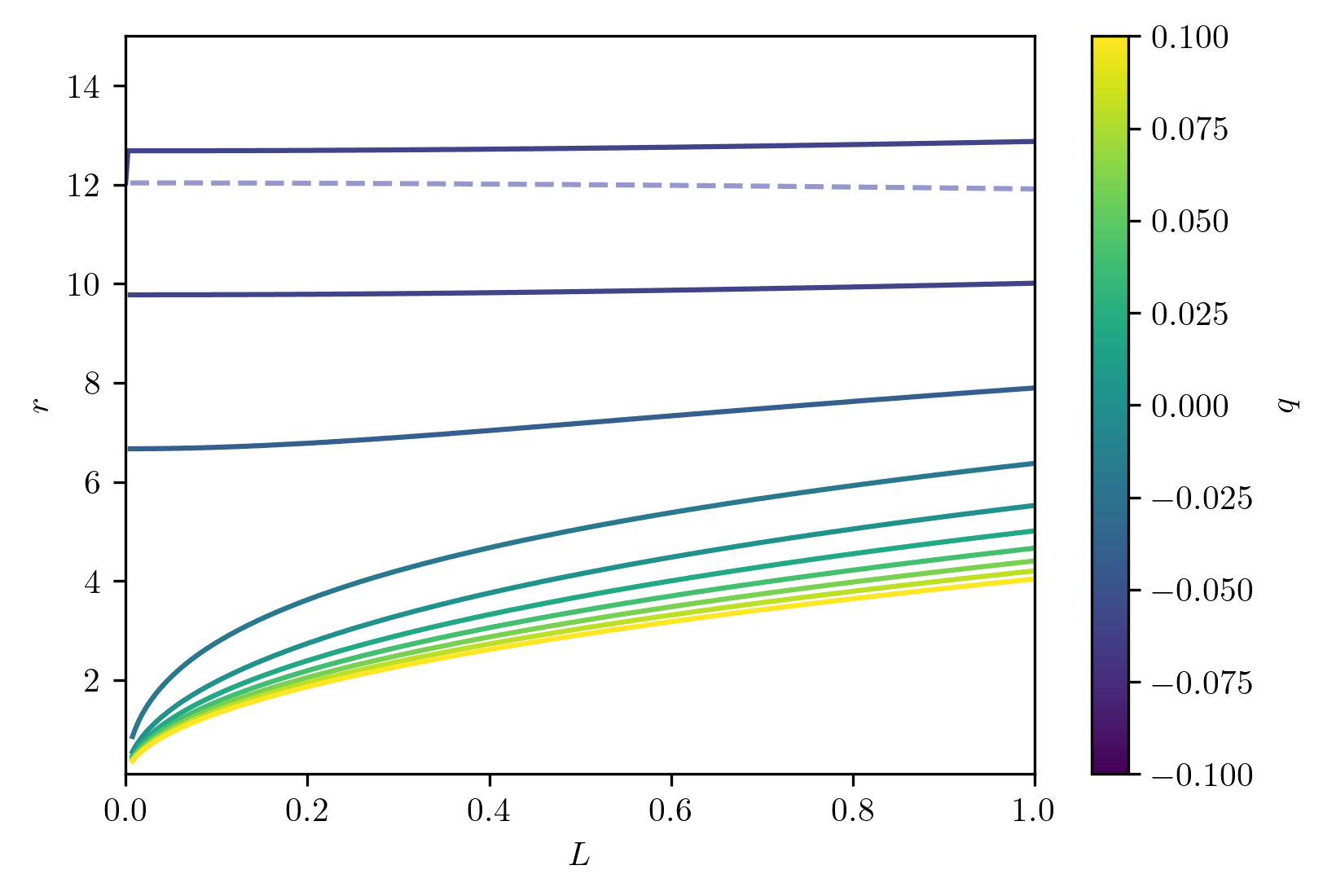}
	\end{subfigure}
    \caption{We show the radius of circular orbits $r$ against the angular momentum $L$ for charged boson star space-times with $Q=60$, for particles with varying charge $q$. Note that the dashed lines indicate that the circular orbits are unstable.}
    \label{fig:st_pts_Q60_cg_overq}
\end{figure}

Fig.~\ref{fig:st_pts_Q60_cg_overq} shows the radius of circular orbits in the space-time of the boson star solution with $Q=60$, for varying particle charge $q$. We now find a richer pattern of circular orbits, some of which are unstable (dashed lines). For $q<-0.06$, no circular orbits are observed. For $q=-0.6$, an additional pair of solutions appears for small $L$, with three static orbits - two stable at $r=9.78$ and $r=12.69$, and one unstable at $r=12.04$. 

\subsubsection{Massless test particles}\label{sec:boson_massless_orbits}
For boson stars, as would be expected, circular orbits for light exist only in limiting cases \cite{grandclement}. We find that for charged boson stars, these appear only in the frozne star limit, which we will discuss below. However, the presence of the boson star does have significant effects on the null geodesics.\\
For massless particles we find for (\ref{eq:S_solution}) that the energy-independent potential is of the form
\begin{equation}
    \tilde{V}_\text{eff}=\frac{\sigma^2NL^2}{r^2} \ .
\end{equation}
Note that if we rescale $E\to\lambda E$ and $L\to\lambda L$, we obtain $\dot{r}^2\to\lambda^2\dot{r}^2$.
However, this is simply a reparameterisation of the affine parameter $\tau$, $\tau\to\frac{\tau}{\lambda}$, and so does not change the geodesics themselves, and only the ratio $L/E$ has physical meaning. This is the impact parameter. In this section, therefore, $\tilde{V}$ will be shown with $L=1$ (other values would simply be a rescaling), and plots of geodesics will be presented in terms of $L/E$.\\

\begin{figure}[!h]
	\centering
	\includegraphics[width=0.5\linewidth]{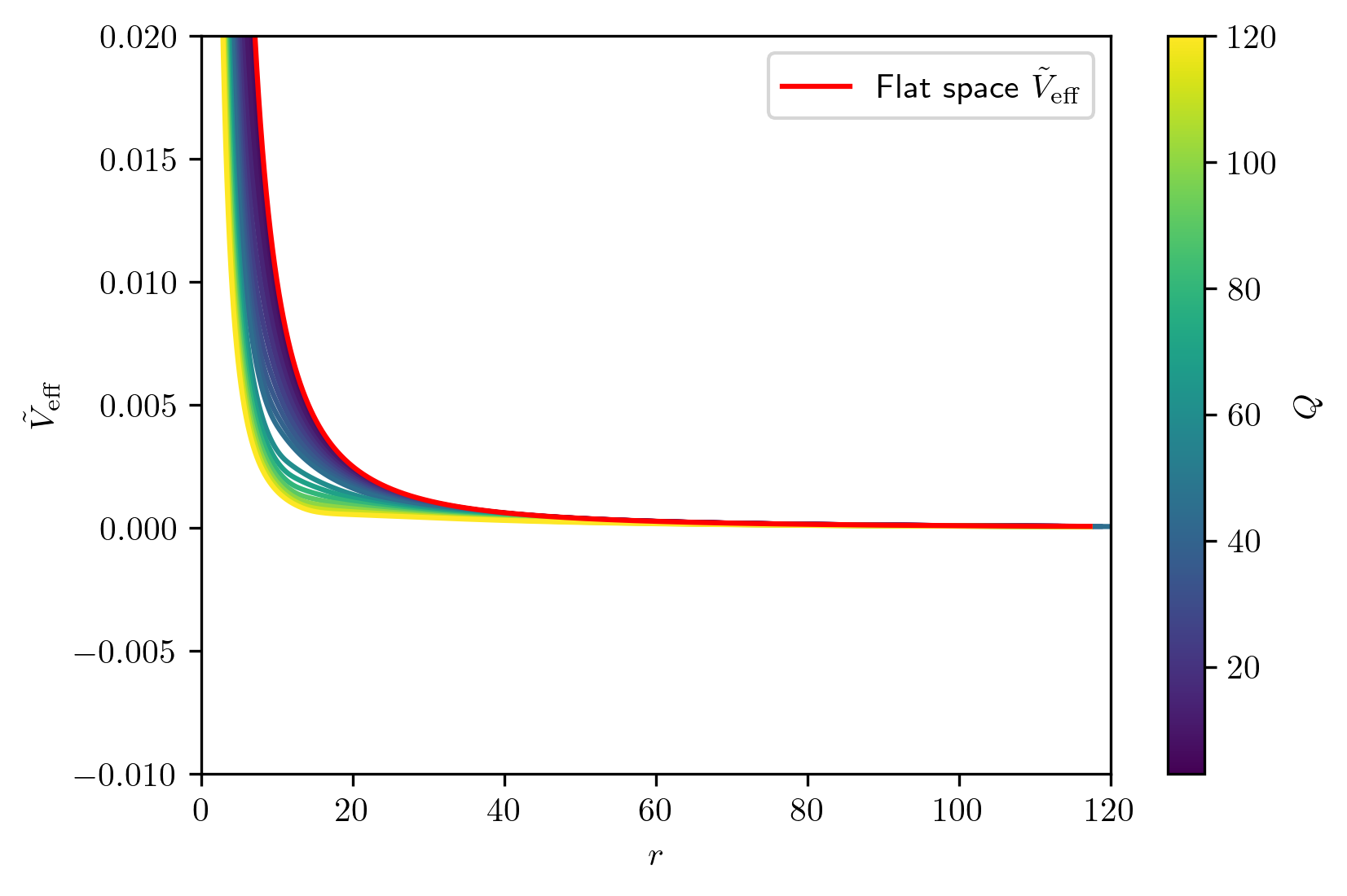}
	\caption{We show the effective potential $\tilde{V}_\text{eff}$ for a particle with $L=1$ in the space-time with varying $Q$ and in flat space (in red).}
	\label{fig:pot_Q120_L1}
\end{figure}

Fig.~\ref{fig:pot_Q120_L1} shows the effective potential $\tilde{V}_\text{eff}$ for a particle with $L=1$ in the boson star space-times for different values of $Q$ well below the frozem state limit and in flat space-time. It can be seen that for large $r$, the effective potentials coincide, but for smaller $r$, the effective potential for boson star space-times has smaller values than that for flat space-time. This means that a massless test particle can approach the $r=0$ axis more closely. As $Q\to3$, the effective potential approaches that for flat space-time.

\begin{figure}[!h]
	\centering
	
	\begin{subfigure}{0.4\textwidth}
		\includegraphics[width=\textwidth]{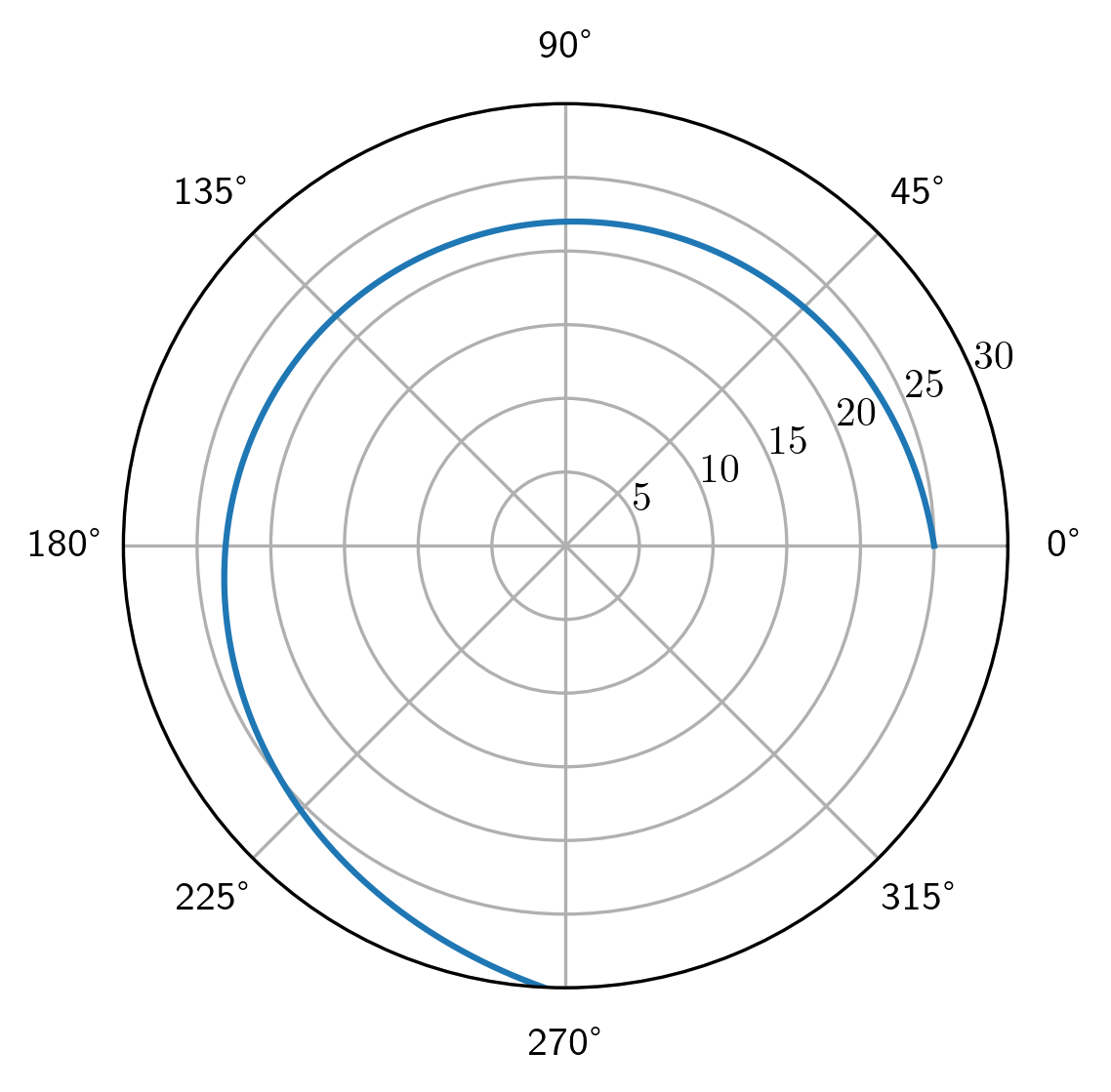}
		\caption{$L/E=43.5$}
		\label{fig:orb_120_lr_0.023}
	\end{subfigure}
	\begin{subfigure}{0.4\textwidth}
		\includegraphics[width=\textwidth]{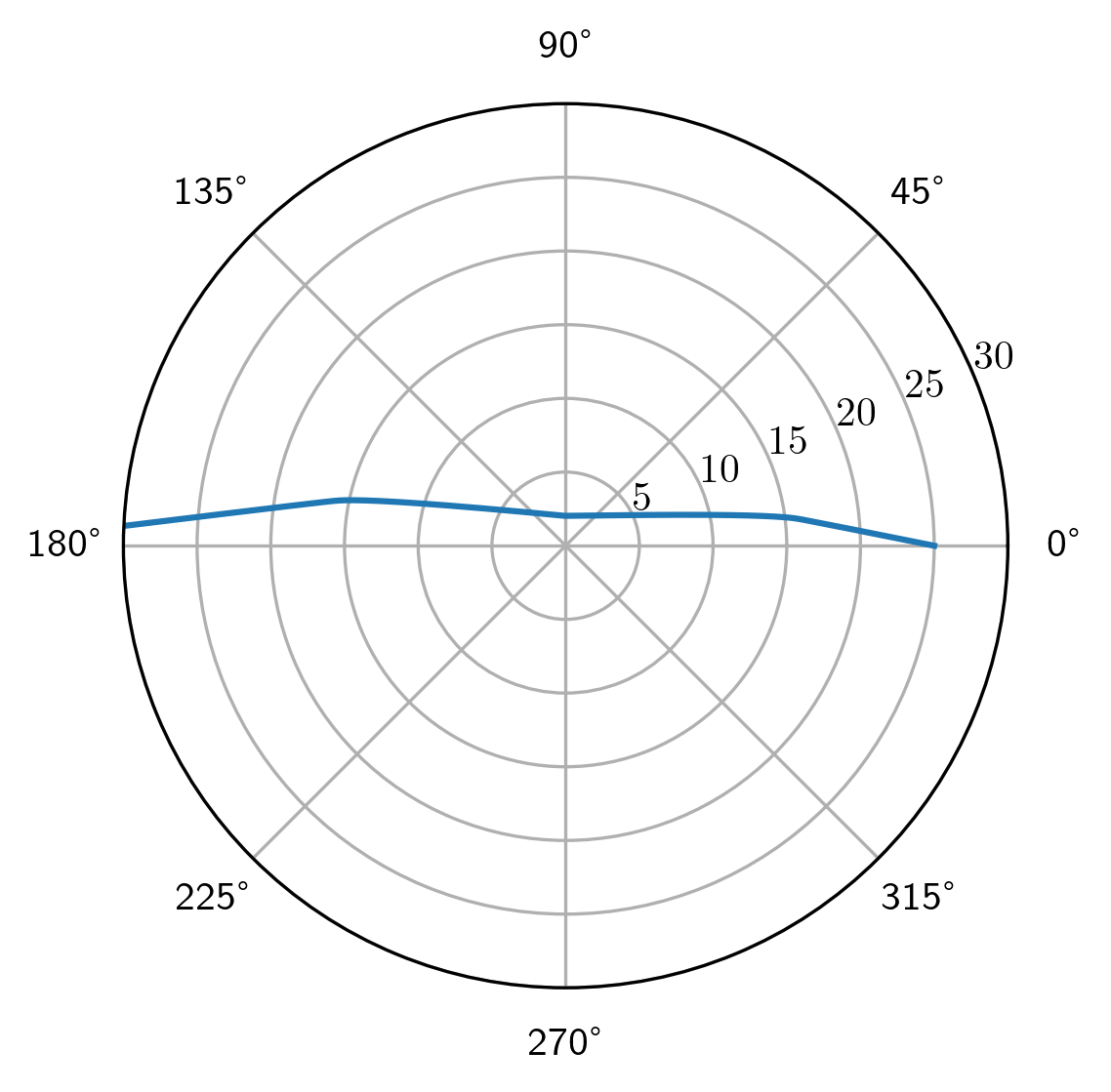}
		\caption{$L/E=5.0$}
		\label{fig:orb_120_lr_0.2}
	\end{subfigure}
	\caption{Almost-circular orbits for $Q=120$ with different values of $L/E$}
	\label{fig:orb_120_lr_shell}
\end{figure}

In Fig.~\ref{fig:orb_120_lr_shell} we show examples of null geodesics for $Q=120$ and for varying initial position $r_0$ and $L/E$. 
It can be seen that a partial orbit occurs around the center of the boson star, but that the gravitational field is not sufficient to form a true circular orbit. For particles with greater initial $\frac{\dd \phi}{\dd r}$, the scalar core of the boson star acts to refract the light.

\subsection{Particle Collision}
To analyse the particle collision in boson stars, we can expand the center-of-mass energy about a general point \(r=\tilde{r}\) and then separately study the \(r\rightarrow0\) limit. Boson stars are globally regular solutions, i.e. possess no horizon and, in principle, the particles can reach $r=0$ for $L=0$. Hence  \(N(0)\) and \(\sigma(0)\) are finite. (\ref{CMEstatic}) can be written as
\begin{equation}
\label{ecmA}
    \frac{E^2_{\text{\tiny{C.M.}}}}{2m_0^2}=1+\frac{\tilde{E}_1 \tilde{E}_2}{N\sigma^2}-\frac{L_1L_2}{r^2}-\frac{1}{N}\sqrt{A_1(r)}\sqrt{A_2(r)} \ ,
\end{equation}
where 
\begin{equation}
    \label{A} A_i(r)=\frac{\tilde{E}_i^2}{\sigma^2}-N\bigg(1+\frac{L^2_i}{r^2}\bigg)\ .
\end{equation}
To analyse the behaviour in the vicinity of an arbitrary point \(r=\tilde{r}\), we introduce the expansion \(r=\tilde{r}+\varepsilon\), where \(\varepsilon\) is a small parameter. The relevant quantities can then be expanded in powers of \(\varepsilon\) as follows:
\[N(r)=N_0+N_1\varepsilon+N_2\frac{\varepsilon^2}{2}+O(\varepsilon^3)\ ,\hspace{0.3cm} \sigma(r)=\sigma_0+\sigma_1\varepsilon+\sigma_2\frac{\varepsilon^2}{2}+O(\varepsilon^3) \ , \hspace{0.3cm} V(r)=V_0+V_1\varepsilon+V_2\frac{\varepsilon^2}{2}+O(\varepsilon^3) \ ,
\]
where for any function \(X\in\{N,\sigma,V\}\), we define \(X_n=X^{(n)}(\tilde{r})\). Furthermore, expansion of (\ref{A}) leads to
\[A_i(r)=A_{i0}+A_{i1}\varepsilon+A_{i2}\frac{\varepsilon^2}{2}+O(\varepsilon^3) \ , \quad A_{i0}=\frac{\tilde{E}^2_{i0}}{\sigma_0^2}-N_0-\frac{N_0L_i^2}{\tilde{r}^2}\]
where \(\tilde{E}_{i0}=E_i+q_iV_0\) and \(A_{in}=A_i^{(n)}(\tilde{r})\) for \(i=1,2\). Since the radial motion is determined through \(\dot{r}_i^2=A_i(r)\), the reality of the radial velocity requires \(A_i(r)\geq0\), so the particle can only access regions where \(A_i(r)\) is non-negative. To examine the behaviour of the radial velocity in the vicinity of \(\tilde{r}\) and identify the conditions required for the expansion to remain real and well defined, we first expand \(\sqrt{A_i(r)}\)~:
\begin{equation}
\sqrt{A_i(r)}=\sqrt{A_{i0}}+\frac{1}{2\sqrt{A_{i0}}}A_{i1}\varepsilon+O(\varepsilon^2) \ .
\end{equation}
Therefore, we require \(A_{i0}>0\). The critical particle must be considered separately from the arbitrary case where we take \(\tilde{r}=r_c\). Assuming that particle \(i=1\) corresponds to the critical particle, its critical orbit must satisfy \(A_{10}=0\) and \(A_{11}=0\) with the following expansion:
\[A_1(r)=A_{12}\frac{(r-r_c)^2}{2}+O((r-r_c)^3) \Rightarrow\sqrt{A_1(r)}=\sqrt{\frac{1}{2}A_{12}} \ |r-r_c|+... \ , \]
where \(A_{12}>0\). Consequently, the critical orbit conditions changes the leading-order behaviour. 

\begin{figure}[!h]
	\centering
	\begin{subfigure}{0.45\textwidth}
		\includegraphics[width=\textwidth]{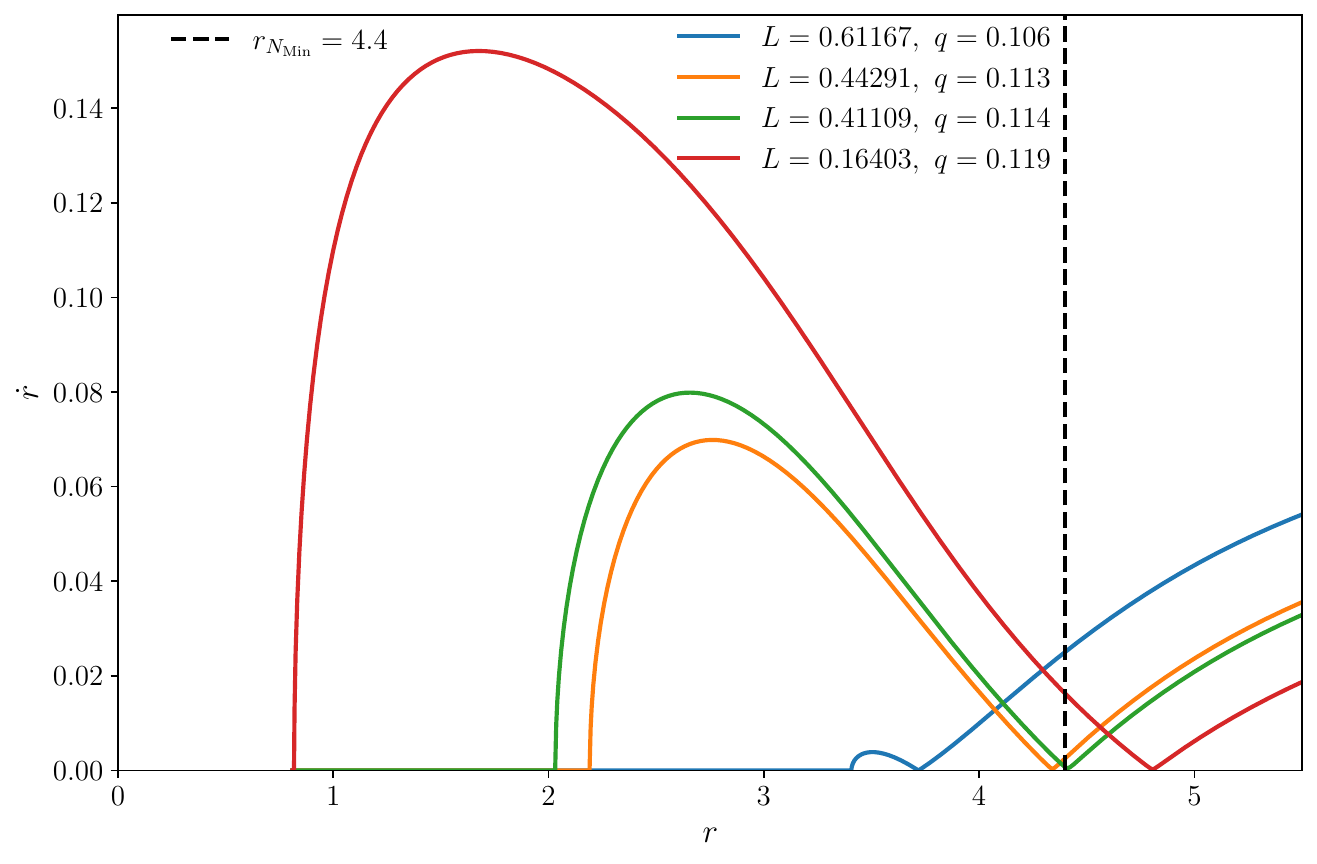}
	\end{subfigure}
    \begin{subfigure}{0.48\textwidth}
		\includegraphics[width=\textwidth]{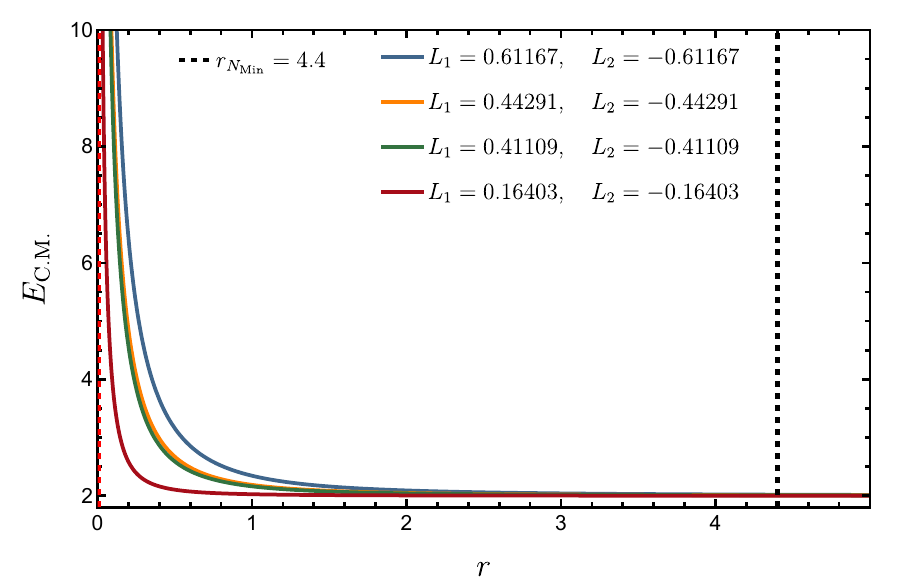}
	\end{subfigure}
    \caption{{\it Left}: We plot the radial velocity \(\dot{r}\) as a function of \(r\) for charged particles moving in the boson star space-time, with electric charge \(Q=3\), \(\alpha=0.012\), and \(g=0.08\), for several values of the angular momentum \(L\). {\it Right}: We plot the center-of-mass energy as a function of \(r\) for collisions between two charged particles. The blue, orange, green, and red solid curves correspond to \(q_i=0.106\), \(q_i=0.113\), \(q_i=0.114\), and \(q_i=0.119\), respectively, where \(i=1,2\). In each case, the two particles have angular momenta of opposite sign.}\label{Q=3}
\end{figure}

We mow consider the expansion around the center, \(r=0\). For a regular boson star, the center does not correspond to the horizon; instead the metric functions remain regular ar \(r=0\), hence the expansion at the center is \cite{Ma:2025, Chicaiza:2026}~: 
\[N(r)=1+N_2\frac{r^2}{2}+O(r^4)\ ,\hspace{0.3cm} \sigma(r)=\sigma_0+\sigma_2\frac{r^2}{2}+O(r^4) \ , \hspace{0.3cm} V(r)=V_0+V_2\frac{r^2}{2}+O(r^4) \ ,
\]
and 
\begin{equation} \tilde{E}_i(r)=\tilde{E}_{i0}+q_iV_2\frac{r^2}{2}+O(r^4) \ .
\end{equation}
The behaviour near the center differs according to whether the particles possess angular momentum. Hence, suppose both particles have non-zero angular momentum; \(L_1\neq0\) and \(L_2\neq0\). Considering one of the square roots \(\sqrt{A_i(r)}\) where
\begin{equation}
N(r)\frac{L_i^2}{r^2}=\frac{L^2_i}{r^2}+O(1) \ ,  
\end{equation}
so the dominant terms as \(r\rightarrow0\) is therefore \(L_i^2/r^2\). The first term \(\tilde{E}_i^2/\sigma^2\) is finite at the center since both \(\tilde{E}_i\) and \(\sigma\) are finite. Hence
\begin{equation}
\frac{\tilde{E}_i^2}{\sigma^2}-N\bigg(1+\frac{L^2_i}{r^2}\bigg)=-\frac{L_i^2}{r^2}+O(1) \ ,
\end{equation}
and \(L_i^2/r^2\rightarrow-\infty\) \(r\rightarrow0\). Consequently, the quantity under the square root becomes negative before the particle reaches \(r=0\). Hence, when \(L_i\neq 0\), the particle is prevented from reaching the center. This behaviour is also evident from the data presented below. Examining only the terms \(L_1L_2/r^2\) does not, by itself, imply a divergence of the center-of-mass energy. The full expansion of \(E_{\text{\tiny{C.M.}}}\) is physically meaningful only when both particles have real radial motion and are therefore able to reach the collision point. Thus, any apparent divergence arising from \(L_1L_2/r^2\) term must be considered together with the radial motion conditions for both particles. For \(L_i\neq0\), the radial equation near the center is~:
\begin{equation}
\dot{r}_i^2=-\frac{L_i^2}{r^2}+O(1)<0 \ .
\end{equation}
So the particle encounters a turning point at some finite radius \(r_{\text{min}}>0\) and never gets into the region where \(r\) is sufficiently small. Hence, the physical trajectory terminates at \(r_{\min}\), rather than reaching \(r=0\). The plot for the center-of-mass energy, however, shows otherwise. The expansion of \(A_i(r)\) in (\ref{A}) near \(r=0\) is :
\begin{equation}\label{A_i}
A_i(r)=-\frac{L^2_i}{r^2}+\bigg(\frac{\tilde{E}_{i0}^2}{\sigma_0^2}-1-\frac{N_2L_{i}^2}{2}\bigg)+O(r^2) \ .
\end{equation}
Defining 
\begin{equation}
 C_i=\frac{\tilde{E}_{i0}^2}{\sigma_0^2}-1-\frac{N_2L_{i}^2}{2} \ ,   
\end{equation}
and (\ref{A_i}) can be written as \(A_i(r)=-L_i^2/r^2+C_i+O(r^2)\). For \(L_i\neq0\), the leading term is negative as \(r\rightarrow0\), so the square roots appearing in the algebraic expression for \(E_{\text{\tiny{C.M.}}}\) become imaginary sufficiently close to the regular center. Thus, the radial motion is not physically admissible in this region. Nevertheless, we can examine the formal asymptotic behavior of the algebraic expression by analytically continuing the square roots. We obtain
\begin{equation}
\sqrt{A_i(r)}=\frac{i|L_i|}{r}-\frac{iC_i}{2|L_i|}r+O(r^3) \ . 
\end{equation}
Consequently, 
\begin{equation}
\sqrt{A_1(r)}\sqrt{A_2(r)}=-\frac{|L_1L_2|}{r^2}+\frac{1}{2}\bigg(\frac{C_1|L_2|}{|L_1|}+\frac{C_2|L_1|}{|L_2|}\bigg)+O(r^2) \ ,
\end{equation}
and the full expansion of (\ref{ecmA}) about \(r=0\) is given by~:
\begin{equation}
\frac{E^2_{\text{\tiny{C.M.}}}}
{2m_0^2}=\frac{|L_1L_2|-L_1L_2}{r^2}+O(1) \ .
\end{equation}
If \(L_1L_2>0\), then \(|L_1L_2|=L_1L_2\) and \(E^2_{\text{\tiny{C.M.}}}/2m_0^2=O(1)\), so there is no divergence in the center-of-mass energy from the angular momentum terms. If \(L_1L_2<0\), then \(|L_1L_2|=-L_1L_2\), and \(E^2_{\text{\tiny{C.M.}}}/2m_0^2\sim 2|L_1L_2|/r^2\). In the cases illustrated in Fig.~\ref{Q=3} and Fig.~\ref{Q=6,25}, we have taken \(L_1=L\) and \(L_2=-L\), therefore \(E_{\text{\tiny{C.M.}}}\sim 2m_0|L|/r\). The plot in Fig.~\ref{Q=3} shows the center-of-mass growing rapidly as \(r\) decreases which is consistent with the formal asymptotic behaviour. However, as discussed earlier, a particle with nonzero angular momentum cannot reach \(r=0\). Therefore, the formal divergence obtained from the asymptotic expansion corresponds to an extrapolation of the algebraic expression into a region that is not physically accessible to the particle. 

The results shown in Fig.~\ref{Q=3} indicate that the occurrence of the BSW effect is sensitive to the particle charge. For \(0.106\leq q \leq 0.113\), the collision takes place at \(r_c<r_{N_{\text{Min}}}\), where $r_{N_{\text{Min}}}$ is the radius at which $N(r)$ has its minimum, while for \(0.114\leq q \leq 0.119\), the collision occurs at \(r_c>r_{N_{\text{Min}}}\). Examining the center-of-mass energy, we find that in all cases the particles are unable to reach \(r=0\). 
\vspace{0.3cm}
\begin{figure}[!h]
	\centering
	\begin{subfigure}{0.45\textwidth}
		\includegraphics[width=\textwidth]{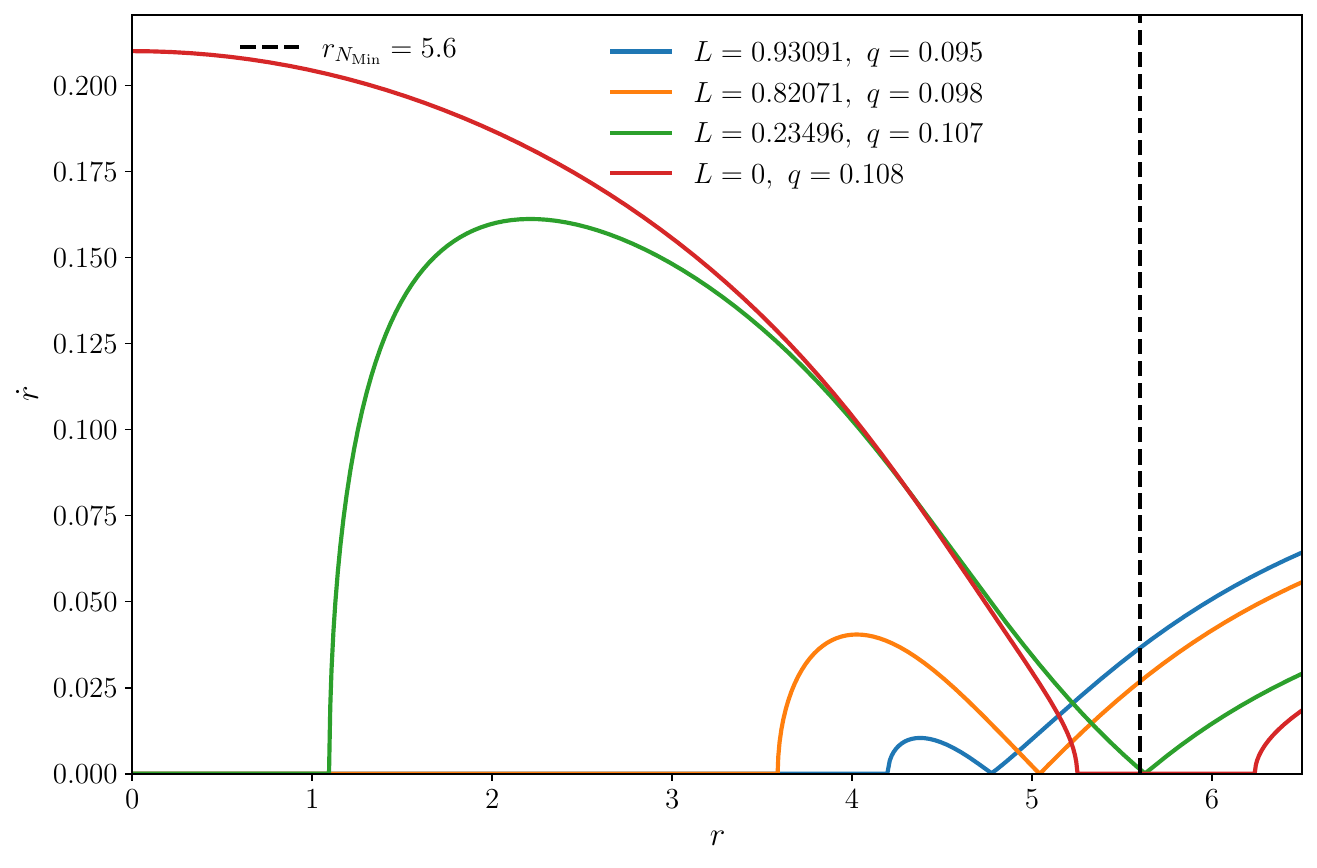}
	\end{subfigure}
    \begin{subfigure}{0.45\textwidth}
		\includegraphics[width=\textwidth]{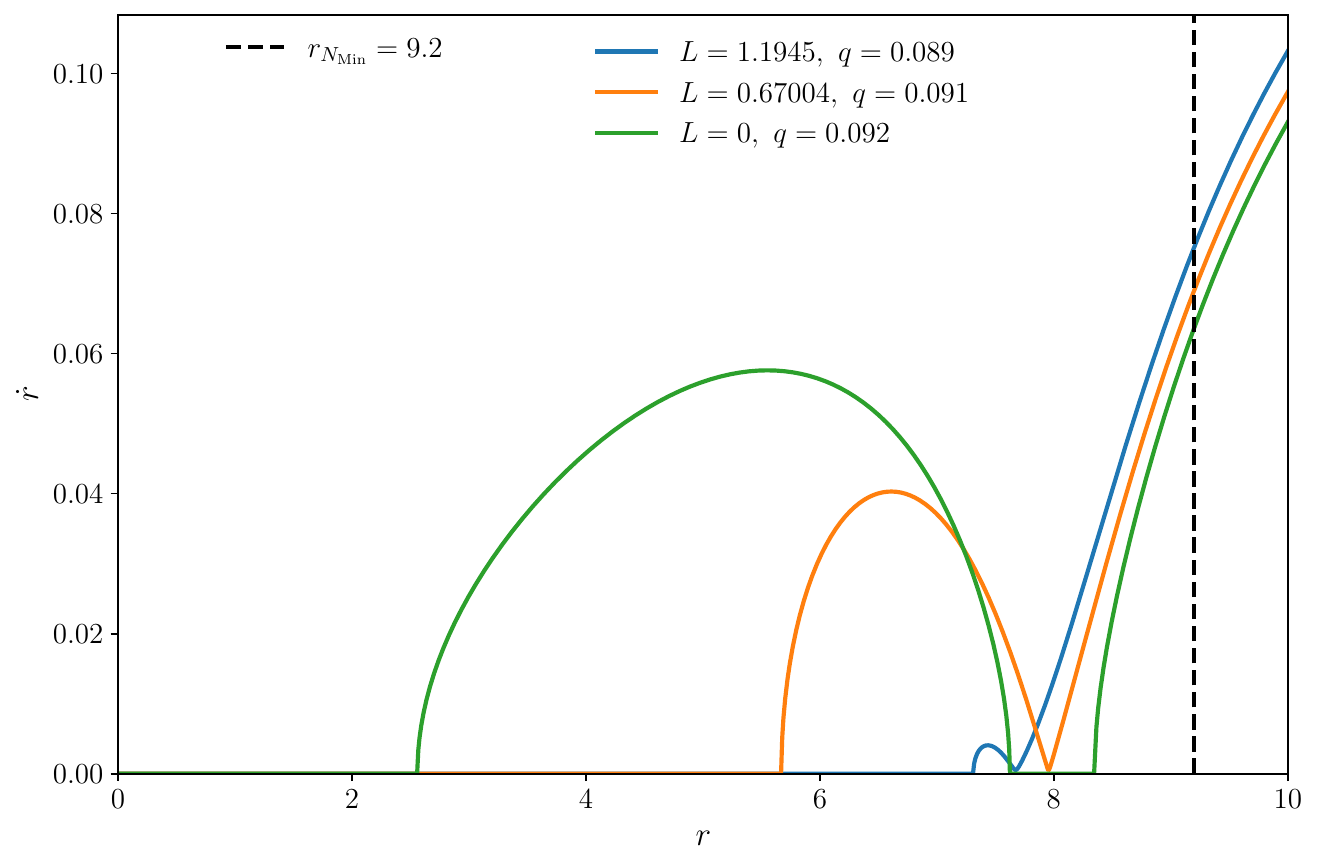}
	\end{subfigure}
    \caption{{\it Left}: We show the radial velocity \(\dot{r}\) as a function of the radial coordinate \(r\) for charged particles in the boson star space-time for \(Q=6\) and a range of angular momentum values \(L\). {\it Right}: Same as left, but for \(Q=25\). For both solutions, we have chosen \(\alpha=0.012\) and \(g=0.08\).
}
    \label{Q=6,25}
\end{figure}

From Fig.~\ref{Q=6,25}, we observe that increasing \(Q\) causes the location of the collision to move progressively further away from the center. For \(Q=6\), collisions are not possible for particle charges \(q>0.107\). The case \(q=0.107\) is particularly notable, as it is the only particle where the collision approaches \(r_{N_{\text{Min}}}\). For all other values of \(q\), the particles are unable to reach the center, and the collision occurs before \(r_{N_{\text{Min}}}\). For \(Q=25\), we find that collisions cannot occur beyond \(r_{N_{\text{Min}}}\). This behaviour is also observed more generally for \(Q>6\), where the collision is always confined to the region \(r_c<r_{N_{\text{Min}}}\). Furthermore, as \(Q\) increases, the collision points occur progressively father from the center, indicating that the collisions are shifted towards larger radial distances. 
Consequently, a physical collision at the regular center \(r=0\) can only be considered for particles with \(L_1=L_2=0\). This highlights a fundamental difference from the standard near-horizon BSW analysis, where the divergence in the center-of-mass energy can arise from the collision of particles approaching the horizon. In the present case, the center-of-mass energy near \(r=0\) is given by~:
\[\frac{E^2_{\text{\tiny{C.M.}}}}{2m_0^2}\bigg|_{r=0}=1+\frac{\tilde{E}_{10}\tilde{E}_{20}}{\sigma_0^2}-\sqrt{\frac{\tilde{E}^2_{10}}{\sigma^2_0}-1}\sqrt{\frac{\tilde{E}^2_{20}}{\sigma^2_0}-1} \ .\]
Thus, for two particles with \(L_1=L_2=0\), \(E_{\text{\tiny{C.M.}}}\)remains finite at the center, provided that the particles are physically able to reach \(r=0\). The condition for a particle to reach the regular center follows directly from the radial equation. For \(L_i=0\), this equation reduces to
\[\dot{r}^2_i=\frac{\tilde{E_i^2}}{\sigma^2}-N \ .\] 
Evaluating this expression at \(r=0\), where \(N(0)=1\) and \(\sigma(0)=\sigma_0\), gives \(\dot{r}^2_i(0)=\tilde{E}^2_{i0}/\sigma^2_0-1\). Therefore, for the particle to reach the center with nonzero radial velocity we require, \(|E_i+q_{i}V_0|>\sigma_0\). 
Since the particles are released from rest at infinity, the particle's charge must satisfy the following condition~: 
\[q_i>\frac{\sigma_0-1}{V_0-V_{\infty}} \ .\]
\begin{figure}[!h]
	\centering
	\includegraphics[width=0.6\linewidth]{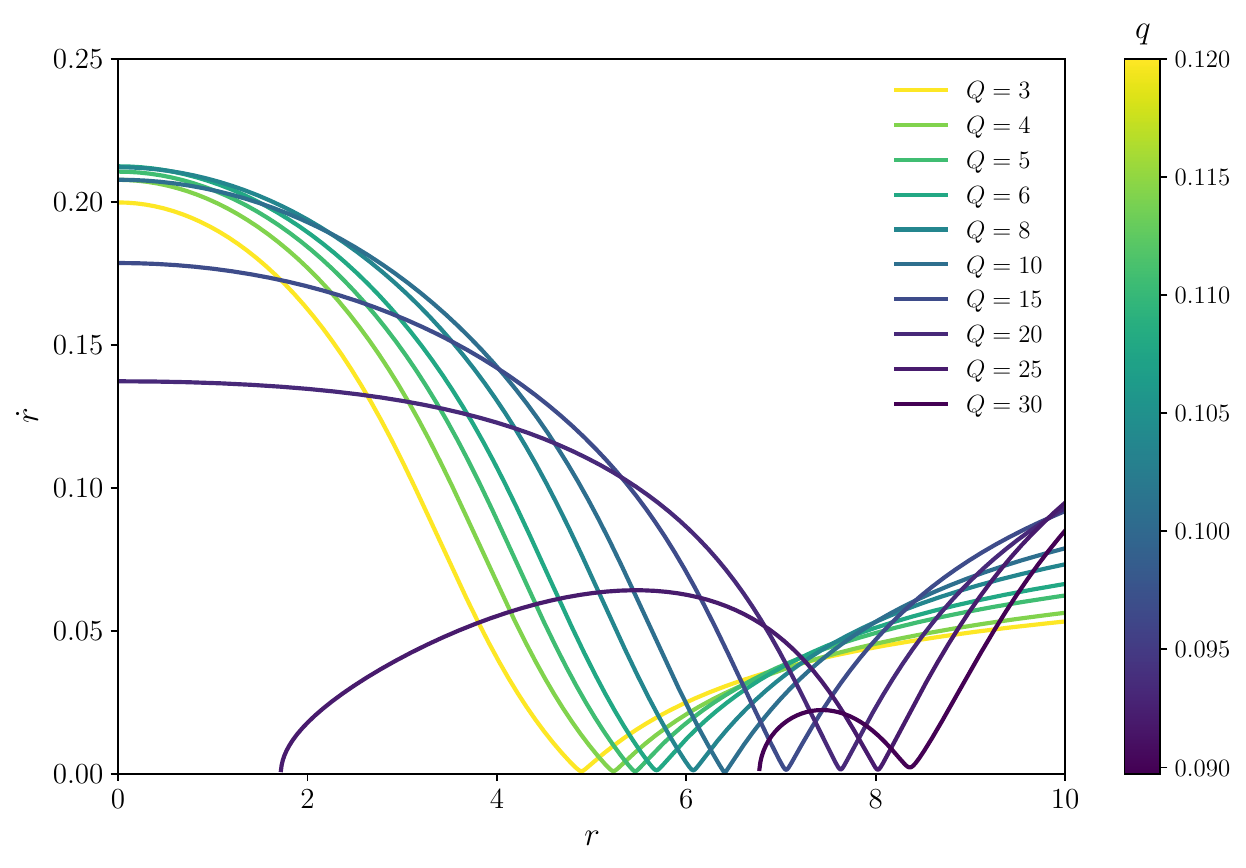}
	\caption{We plot the radial velocity \(\dot{r}\) as a function of \(r\) for charged particles moving in the boson star space-time with \(\alpha=0.012\) and \(g=0.08\), considering \(Q=3,4,5,6,8,10,15,20,25\) and 30, for particles with \(L=0\). Note that for each $Q$ the colour-bar to the right indicates the value of $q$ necessary for $\dot{r}$ to become zero.}
	\label{L=0}
\end{figure}
\newline Fig.~\ref{L=0} shows that in all cases, the collision occurs far from the center. For \(Q=25\) and \(Q=30\), the particle trajectories do not reach \(r=0\). Nevertheless, in each case, the collision takes place very close to the value of $r$ at which $N(r)$ has its minimum, $r=r_{N_{\text{Min}}}$. Here the center-of-mass energy attains its largest value. However, for \(Q>30\), we are unable to obtain unstable circular orbits, and consequently, the equivalent of the BSW effect is no longer observed.

\section{Particle motion in frozen star space-times} \label{sec:frozen_stars}
\begin{figure}[!h]
	\centering
	\includegraphics[width=0.6\textwidth]{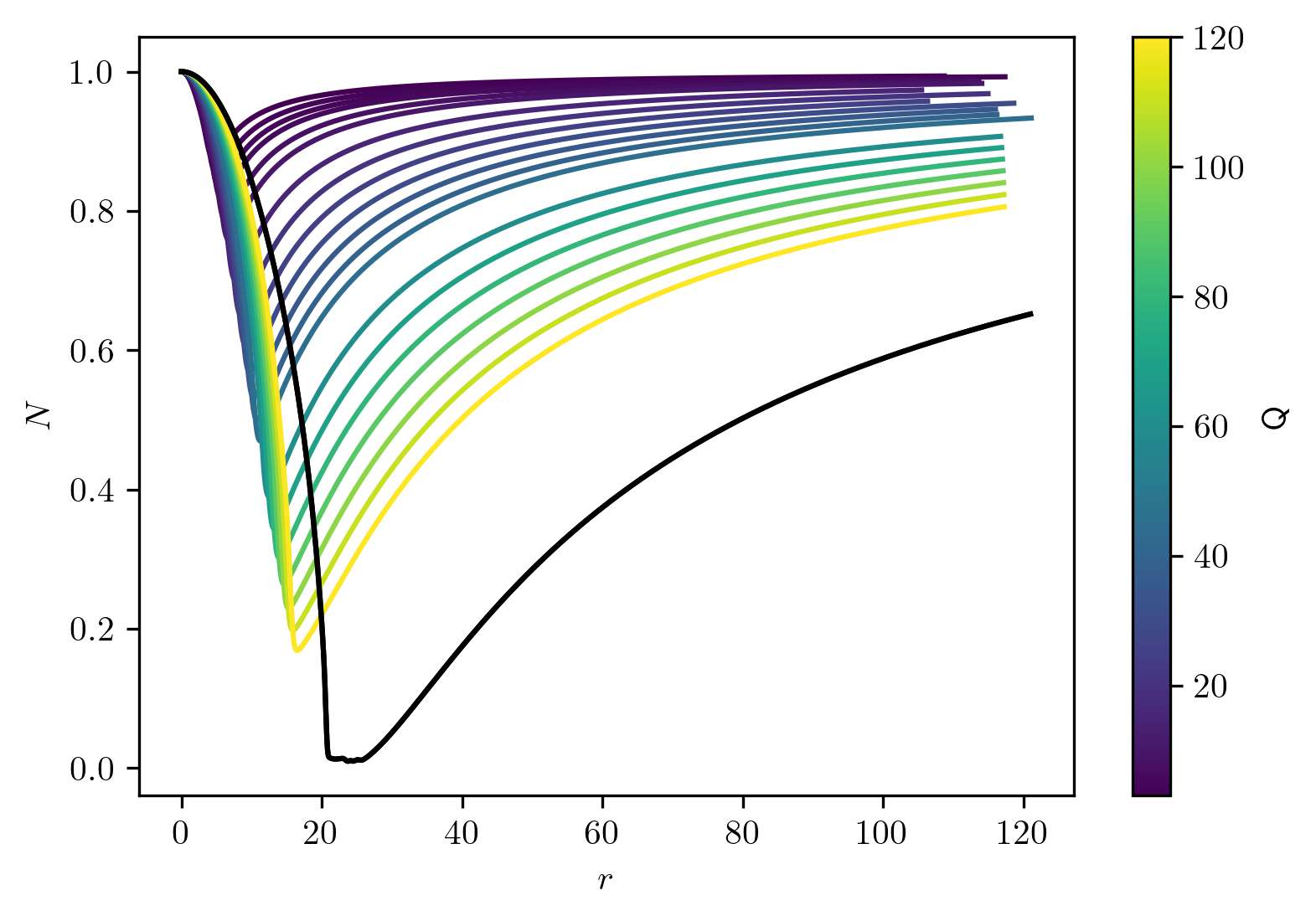}
    \caption{We show the metric funcion $N(r)$ for various non-limiting boson star solutions given in terms of $Q$, compared with the frozen star solution $k=2$ (in black).}
    \label{fig:N_new_all}
\end{figure}

We now consider the limiting case of frozen boson stars. These solutions have been originally discussed in \cite{Brihaye:2025dlq} and again are only given numerically. We have followed the same strategy as that for boson stars discussed above.
Frozen boson stars exist only for very specific choices of the parameters. Here, we consider the case $Q=213$, $g=0.08$ and $\alpha=0.0012$.
In fact, frozen states of boson stars form a discrete tower of solutions distinguished by the number of nodes of the scalar field function. This corresponds to the interval of the radial coordinate $r$ in which the quantity $\frac{1}{N}-\frac{q^2V^2}{\sigma^2N^2}$ becomes negative \cite{Brihaye:2025dlq}. We denote the number of such nodes with $k$ and consider here solutions with $k=2$, $4$, $14$ and $18$. In Fig.~\ref{fig:N_new_all} we show the metric function $N(r)$ and demonstrate how the frozen star limit is approached. This is for $k=2$ (compare also \cite{Brihaye:2025dlq}).

\subsection{Defining the shell}
To define the shell interpolating between the de Sitter interior and the RN exterior, we have taken two different approaches. 
\begin{itemize}
\item The first approach is to find the values of $r$ at which $N(r)=0.02$. We define the shell to be where $N(r) < 0.02$. 
Table \ref{tab:shell_edges} shows these values for the solutions with different $k$.

\begin{center}
	\begin{tabular}{||c | c c||} 
		\hline
		$k$ &  Inner Shell & Outer Shell \\ [0.5ex] 
		\hline\hline
		2 & 20.907 & 26.981 \\ 
		\hline
		4 & 21.096 & 27.034  \\
		\hline
		14 & 21.790 & 27.152 \\
		\hline
		18 & 21.669 & 27.139 \\
		\hline
	\end{tabular}
    \captionof{table}{The values of $r$ for which $N=0.02$ for each solution}\label{tab:shell_edges}
\end{center}
In Fig.~\ref{fig:2_shell_bounds} we indicate these locations on a plot of the metric function $N(r)$ for the frozen star solution $k=2$ with the location of the shell boundaries, the ISCO for $k=2$ (see below) and the equivalent extremal RN horizon, i.e. the horizon of a charged black hole with mass and charge equal to that of the frozen star (see below). 
\
\begin{figure}[!h]
	\centering
	\begin{subfigure}{0.4\textwidth}
		\includegraphics[width=\textwidth]{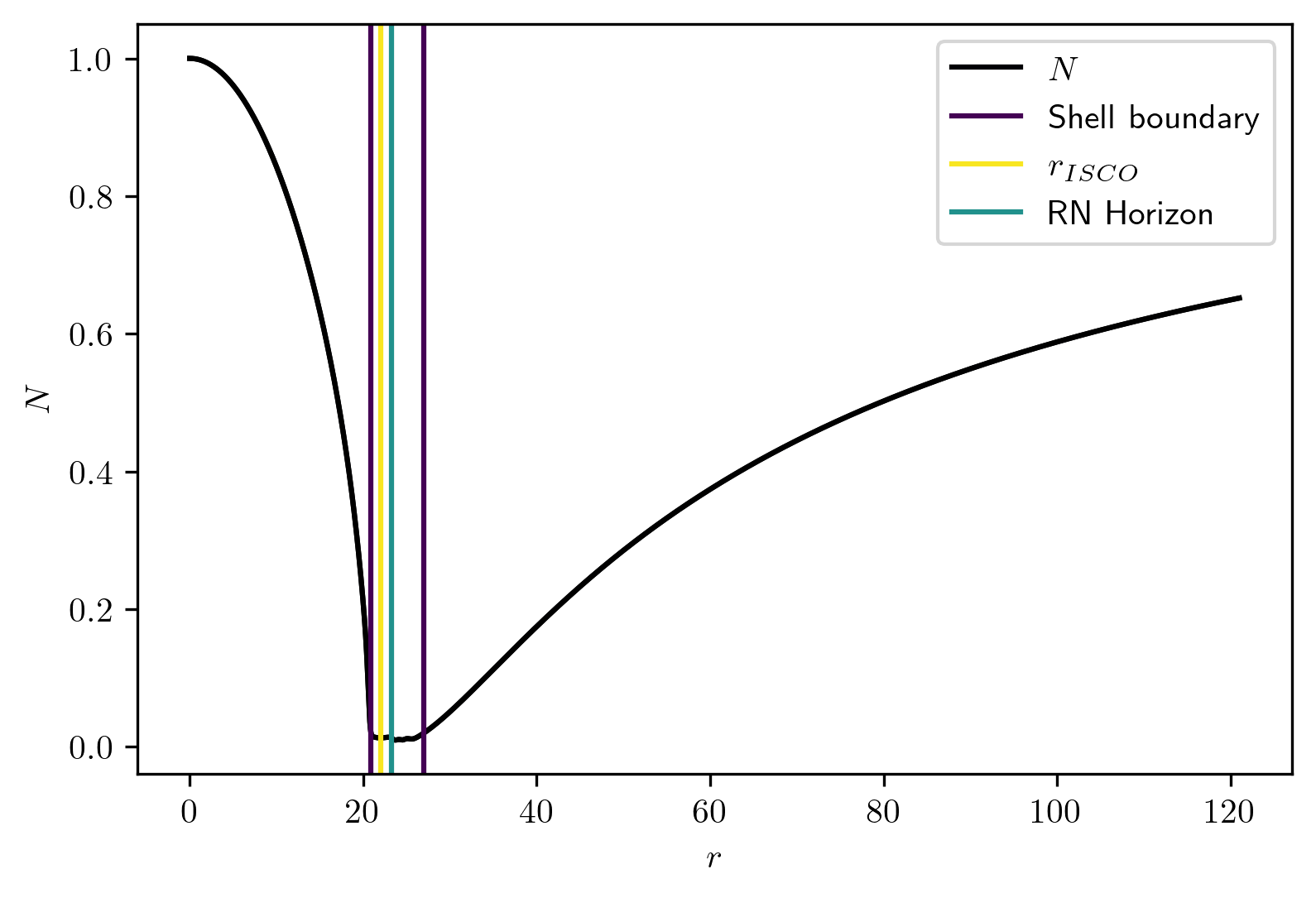}
	\end{subfigure}
	\begin{subfigure}{0.4\textwidth}
		\includegraphics[width=\textwidth]{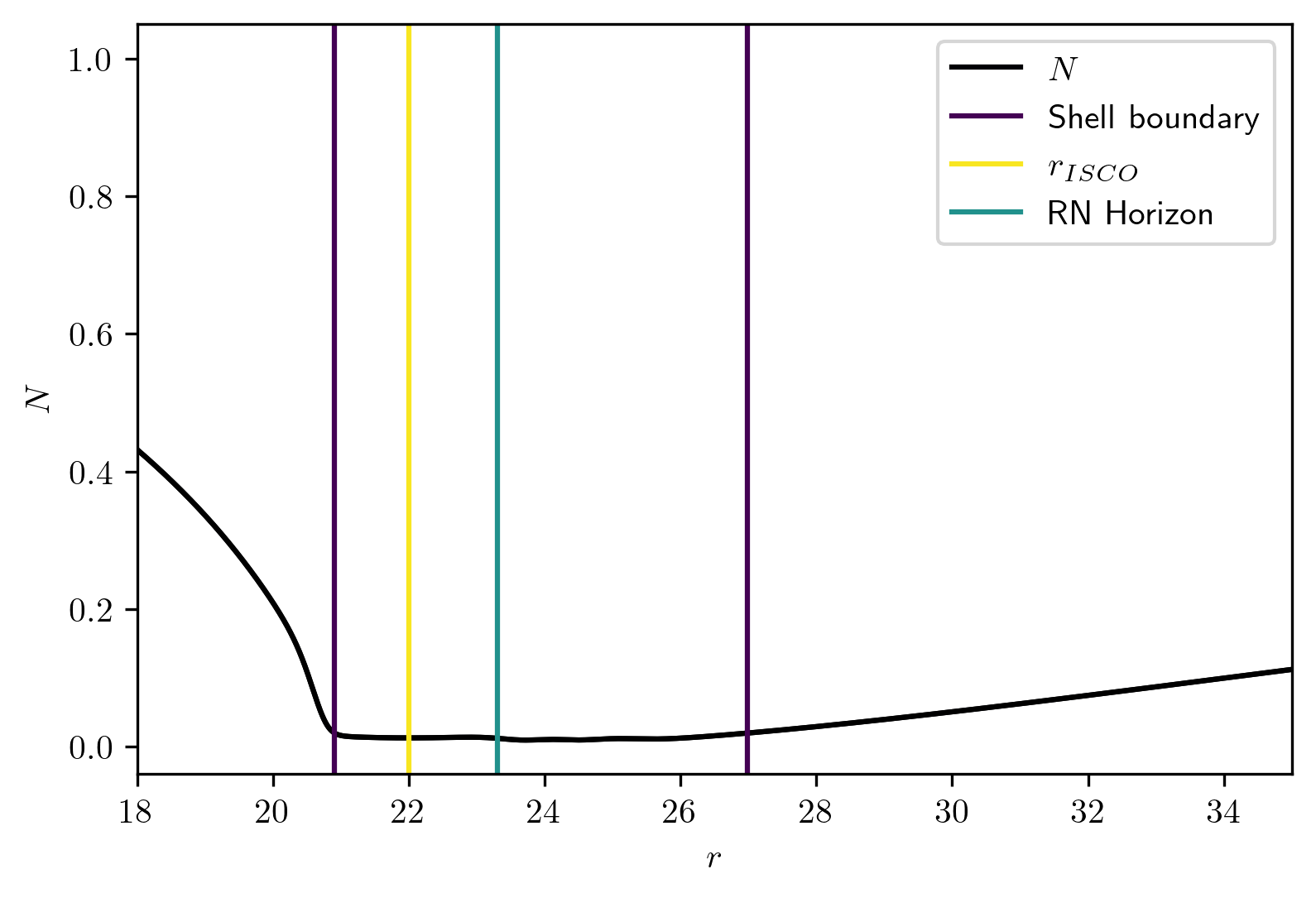}	
	\end{subfigure}
    \caption{The metric function $N(r)$ for $Q=213$, $k=2$, with the location of the shell boundaries (purple), the ISCO (yellow) and the horizon for the equivalent extremal RN  black hole (turquoise).}
    \label{fig:2_shell_bounds}
\end{figure}

\item The second approach is based on examining the behaviour of the energy density and pressure. By analysing these quantities as functions of the radial coordinate, we can identify the region where the matter distribution changes and thereby determine the location of the shell. Our configuration consists of an interior dominated by the scalar field, an intermediate transition shell and the asymptotic RN exterior. The exterior is not vacuum because the Maxwell field remains \(V'(r)\neq0\) with energy density \(\rho_{\text{RN}}=Q^2/(2r^4)\). To locate the shell of a charged  frozen boson star, we want to find the radius where the scalar contribution disappears. We hence match the remaining energy density and pressure to the RN electromagnetic field and define the outer shell as where the solution becomes effectively RN. From the stress tensor, we identify the electromagnetic and scalar field energy densities:
\begin{equation}
\rho_{\text{EM}}(r)=\frac{V'^2}{2\sigma^2} \ \ , \ \ \rho_{\Psi}= N {{\psi}}'^2 + \frac{(\omega - q V)^2 {{\psi}}^2}{N \sigma^2} +  U({\psi}) \ ,
\end{equation}
such that \(\rho=\rho_{\text{EM}}+\rho_{\Psi}\). The outer shell is where
\[\rho_{\Psi}\rightarrow0 \ , \ \ \ \rho\rightarrow \rho_{\text{RN}} \ , \ \ \ p_r\rightarrow p^{\text{RN}}_r=-\rho_{\text{RN}} \ , \ \ \ p_t\rightarrow p_t^{\text{RN}}=\rho_{\text{RN}}  \ .\]
From our numerical data, \(N(r)\) is given by (13) and (26) is the RN mass \(M\) where for the extremal RN, the horzion radius is \(r_+=M\). We then measure how far our solution is from RN and define the energy density, radial pressure and tangential pressure difference, respectively, via
\[\Delta_{\rho}=\frac{|\rho-\rho_{\text{RN}}|}{\rho_{\text{RN}}} \ , \ \ \ \Delta_{p_r}=\frac{|p_r+\rho_{\text{RN}}|}{\rho_{\text{RN}}} \ , \ \ \ \Delta_{p_t}=\frac{|p_t-\rho_{\text{RN}}|}{\rho_{\text{RN}}} \ .\]
This quantity tells us how close our numerical solution of the frozen boson star is to the RN solution. Combining the above into one RN matching function 
\[\Delta_{\text{RN}}=\sqrt{\Delta^2_{\rho}+\Delta^2_{p_r}+\Delta^2_{p_t}}\]
allows us to locate the radius of the outer shell with a tolerance of \(\Delta_{\text{RN}}<10^{-3}\). This helps in identifying at which $r$ the frozen star transitions into the extremal RN exterior. The inner region is  where the scalar field dominates and fulfils the equation of state \(\rho=p_t=-p_r\). Hence we only look at the energy-momentum generated by our scalar solution itself. To find the inner region of the shell, we determine the first radius where the shell condition holds and use the same tolerance of \(10^{-3}\). In Fig.~\ref{fig:energy_density1} we show the components of the energy-momentum
tensor for $k=2$ and $k=4$, while in Fig.~\ref{fig:energy_density2} we give these components for $k=14$ and $k=18$. We indicate the inner and outer shells as well as the location of the horizon of the extremal RN solution with same mass.

\begin{figure}[!h]
	\centering
	\begin{subfigure}{0.4\textwidth}
		\includegraphics[width=\textwidth]{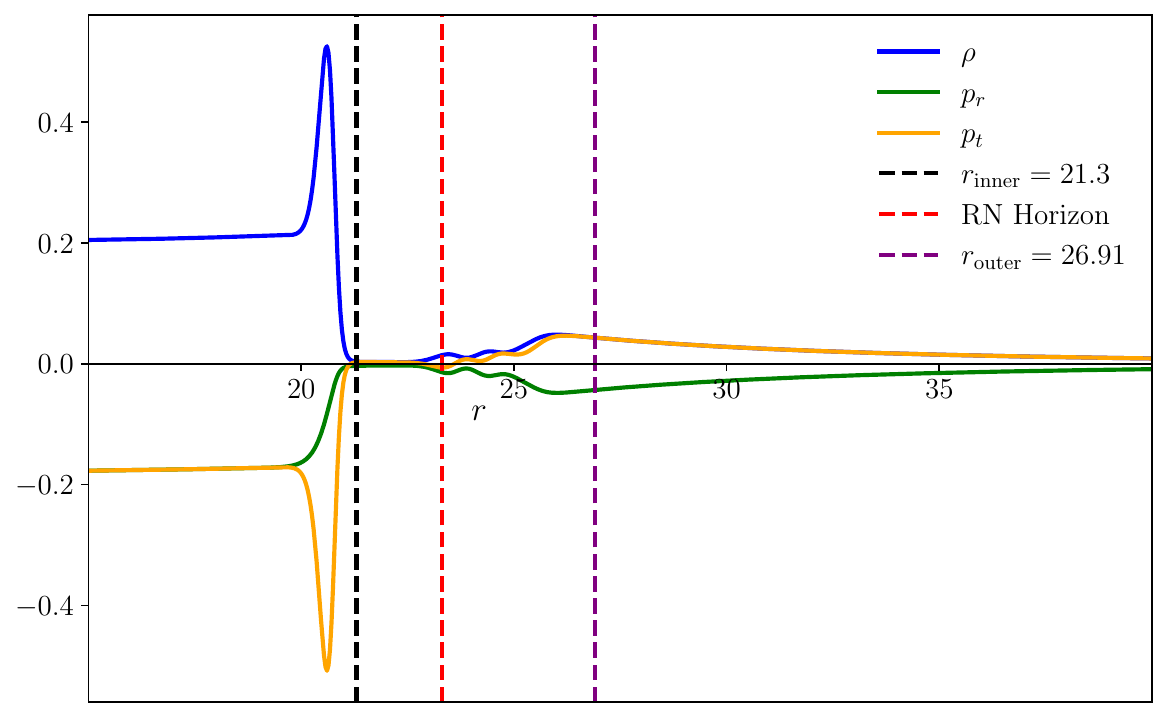}
	\end{subfigure}
    \begin{subfigure}{0.4\textwidth}
		\includegraphics[width=\textwidth]{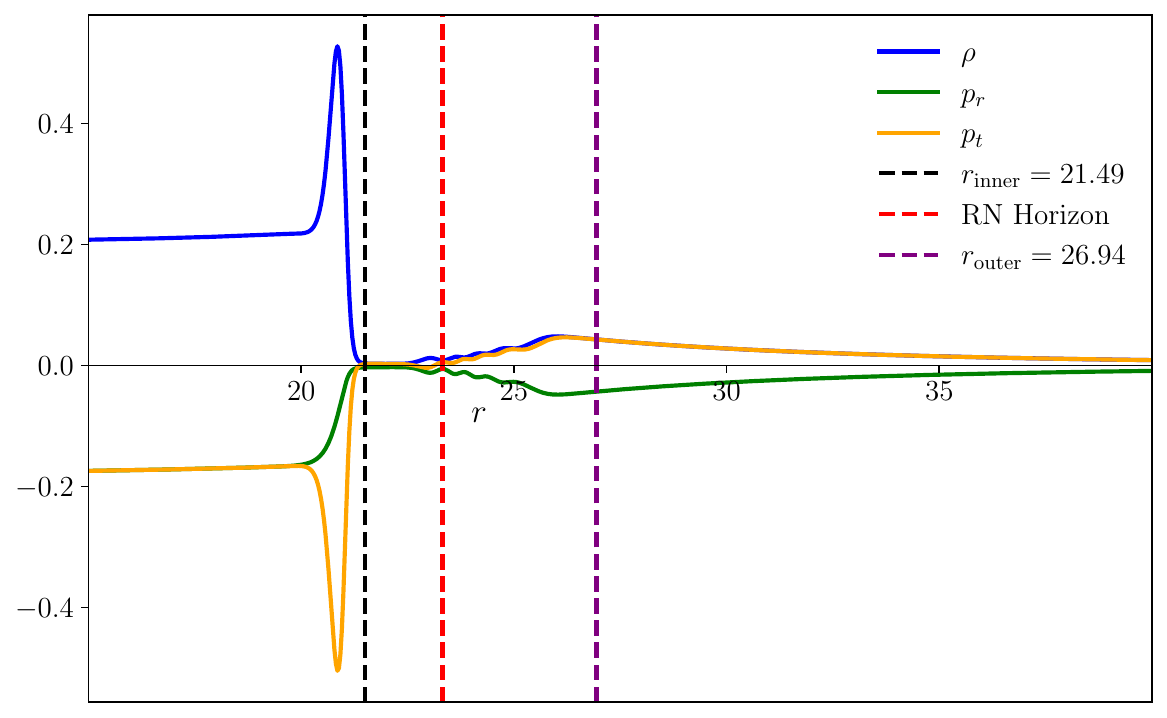}
	\end{subfigure}
    \caption{{\it Left}: We show the energy density \(\rho\) and pressure \(p\) as functions of the radial coordinate \(r\) for the boson star space-time with \(Q=213\) and node number \(k=2\). The inner and outer boundaries of the shell are indicated by black and purple dashed lines respectively. The red dashed line denotes the radial location associated with RN horizon, at which \(N(r)\rightarrow0\). {\it Right}: Same as left, but for \(Q=213\) and \(k=4\).}
    \label{fig:energy_density1}
\end{figure}

\begin{figure}[!h]
	\centering
	\begin{subfigure}{0.4\textwidth}
		\includegraphics[width=\textwidth]{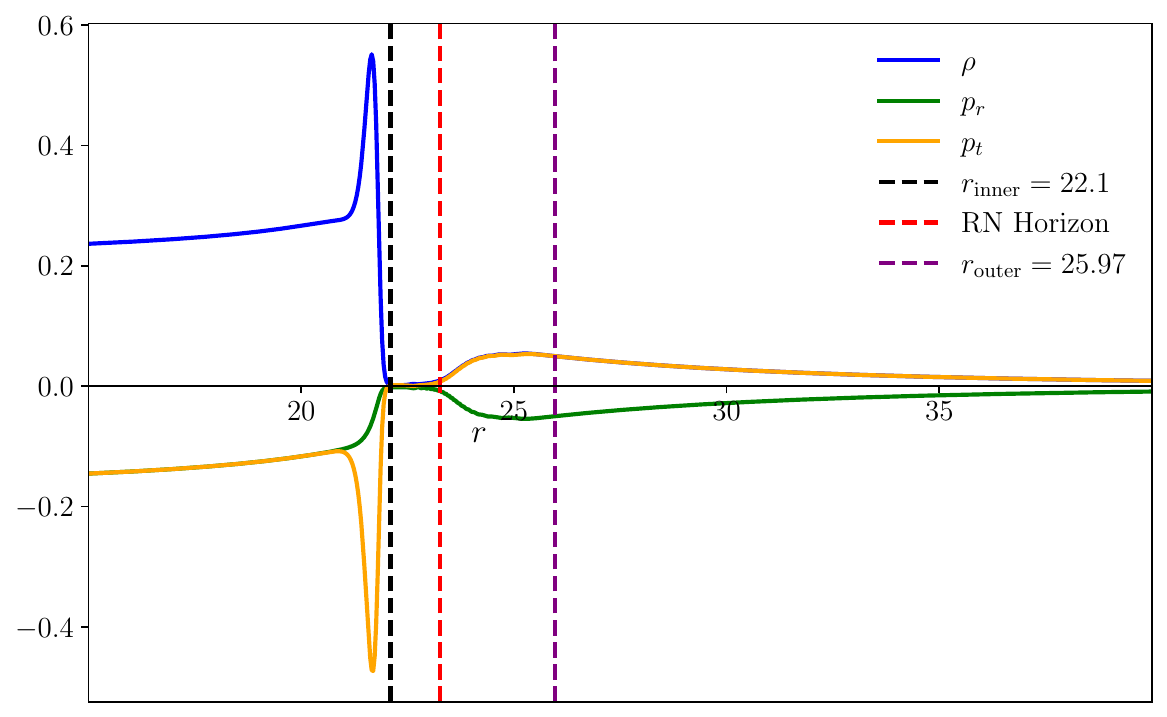}
	\end{subfigure}
    \begin{subfigure}{0.4\textwidth}
		\includegraphics[width=\textwidth]{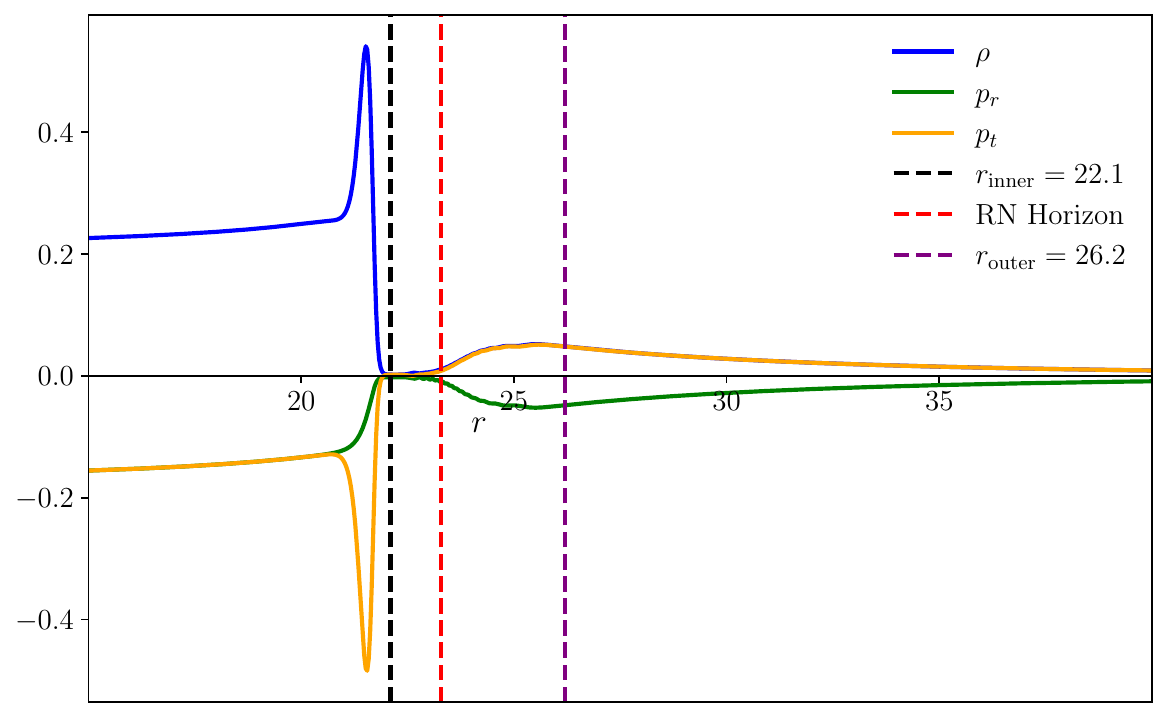}
	\end{subfigure}
    \caption{{\it Left}:  We show the energy density \(\rho\) and pressure \(p\) as functions of the radial coordinate \(r\) for the boson star space-time with \(Q=213\) and node number \(k=14\). The inner and outer boundaries of the shell are indicated by black and purple dashed lines respectively. The red dashed line denotes the radial location associated with RN horizon, at which \(N(r)\rightarrow0\). {\it Right}: Same as left, but for \(Q=213\) and \(k=18\).}
    \label{fig:energy_density2}
\end{figure}

\end{itemize}

As can be seen above, both methods given more or less the same values for the radius of the outer and the inner shell. Interestingly, we find that the inner shell radius is smaller than the radius of the corresponding extremal RN solution.

\subsection{Circular orbits}
\subsubsection{Massive uncharged test particles}\label{sec:orbits_fs_massive_uncharge}
\begin{figure}[h!]
	\centering
	\begin{subfigure}{0.4\textwidth}
		\includegraphics[width=\textwidth]{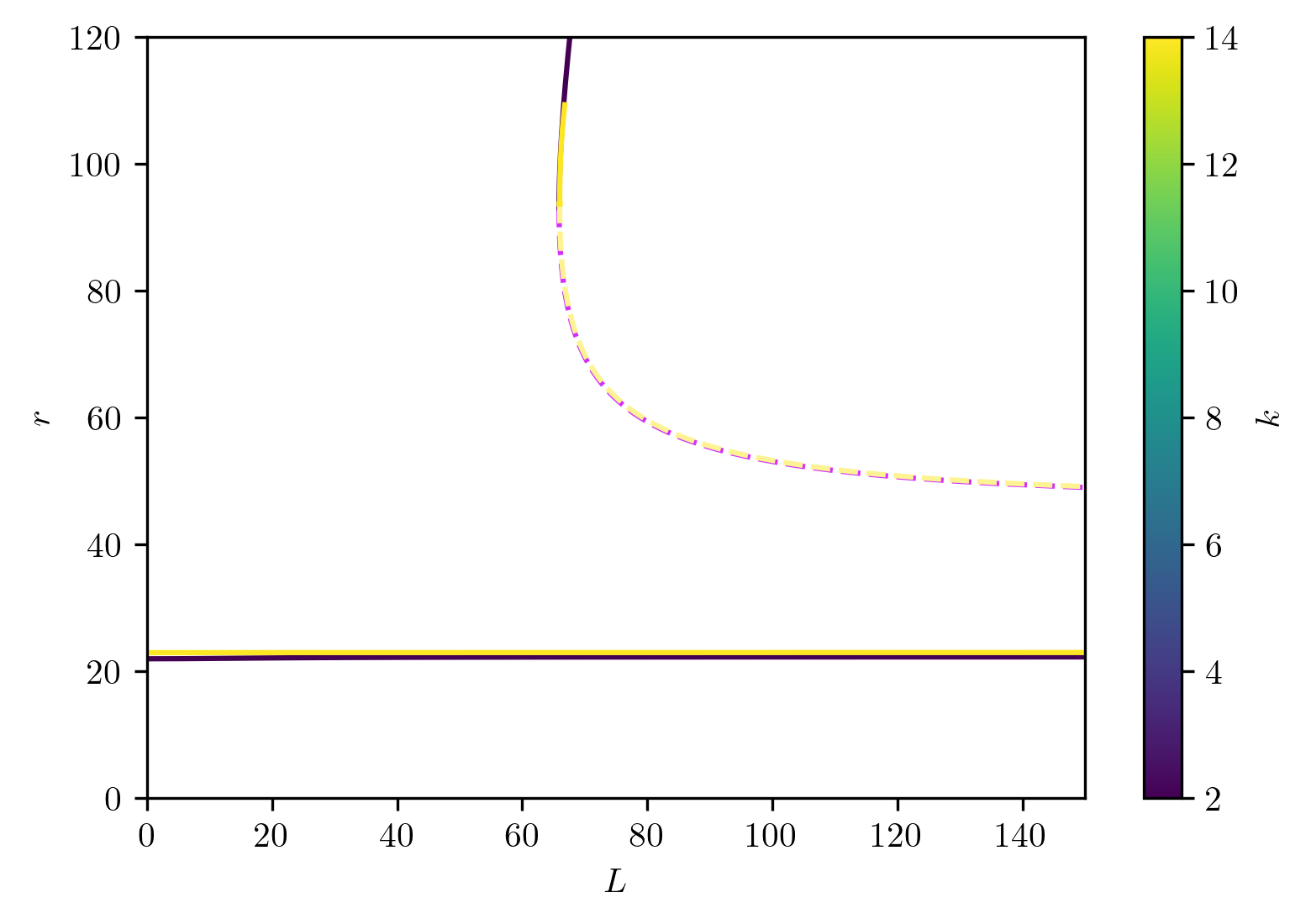}
	\end{subfigure}
	\begin{subfigure}{0.4\textwidth}
		\includegraphics[width=\textwidth]{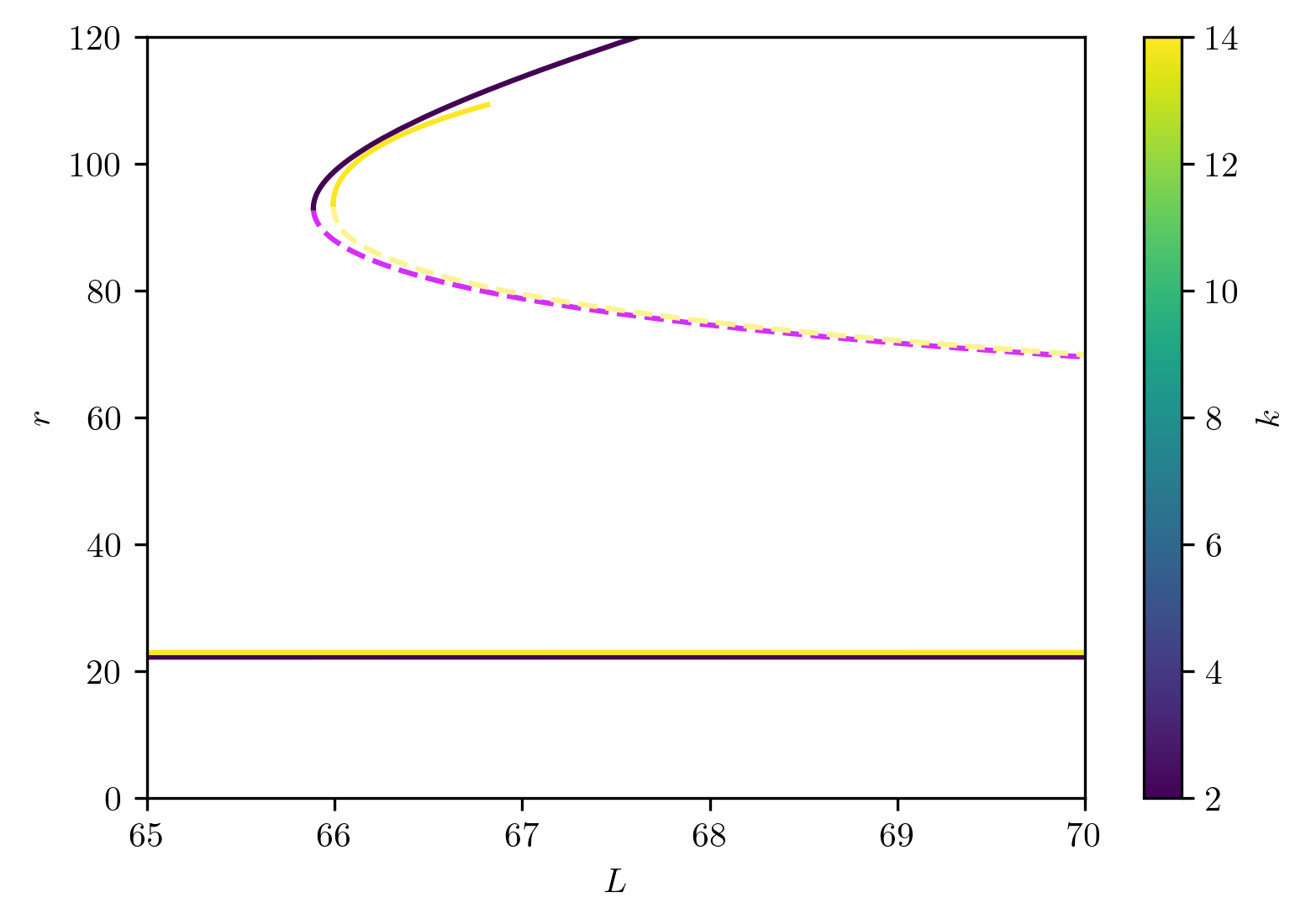}
	\end{subfigure}
    \caption{Stable (solid) and unstable (dashed) circular orbits for massive particles in the space-time of frozen boson star solutions with different $k$. The curves can
    hardly be distinguished. }
    \label{fig:st_pts_Q213}
\end{figure}

Fig.\ref{fig:st_pts_Q213} shows the radius of circular orbits in dependence of the angular momentum $L$ of the particle for the motion in the space-time of the frozen star for all values of $k$. The solid lines indicate stable circular orbits, while the dashed lines are unstable circular orbits The different curves can hardly be distinguished and resemble closely the corresponding curves in black hole space-times. There are two parts: the curve consisting of paired minima and maxima, found at $r>40$, and a curve of individual minima found at $22<r<23$, i.e. within the shell. 
We therefore obtain one stable orbit within the shell for all $L$ and, for sufficiently large $L$ (with the `outer' ISCO having angular momentum $65.8<L_\text{OI}<66.0$ for all four solutions), an additional unstable-stable pair of circular orbits analogous to those seen in black hole space-times. When comparing with Fig.\ref{fig:st_pts_Q_range} we see that this resembles the expected limit as $Q$ becomes large, with the two parts beginning to form, albeit still connected, for the $Q=120$ case.

\begin{figure}[!h]
	\centering
	\begin{subfigure}{0.4\textwidth}
		\includegraphics[width=\textwidth]{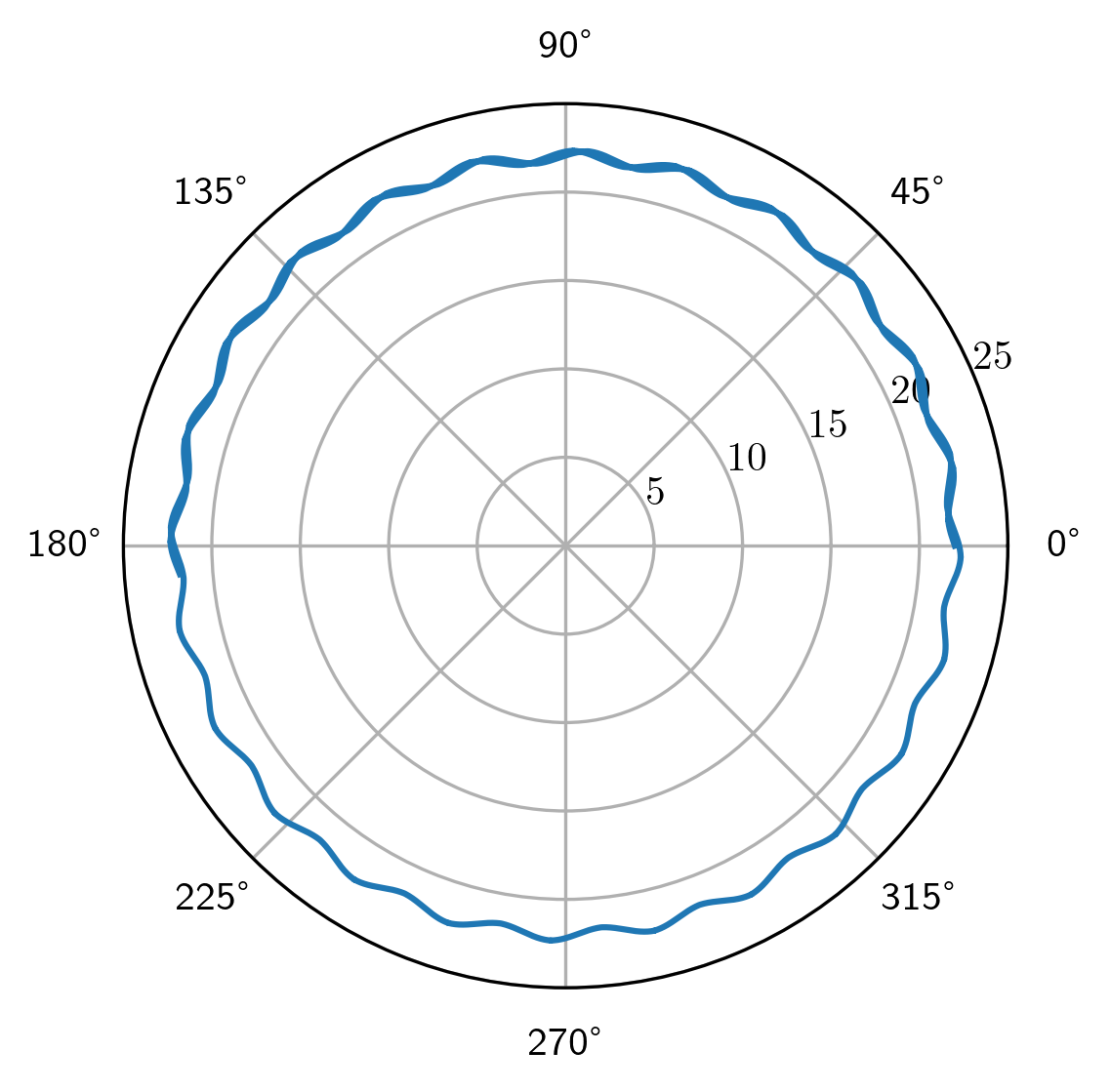}
		\caption{$L=1, E=0.049$}
		\label{fig:orb2_l1_1}
	\end{subfigure}
	\begin{subfigure}{0.4\textwidth}
		\includegraphics[width=\textwidth]{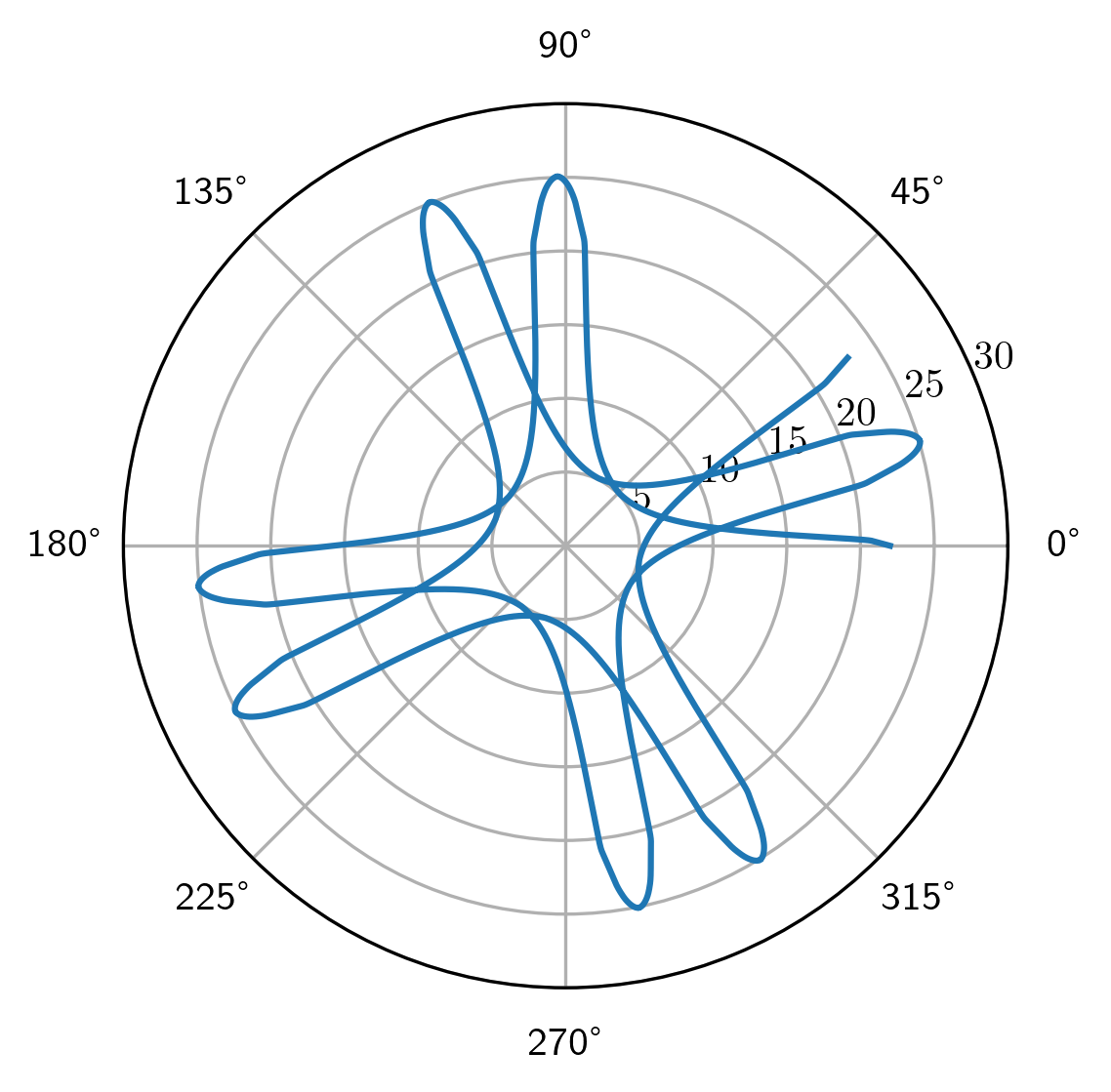}
		\caption{$L=1,E=0.085$}
		\label{fig:orb2_l1_2}
	\end{subfigure}
    	\begin{subfigure}{0.4\textwidth}
		\includegraphics[width=\textwidth]{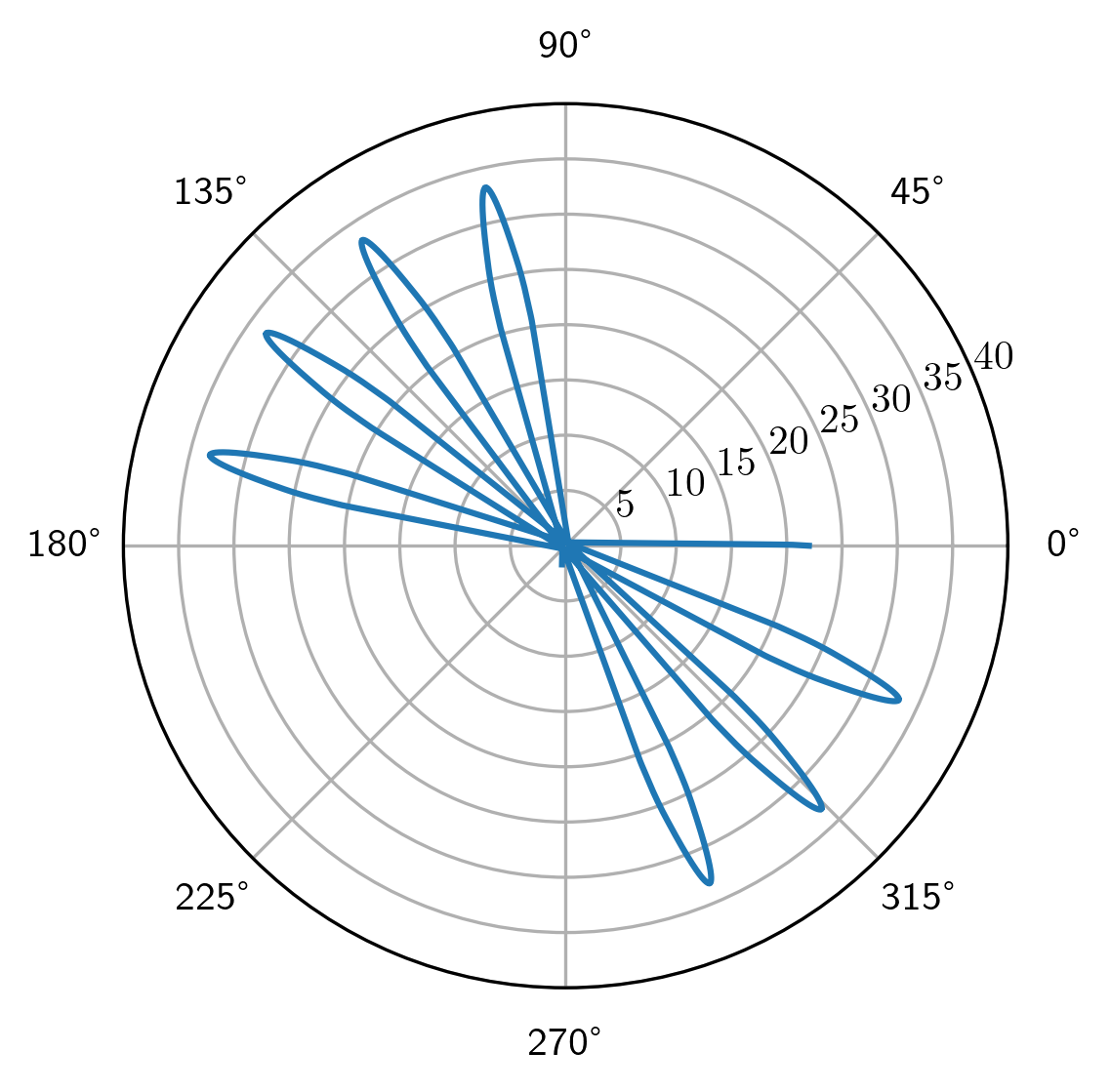}
		\caption{$L=1,E=0.3$}
		\label{fig:orb2_l1_3}
	\end{subfigure}
	\caption{We show orbits for the frozen star solution with $k=2$, for particles with $L=1$}
    \label{fig:orb2_l1}
\end{figure}

Fig.~\ref{fig:orb2_l1} shows geodesic motion around the $k=2$ solution with $L=1$. For this $L$, a circular orbit exists at $r=22.0$. Geodesic particles can be made to oscillate around this or between $r_0>22$ and the origin. In Figs.~\ref{fig:orb2_l1_1} and ~\ref{fig:orb2_l1_2}, the particle turns away from the origin back towards the shell before reaching the origin, while in Fig.~\ref{fig:orb2_l1_3} the particle passes close to the origin to reach the shell diametrically opposite and only then changes direction. Note that particles with non-zero angular momentum can pass arbitrarily close to the origin given sufficient energy, but will never pass through it.

\begin{figure}[!h]
	\centering
	\begin{subfigure}{0.4\textwidth}
		\includegraphics[width=\textwidth]{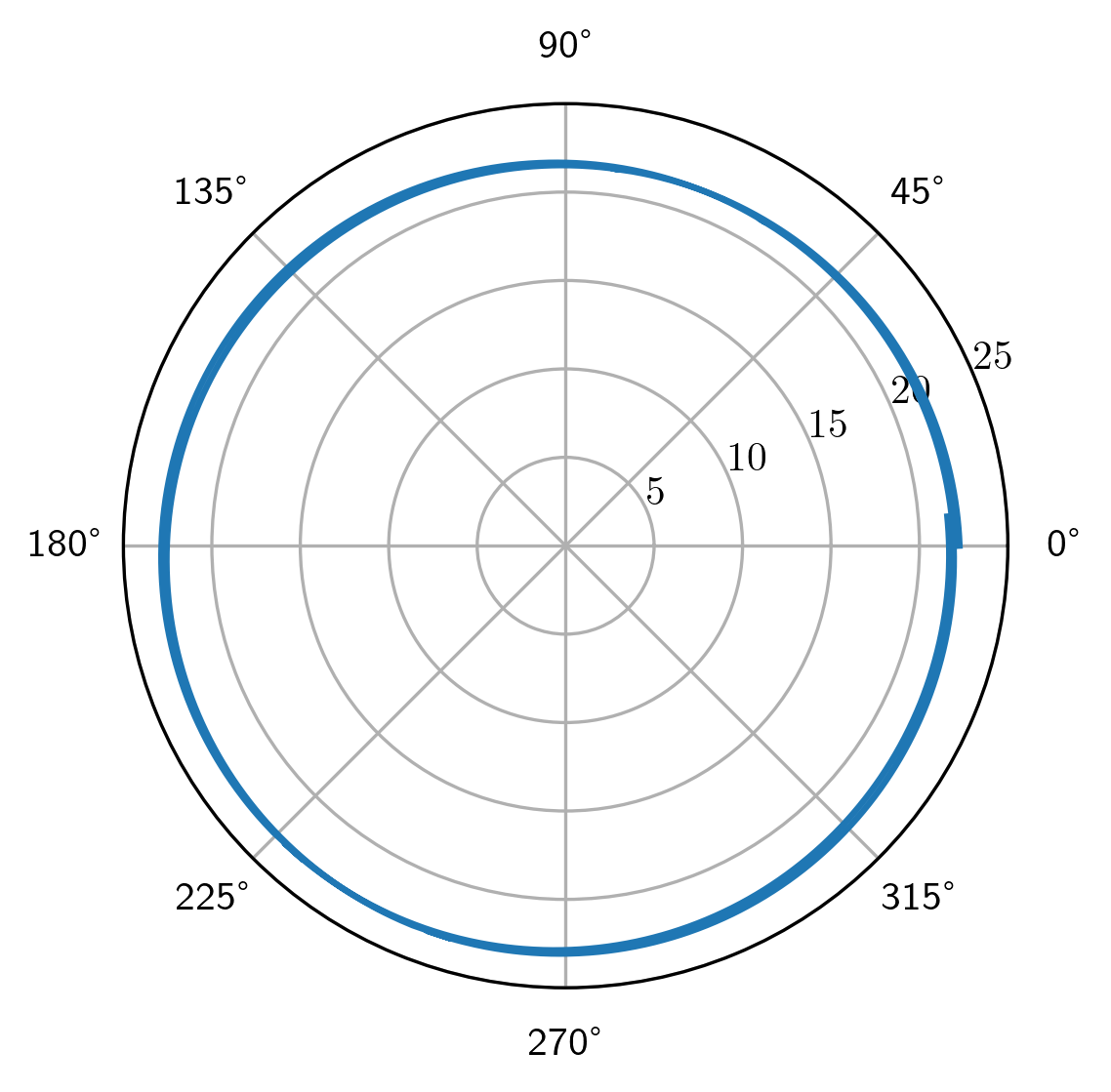}
		\caption{$L=66, E=0.16$}
		\label{fig:orb2_l66_1}
	\end{subfigure}
	\begin{subfigure}{0.4\textwidth}
		\includegraphics[width=\textwidth]{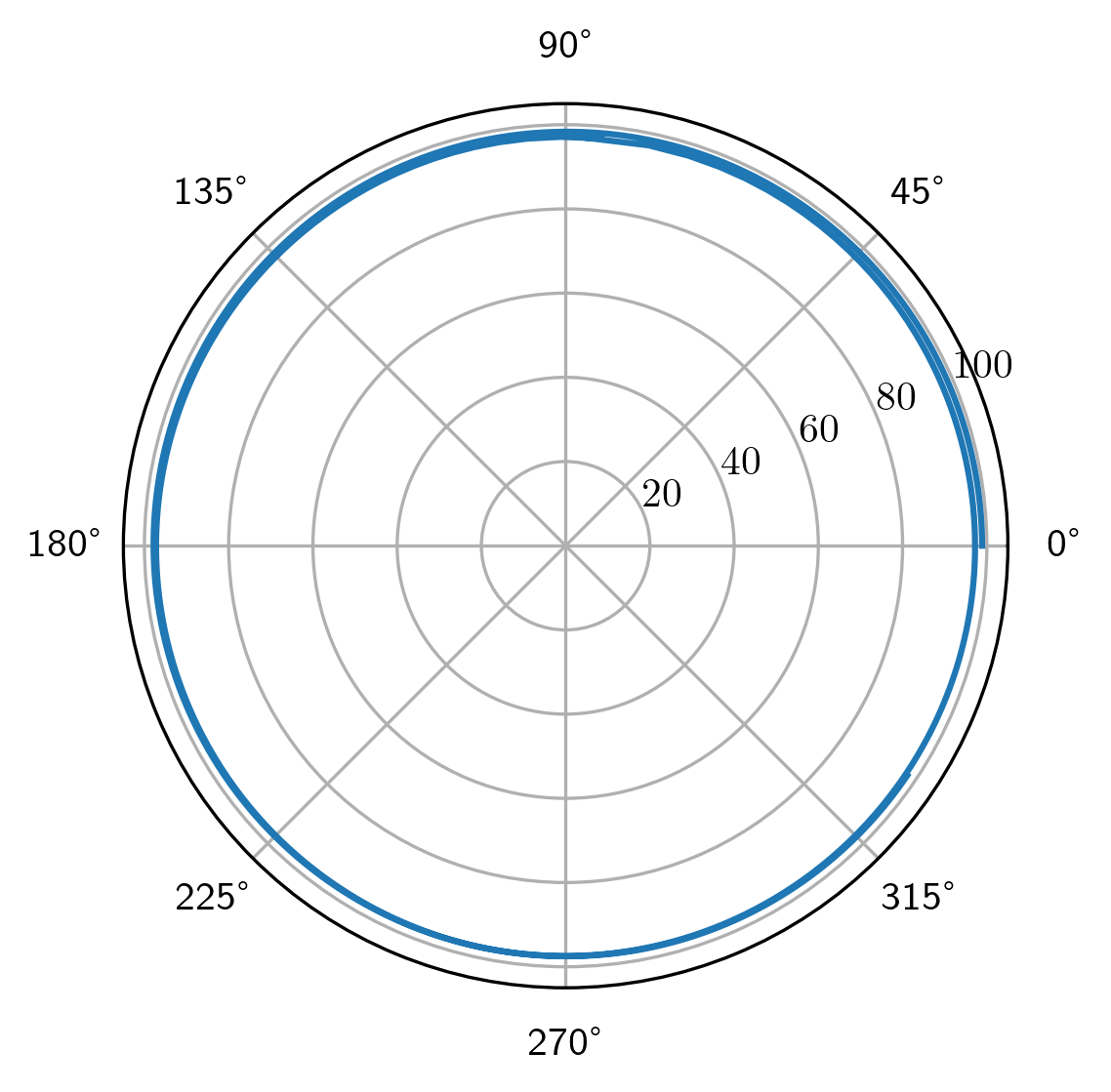}
		\caption{$L=66,E=0.91896$}
		\label{fig:orb2_l66_2}
	\end{subfigure}
    	\begin{subfigure}{0.4\textwidth}
		\includegraphics[width=\textwidth]{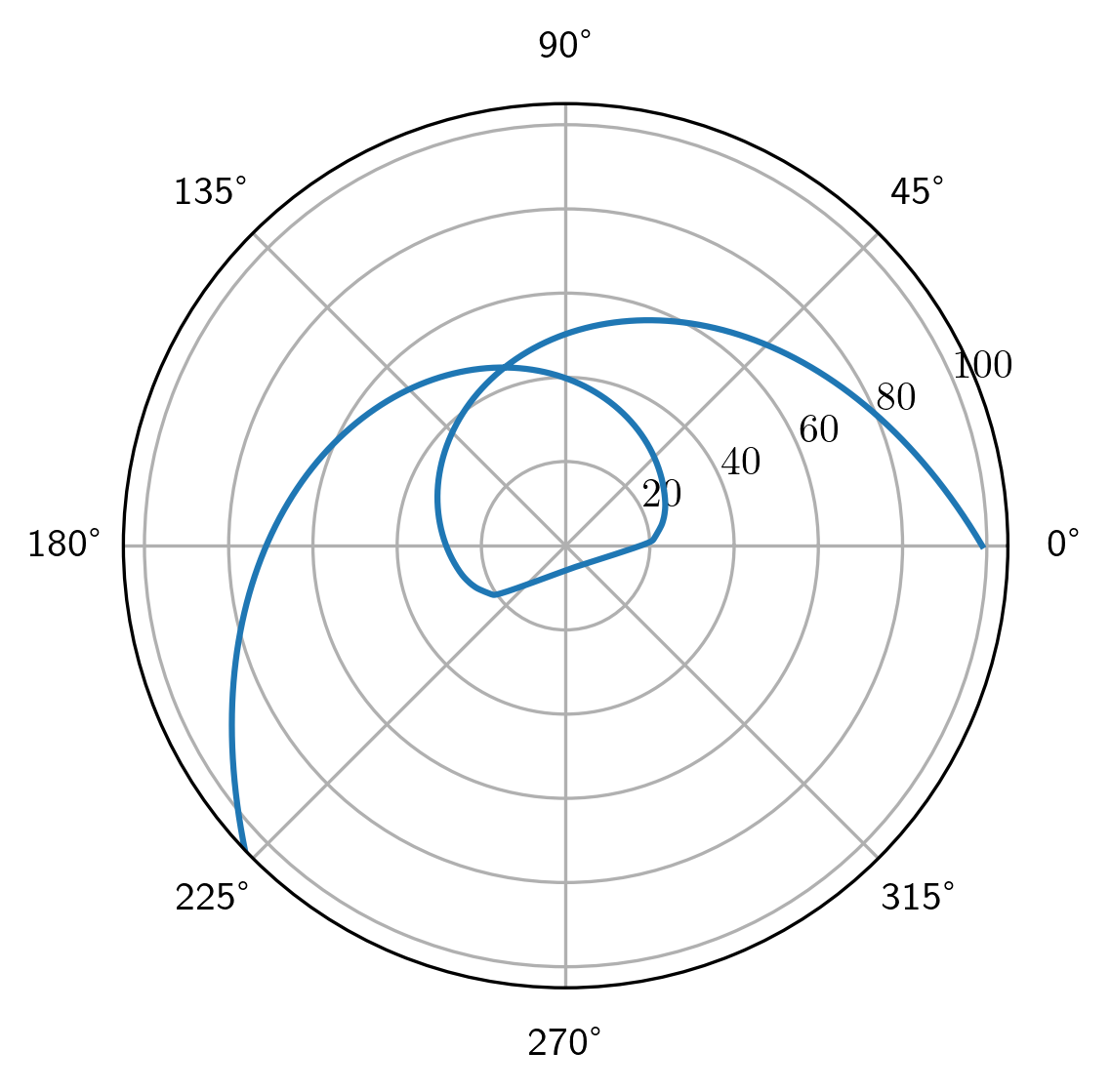}
		\caption{$L=66,E=1.0$}
		\label{fig:orb2_l66_3}
	\end{subfigure}
	\caption{We show orbits for the frozen star solution with $k=2$ with angular momentum $L=66$}
    \label{fig:orb2_l66}
\end{figure}

Fig.\ref{fig:orb2_l66} shows geodesic motion around the $k=2$ solution with $L=66$, i.e. for much larger angular momentum. For this $L$, stable circular orbits exist at $r=22.27$ (see Fig.\ref{fig:orb2_l66_1}) and $r=98.89$ (Fig.\ref{fig:orb2_l66_2}.) Fig.\ref{fig:orb2_l66_3} shows the behaviour when a particle of high energy ($E=1$) approaches the origin: after encountering and passing through the shell, it travels in a straight line through the central region until it emerges from the shell and resumes pseudo-parabolic motion.

\begin{figure}[!h]
	\centering
	\begin{subfigure}{0.4\textwidth}
		\includegraphics[width=\textwidth]{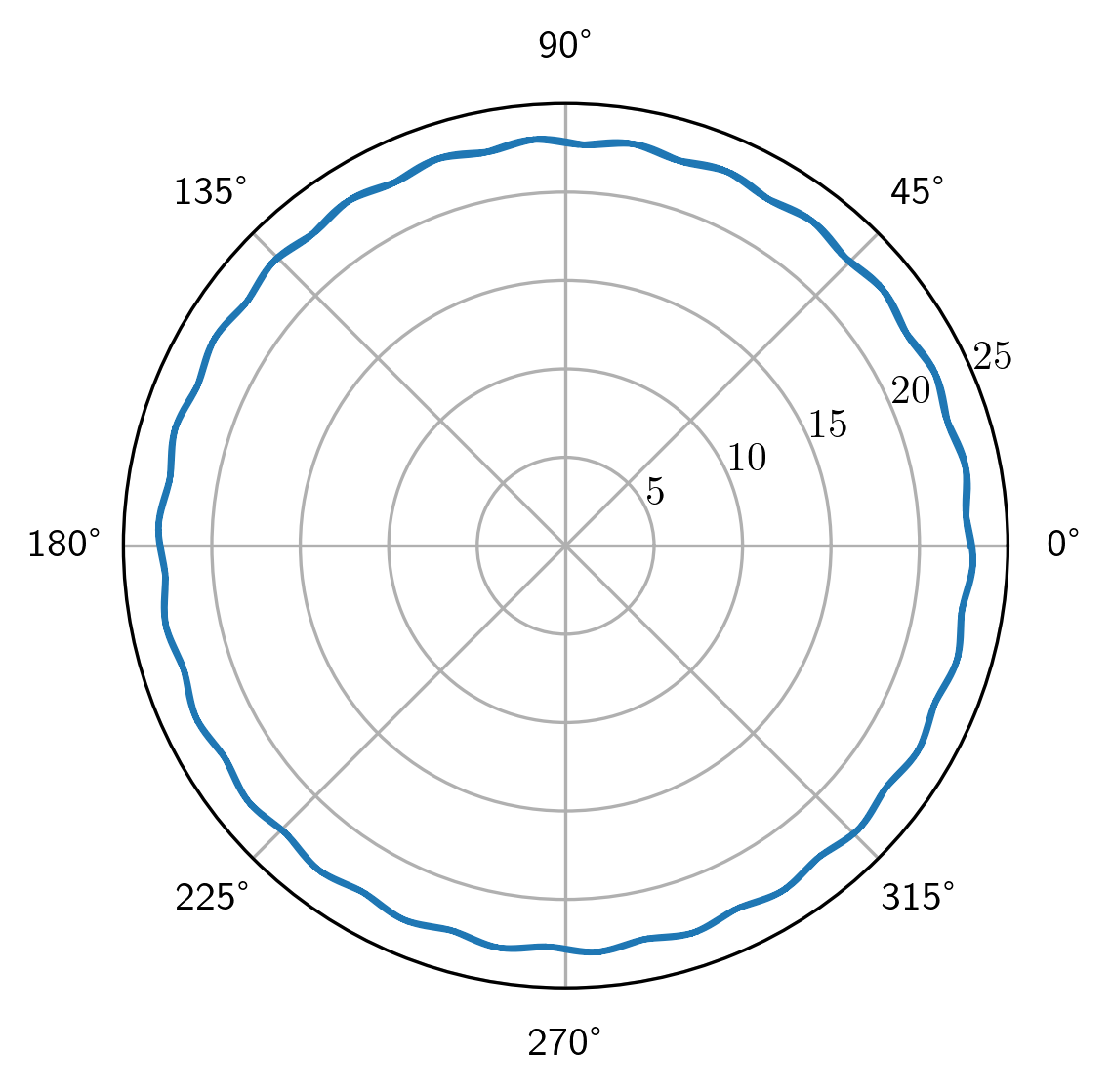}
		\caption{$L=1, E=0.021$}
		\label{fig:orb18_l1_1}
	\end{subfigure}
	\begin{subfigure}{0.4\textwidth}
		\includegraphics[width=\textwidth]{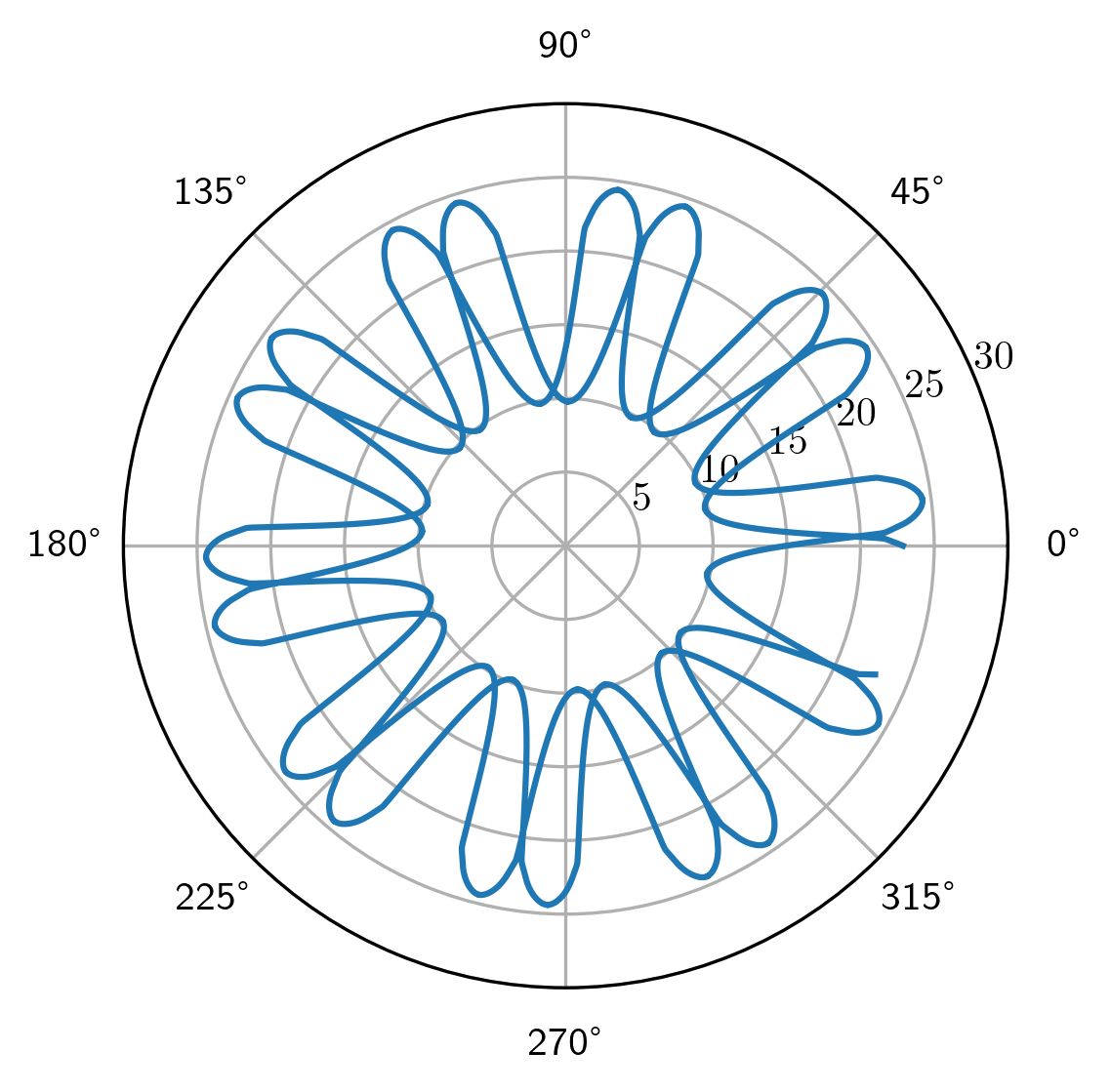}
		\caption{$L=1,E=0.05$}
		\label{fig:orb18_l1_2}
	\end{subfigure}
    	\begin{subfigure}{0.4\textwidth}
		\includegraphics[width=\textwidth]{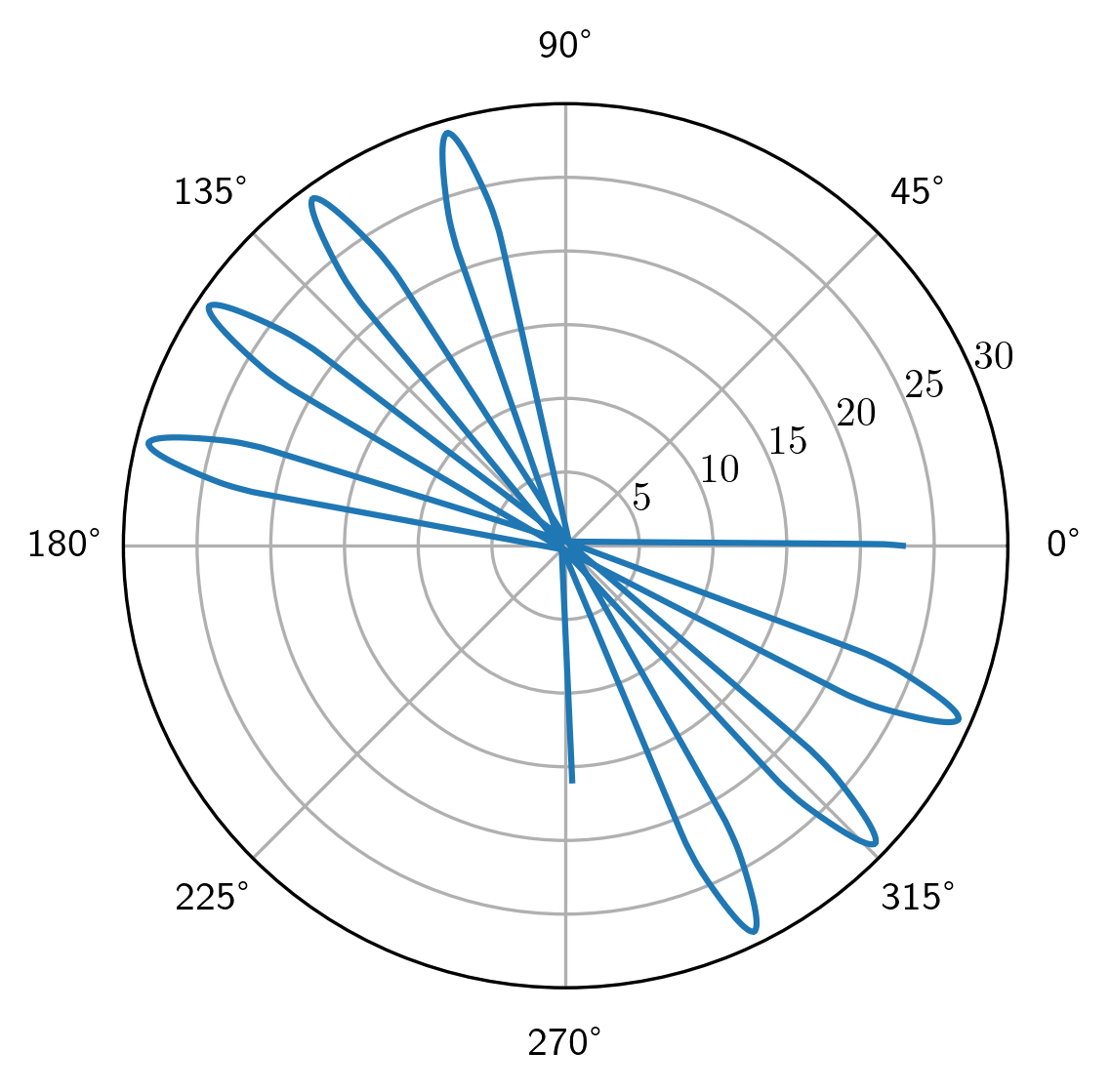}
		\caption{$L=1,E=0.2$}
		\label{fig:orb18_l1_3}
	\end{subfigure}
	\caption{We show orbits for the frozen star solution with $k=18$ with angular momentum $L=1$}
    \label{fig:orb18_l1}
\end{figure}

To compare this with the space-times that have larger $k$, we shown in Fig.\ref{fig:orb18_l1} massive particle motion around a $k=18$ frozen star with $L=1$ for $E=0.021$, $E=0.05$ and $E=0.2$. We see again a range of oscillatory behaviour, from small oscillations around the stable orbit to large oscillations approaching the origin.

\begin{figure}[!h]
	\centering
	\begin{subfigure}{0.4\textwidth}
		\includegraphics[width=\textwidth]{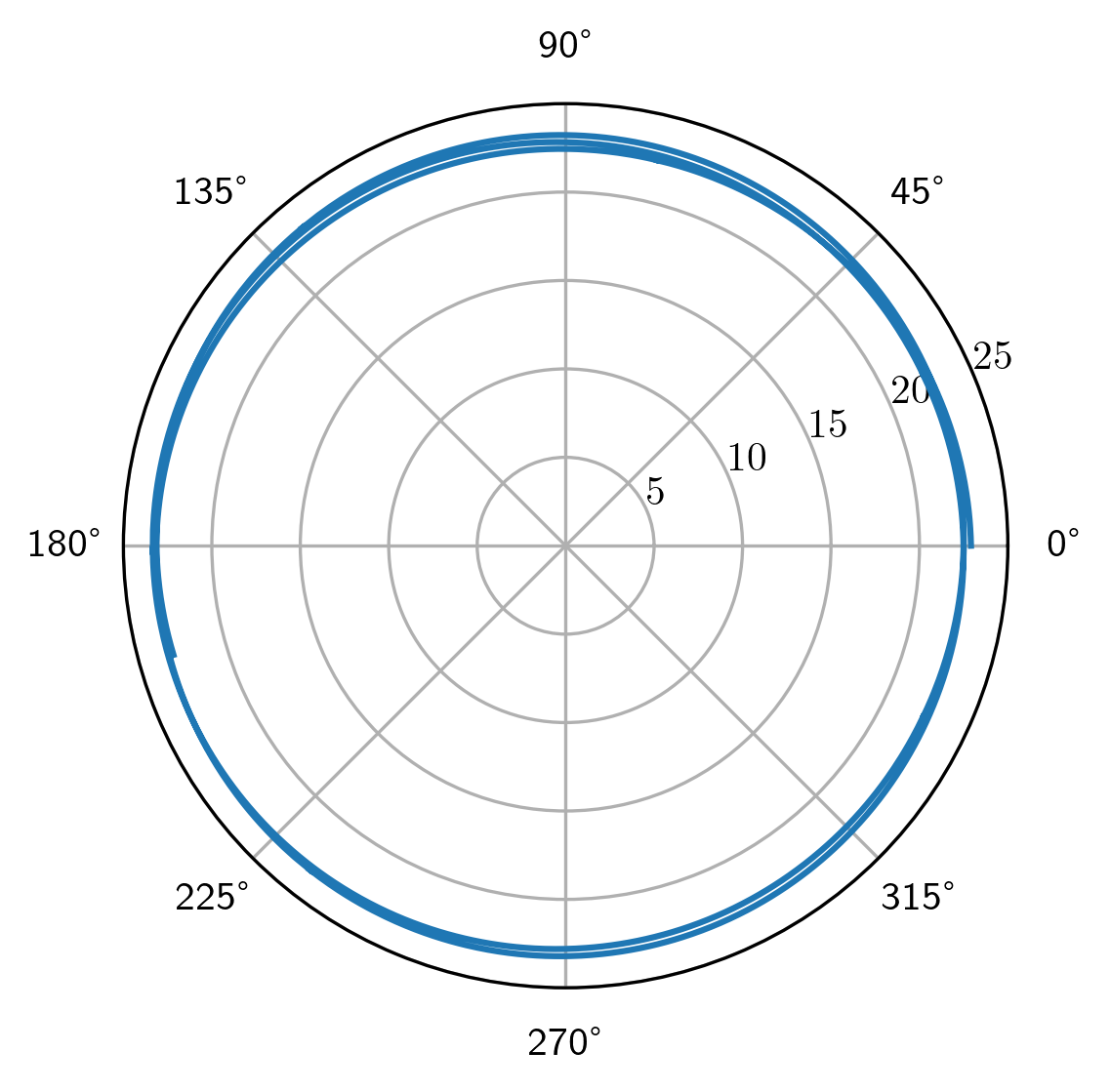}
		\caption{$L=66, E=0.07$}
		\label{fig:orb18_l66_1}
	\end{subfigure}
	\begin{subfigure}{0.4\textwidth}
		\includegraphics[width=\textwidth]{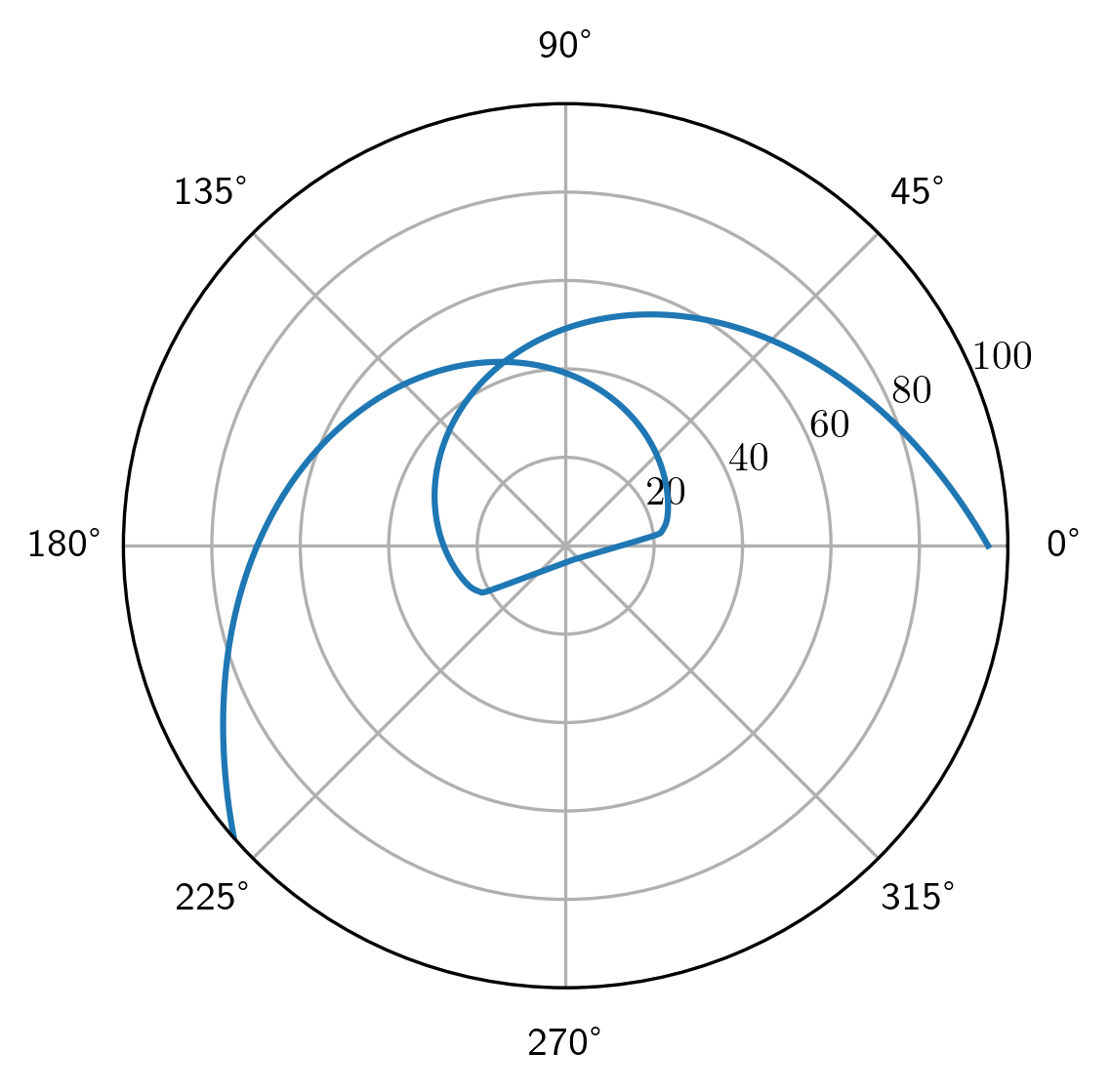}
		\caption{$L=66,E=1.0$}
		\label{fig:orb18_l66_2}
	\end{subfigure}
	\caption{We show orbits for the frozen star solution $k=18$, for particles with $L=66$}
    \label{fig:orb18_l66}
\end{figure}

Finally, for large $L$, $L=66$, we show in Fig.\ref{fig:orb18_l66} the motion of massive particles around $k=18$ frozen star solutions with energies $E=0.07$ and $E=1.0$, respectively. We see again the stable orbit within the shell, and in Fig.\ref{fig:orb18_l66_2} the same pseudo-parabolic motion with a straight line through the interior of the shell.

Let us now discuss the ISCOs. These are again obtained for $L\rightarrow 0$ and hence constitute static orbits. In
Table \ref{tab:shell_ISCO} we show the radius $r_\text{SI}$ and energy  $E_\text{SI}$ of particles on the ISCO within the shell (SI) for these frozen star solutions. We note the very small energies of particles. Since the particles are at rest, the energy is purely potential.

\begin{center}
	\begin{tabular}{||c | c c||} 
		\hline
		$k$ &  $r_\text{SI}$ & $E_\text{SI}$ \\ [0.5ex] 
		\hline\hline
		2 & 22.001 & 0.04852 \\ 
		\hline
		4 & 22.289 & 0.04111  \\
		\hline
		14 & 22.963 & 0.01633 \\
		\hline
		18 & 22.877 & 0.02065 \\ [1ex] 
		\hline
	\end{tabular}
    \captionof{table}{The values of the radius, $r_\text{SI}$, and energy, $E_\text{SI}$, of the ISCO occuring within the shell}\label{tab:shell_ISCO}
\end{center}

Table \ref{tab:outer_ISCO} shows the radius $r$ and angular momentum $L$ of the outer ISCO (OI), corresponding to the one we would find in the RN space-time, for frozen stars. The presence of the orbit in the shell means that in these space-times, this is not the innermost stable circular orbit, but we use the term outer ISCO to denote the stable orbit of smallest radius.
\begin{center}
	\begin{tabular}{||c | c c c||} 
		\hline
		$k$ &  $r_\text{SI}$ & $L_\text{OI}$ & $E_\text{SI}$ \\ [0.5ex] 
		\hline\hline
		2 & 93.110 & 65.887 & 0.91846 \\ 
		\hline
		4 & 93.170 & 65.917 & 0.91849 \\
		\hline
		14 & 93.743 & 65.993 & 0.91860 \\
		\hline
		18 & 93.061 & 65.975 & 0.91854 \\ [1ex] 
		\hline
	\end{tabular}
    \captionof{table}{The values of the radius, $r_\text{OI}$, angular momentum, $L_\text{OI}$, and energy, $E_\text{OI}$, of the ISCO occuring outside of the shell}\label{tab:outer_ISCO}
\end{center}

In order to be able to compare further with the RN solution, we note that the mass $M$ of the frozen star solutions with $Q=213$ is $M\approx 23.3$ for all values of $k$. We find that the radius of the ISCO inside the shell is $r_{ISCO}\approx 22.00$ for
$k=2$, $r_{ISCO}\approx 22.29$ for $k=4$, $r_{ISCO}\approx 22.96$ for $k=14$, and $r_{ISCO}\approx 22.88$ for $k=18$. Noticeably, this has $M>r_\text{ISCO}$. For extremal black holes, a stable null orbit occurs on the horizon $r_h=M$ \cite{Pradhan:2011}, whereas here we have a stable timelike orbit slightly within $r=M$.

\begin{figure}[!h]
    \centering
    \includegraphics[width=0.6\textwidth]{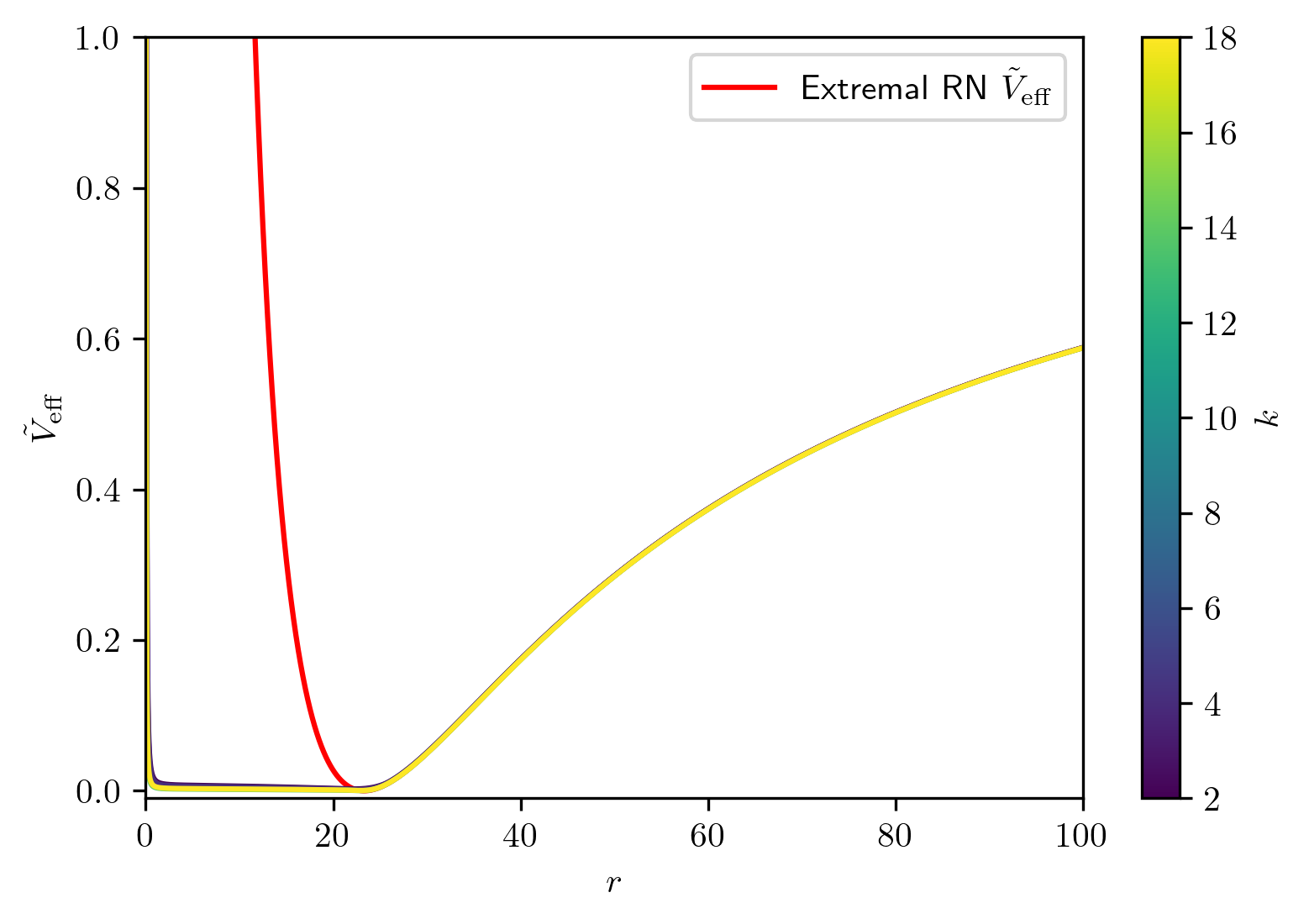}
    \caption{We show the energy-independent effective potential for the frozen star solutions, with $L=1$ and for massive particles, compared with that for extremal Reissner-Nordstr\"om with $M=23.3$ shown in red. Note that for $r<23.3$, the results from the extremal Reissner-Nordstr\"om black hole should be interpreted carefully, since these lie within the event horizon; they are included here for clarity rather than a direct physical comparison.}
    \label{fig:FS_Potential}
\end{figure}

Fig.\ref{fig:FS_Potential} shows the energy-independent effective potential for the frozen star solutions, with $L=1$ and for massive particles, compared with that for extremal Reissner-Nordstr\"om with $M=23.3$ shown in red. For $r<23.3$, corresponding to the interior of the black hole, the effective potential is nearly flat for the frozen star solutions (with the exception of the origin, due to the non-zero angular momentum.) Outside of $r=23.3$, however, the effective potentials for all four frozen star solutions very closely coincide with that for RN.

\begin{figure}[!h]
	\centering
	\begin{subfigure}{0.4\textwidth}
		\includegraphics[width=\textwidth]{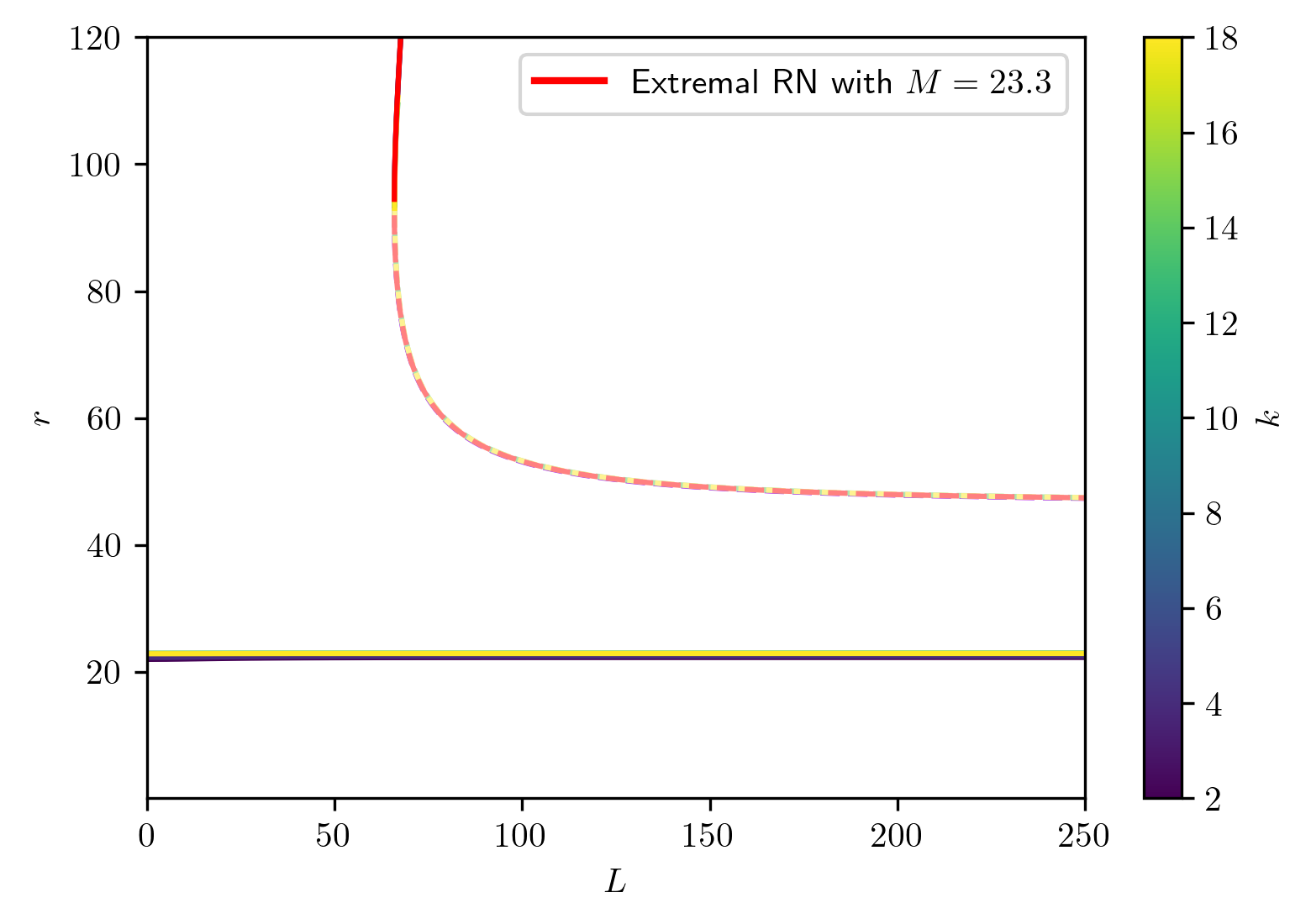}
	\end{subfigure}
	\begin{subfigure}{0.4\textwidth}
		\includegraphics[width=\textwidth]{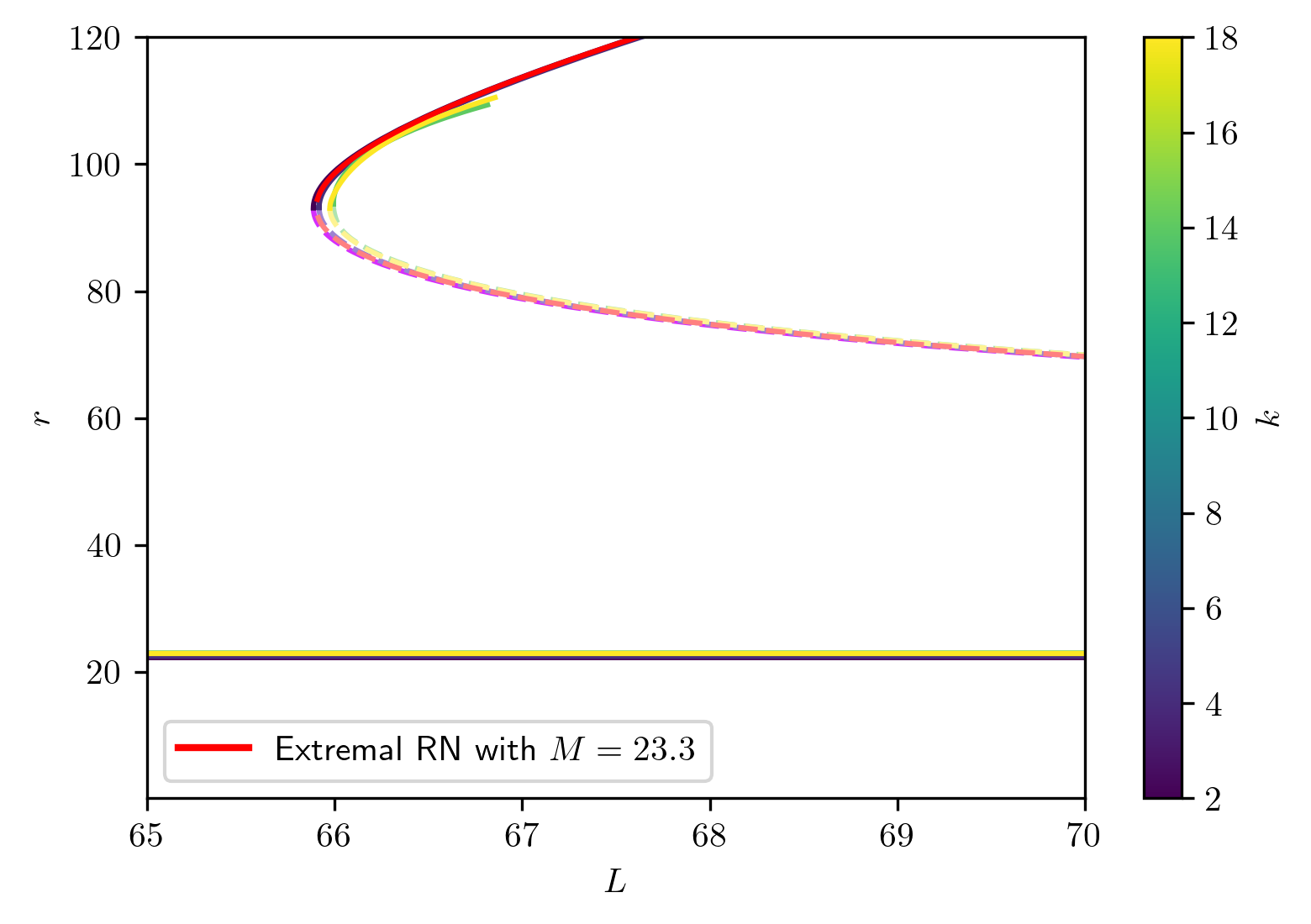}	
	\end{subfigure}
    \caption{The radii of circular orbits of frozen boson stars with $k=2$, $4$, $14$, $18$ and of the extremal RN black hole with same mass.}
    \label{fig:comp_st_pts}
\end{figure}

In Fig.\ref{fig:comp_st_pts} we compare the radii of circular orbits in  an extremal Reissner-Nordstr\"om black hole space-time  with mass $M=23.3$ with those for the frozen stars. It is clear that they match outside the shell as here the space-time is RN.

\subsubsection{Massive charged test particles}\label{sec:fs_charged}
\begin{figure}[!h]
    \centering
    \includegraphics[width=0.6\textwidth]{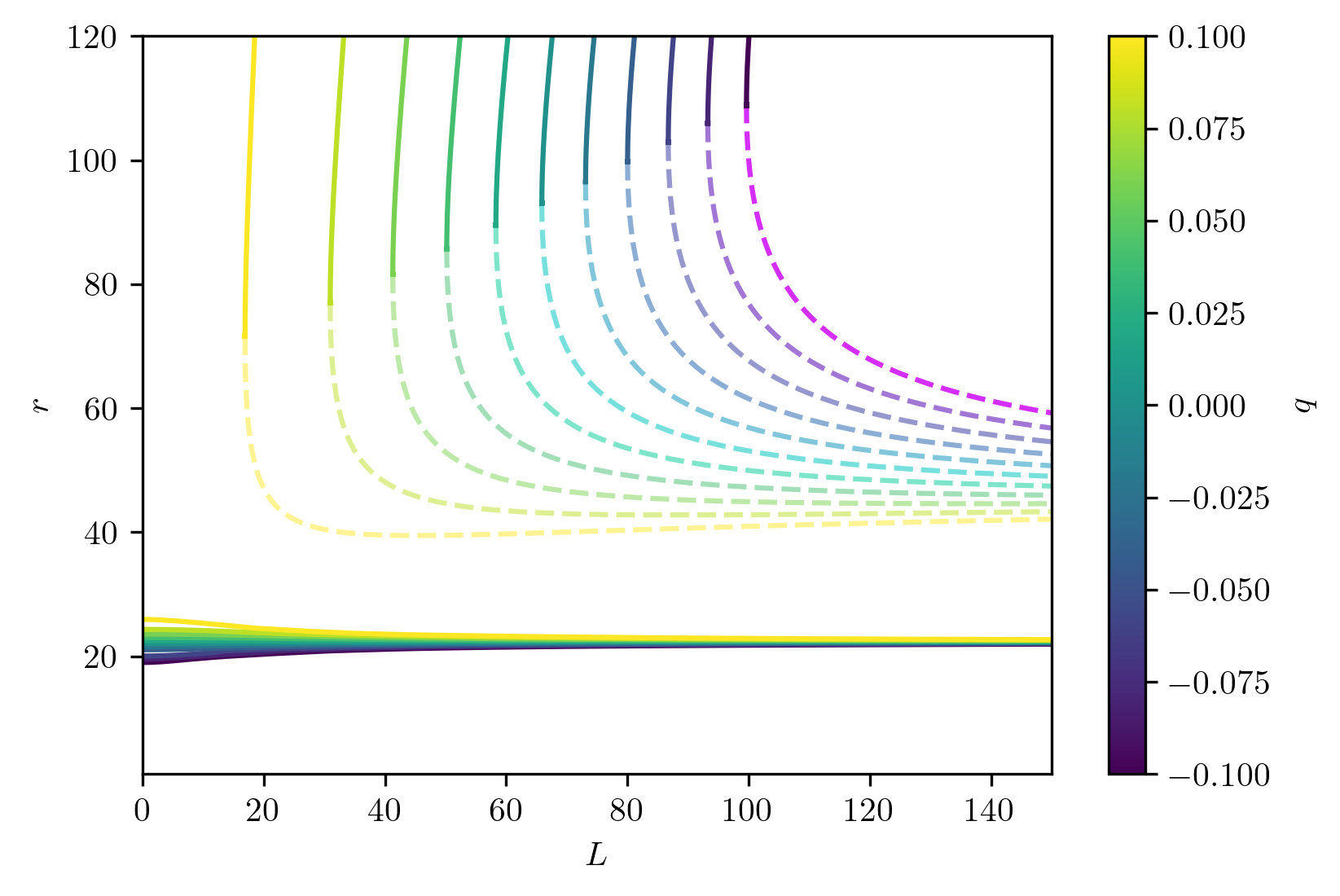}
    \caption{Radius of the circular orbits $r$ against angular momentum $L$ in the space-time of the $k=2$ frozen star, for charged particles with varying $q$}
    \label{fig:st_pts_Q213_charge}
\end{figure}

Fig.~\ref{fig:st_pts_Q213_charge} shows the location and angular momentum of stationary points for $k=2$, for charged particles with varying $q$. We see the same qualitative behaviour for all charges. Positively charged particles have the orbit near the shell occurring at greater radius than for neutral or negatively charged particles, but have their outer ISCO at smaller radius and angular momentum. Negatively charged particles show the reversed behaviour. For the outer pair of orbits, this behaviour mirrors that in the RN space-time.
Note that for particles with charge of sufficient magnitude, the inner stable orbit that for neutral particles occurred within the shell can now, for small $L$, occur on either side of the shell.

\begin{figure}[!h]
	\centering
	\begin{subfigure}{0.4\textwidth}
		\includegraphics[width=\textwidth]{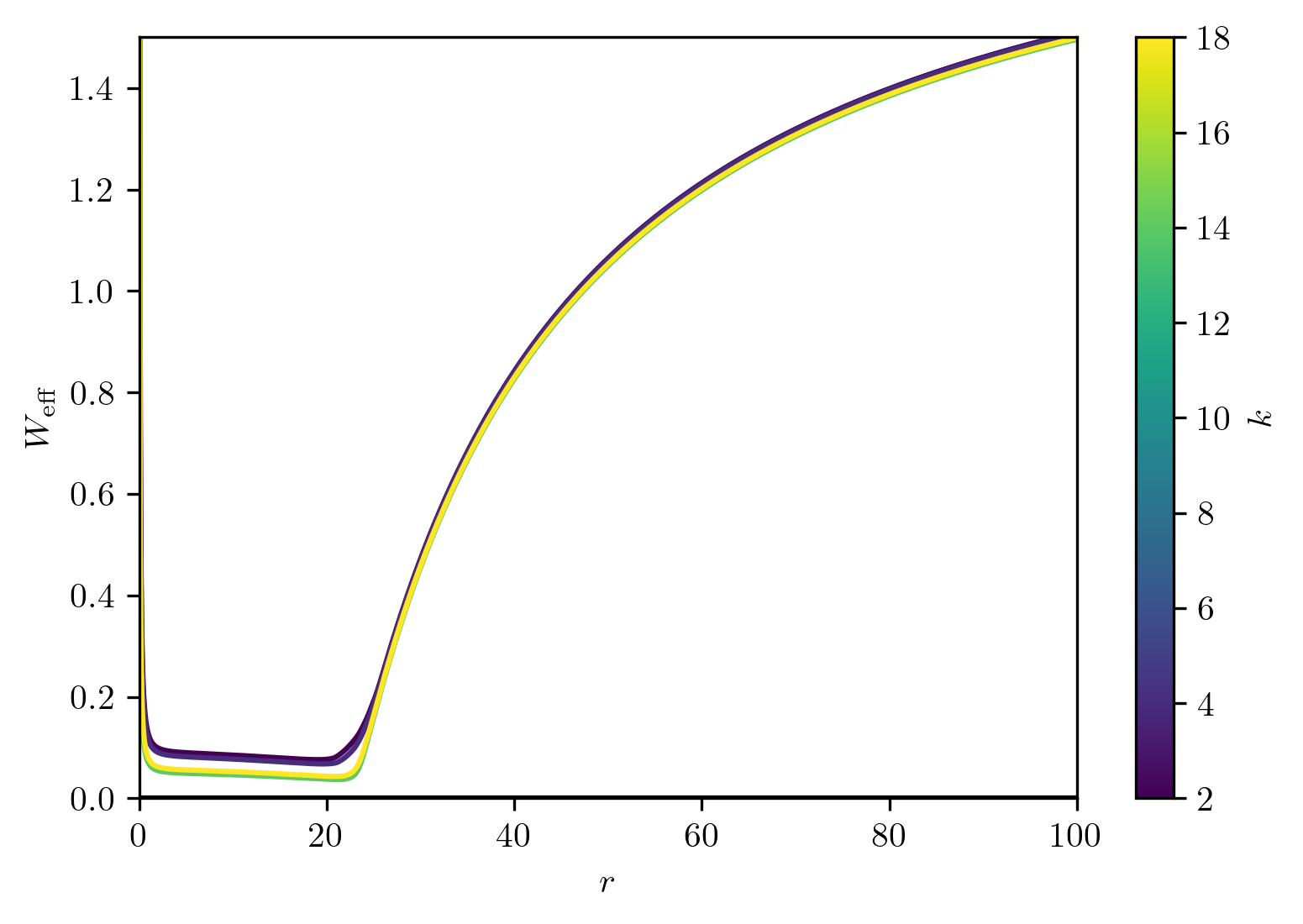}
		\caption{$q=-0.1$}
		\label{fig:fs_pot_qm0.1}
	\end{subfigure}
	\begin{subfigure}{0.4\textwidth}
		\includegraphics[width=\textwidth]{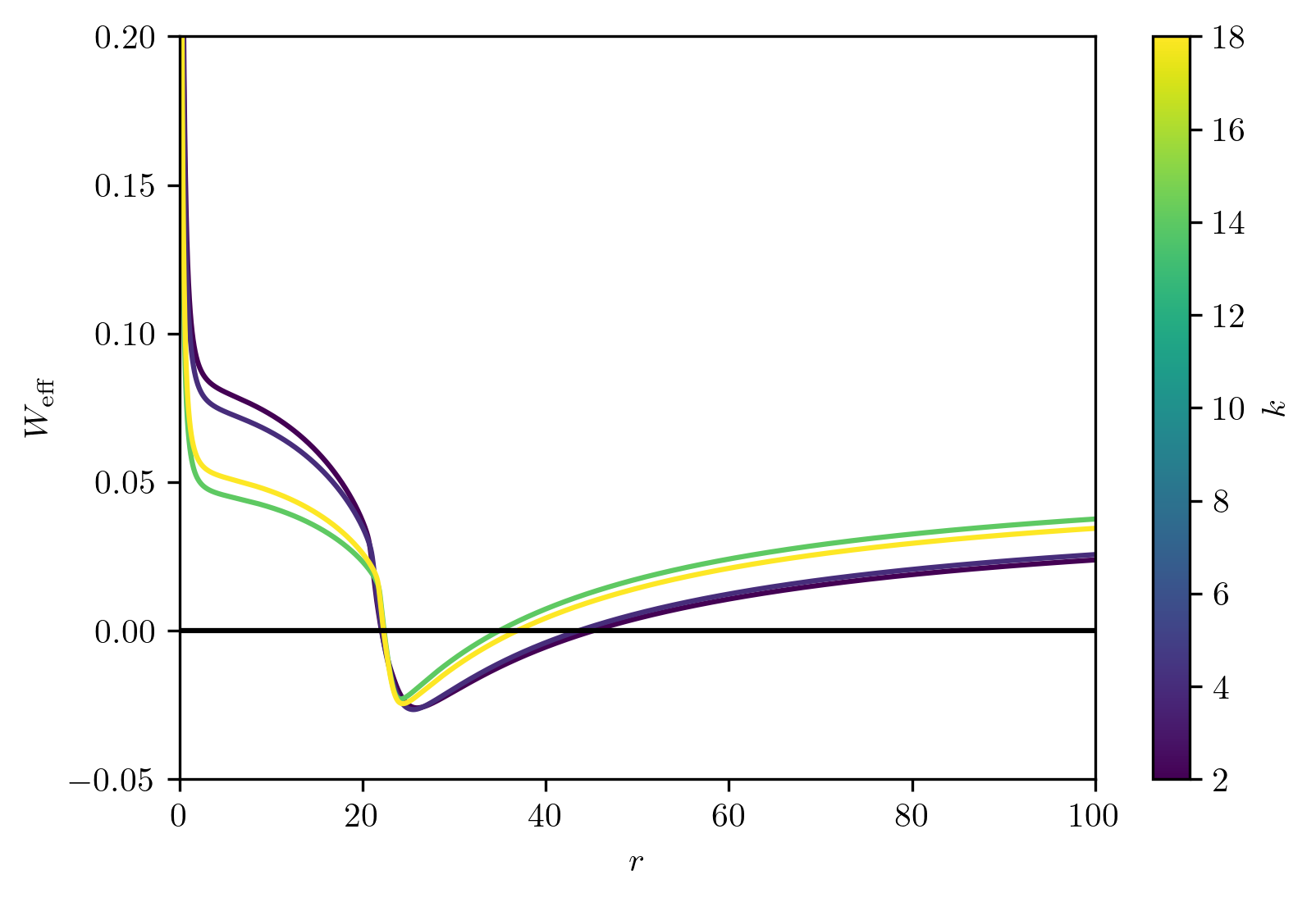}
		\caption{$q=+0.1$}
		\label{fig:fs_pot_qp0.1}
	\end{subfigure}
	\caption{We show the modified potential $W_\text{eff}$ for massive particles with charge $q=\pm0.1$ and angular momentum $L=1$ in frozen star space-times}
    \label{fig:fs_pot_qpm0.1}
\end{figure}

However, not all of the theoretical circular orbits within the shell are actually achievable. Fig.~\ref{fig:fs_pot_qpm0.1} shows the modified potential $W_\text{eff}$ for particles with charge $q=\pm0.1$. While for $q=-0.1$ (see Fig.~\ref{fig:fs_pot_qm0.1}) $W_\text{eff}$ appears as expected, for $q=+0.1$ (Fig.~\ref{fig:fs_pot_qp0.1}) we find that the circular orbits occur at $W_\text{eff}<0$. Since circular orbits for charged particles require $E=W_\text{eff}$, and $E$ cannot be negative, no particle can actually remain in this circular orbit. 

For $q=-0.1$, we see a very broad minimum in the shell, with $W_\text{eff}$ remaining small in the interior of the shell (away from the origin), which would allow the oscillations discussed above for uncharged particles in frozen star space-times, as well as for uncharged particles in boson star space-times.

For $q=+0.1$, we see that $W_\text{eff}$ is monotonically decreasing for $r$ below the radius of the circular orbit, but monotonically increasing for $r$ greater. This means that all particles will either escape to infinity, or oscillate around the circular orbit with wide variations in radius -  smaller oscillations would again require $E<0$.

\begin{figure}[!h]
    \centering
    \includegraphics[width=0.6\textwidth]{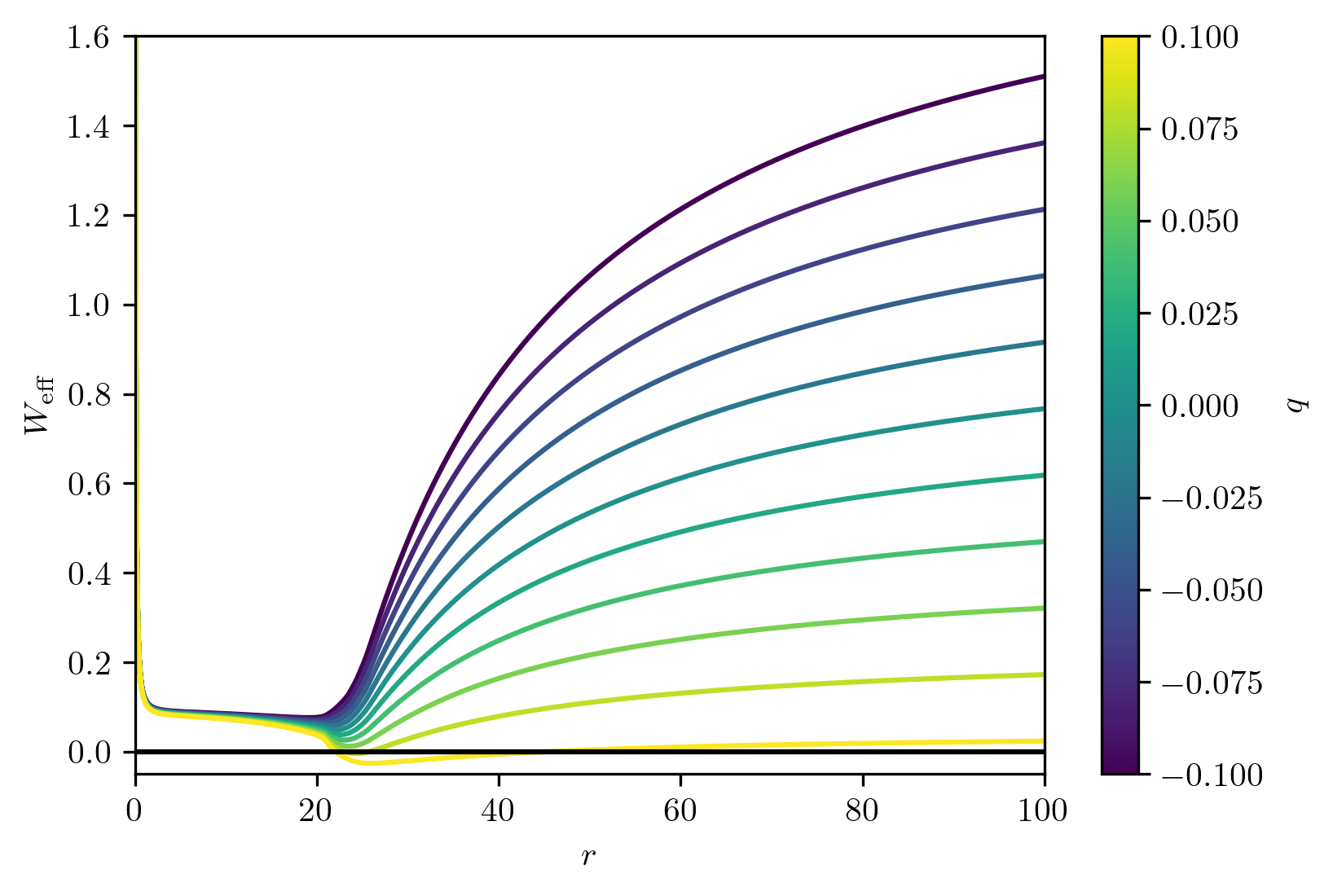}
    \caption{We show the modified potential $W_\text{eff}$ for massive particles with charge $q\in[-0.1,+0.1]$ and angular momentum $L=1$ in the frozen star space-time with $k=2$}
    \label{fig:fs_pot_qk2}
\end{figure}

Fig.~\ref{fig:fs_pot_qk2} shows $W_\text{eff}$ for $q\in[-0.1,+0.1]$ for the space-time with $k=2$. We see that in the interior of the shell, the charge of the particle has little effect on the modified potential. Outside of the shell, however, the value of the effective potential varies considerably. It is significantly easier for particles with positive charge to escape to infinity due to the repulsive electromagnetic force.

\begin{figure}[!h]
	\centering
	\begin{subfigure}{0.4\textwidth}
		\includegraphics[width=\textwidth]{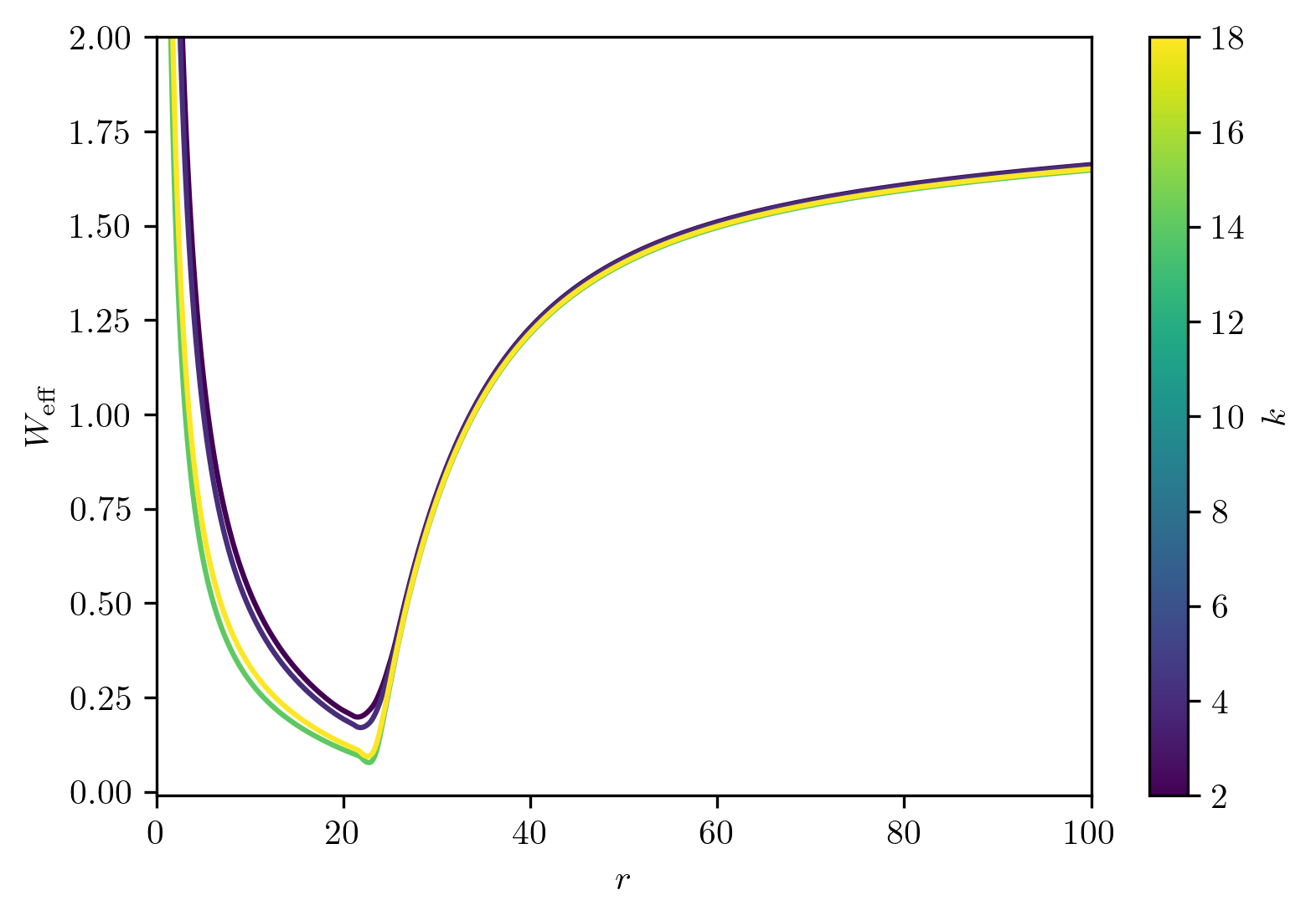}
		\caption{$q=-0.1$}
		\label{fig:fs_pot_qm0.1_L66}
	\end{subfigure}
	\begin{subfigure}{0.4\textwidth}
		\includegraphics[width=\textwidth]{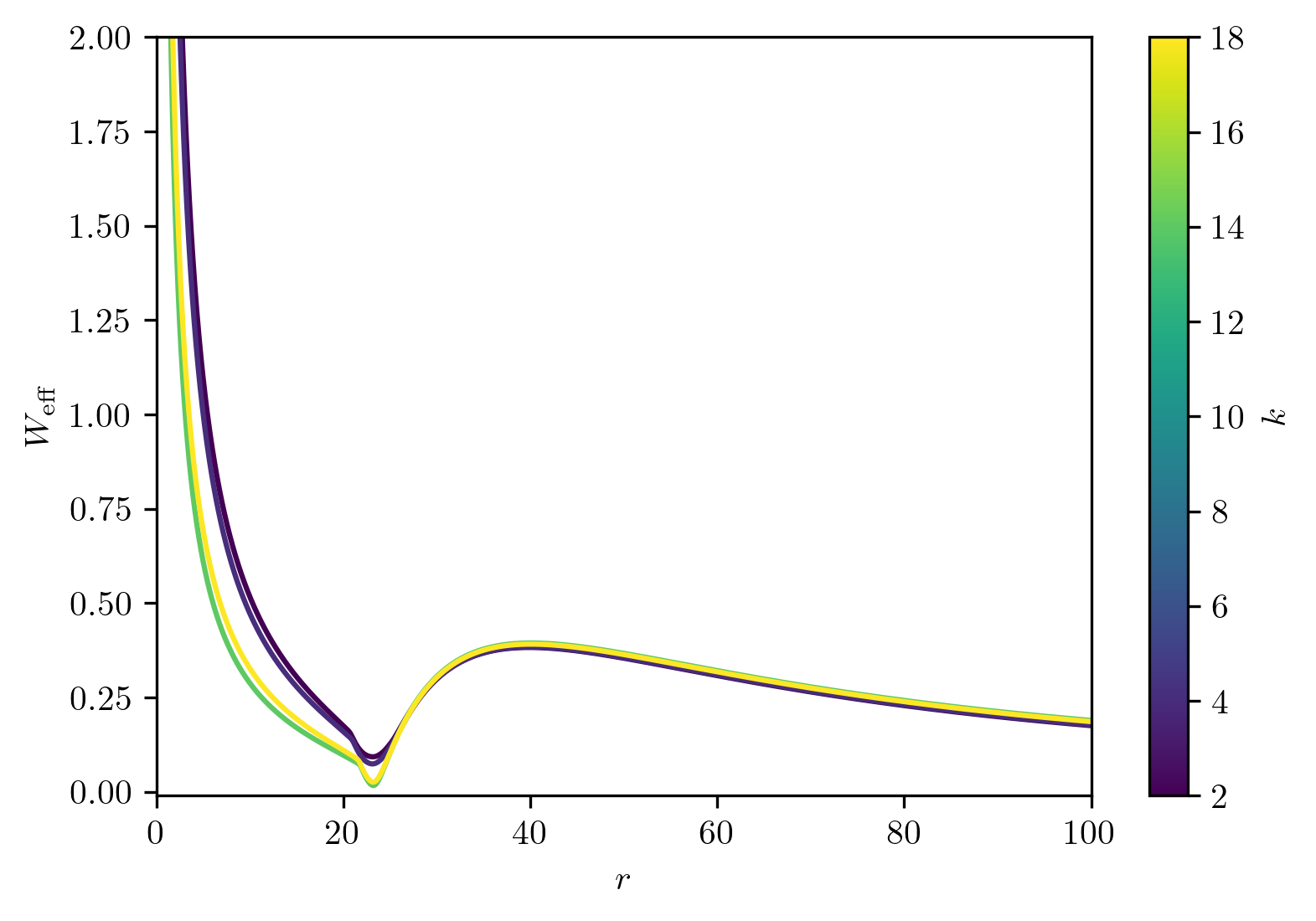}
		\caption{$q=+0.1$}
		\label{fig:fs_pot_qp0.1_L66}
	\end{subfigure}
	\caption{We show the modified potential $W_\text{eff}$ for massive particles with charge $q=\pm0.1$ and angular momentum $L=66$ in frozen star space-times}
    \label{fig:fs_pot_qpm0.1_L66}
\end{figure}

In Fig.~\ref{fig:fs_pot_qpm0.1_L66} we show $W_\text{eff}$ for particles with charge $q=\pm0.1$ and angular momentum $L=66$. 

There are three significant differences between this case and $L=1$:
\begin{enumerate}
    \item $W_\text{eff}>0$ for both positive and negatively charged particles, and hence the circular orbits at the minimum can now occur
    \item The minimum within the shell has narrowed, and hence particles will not oscillate around the stationary point with such frequency
    \item For $q=+0.1$, we now see a maximum, representing an unstable orbit. We would also expect to see a minimum, representing a stable orbit; however this would occur at $r>120$, and hence outside the region of space-time we have modelled.
\end{enumerate}

\begin{figure}[!h]
    \centering
    \includegraphics[width=0.6\textwidth]{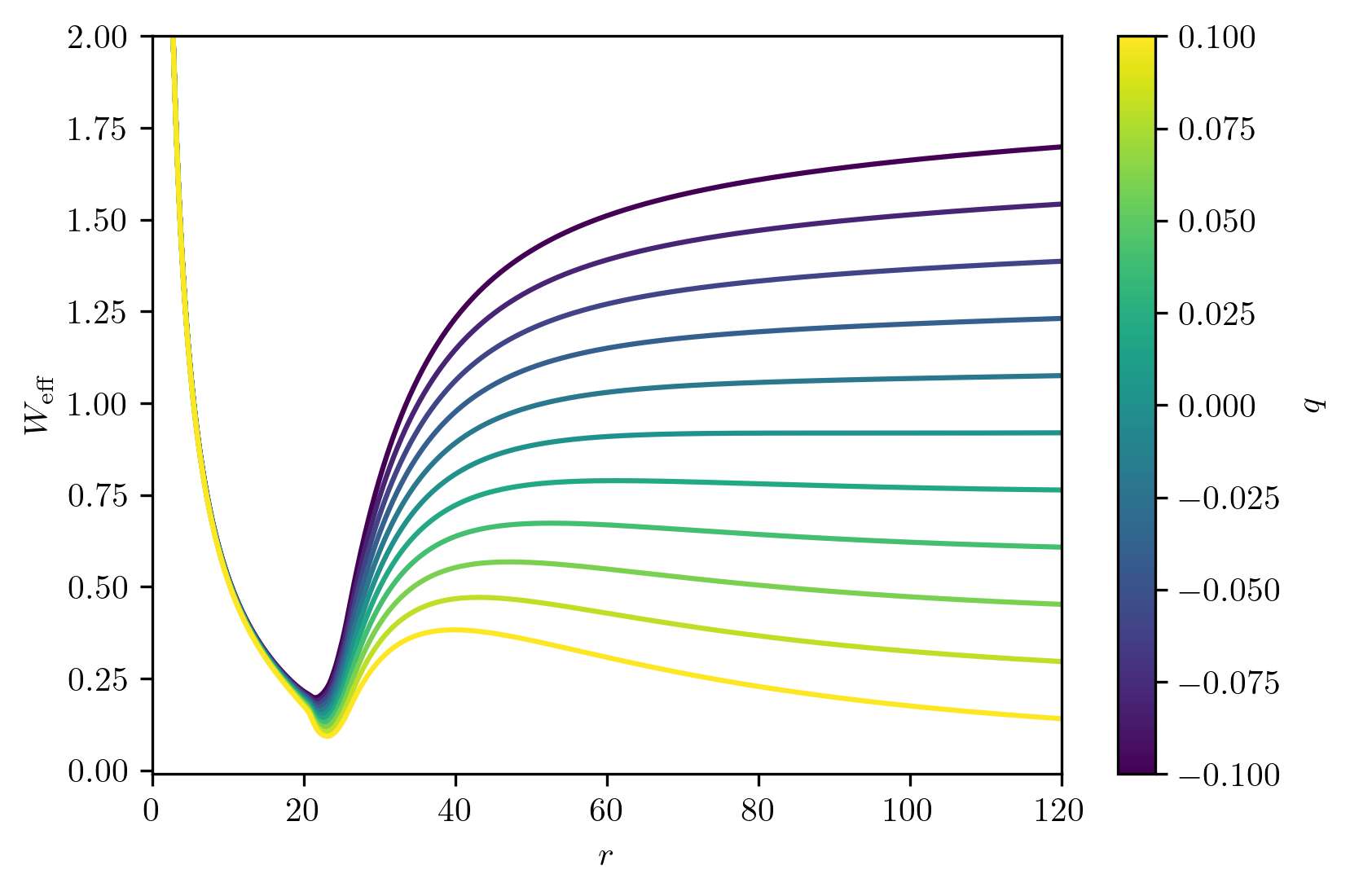}
    \caption{We show the modified potential $W_\text{eff}$ for massive particles with charge $q\in[-0.1,+0.1]$ and angular momentum $L=66$ in the frozen star space-time with $k=2$}
    \label{fig:fs_pot_qk2_L66}
\end{figure}

\subsubsection{Massless test particles}
Table \ref{tab:FS_light_radius} shows the location of stable and unstable circular orbits for massless particles ($\delta=0$). As discussed above, the angular momentum $L$ is not a relevant independent quantity, and hence we observe only one stable circular orbit.

\begin{center}
	\begin{tabular}{||c | c c||} 
		\hline
		$k$ &  $r_\text{stable}$ & $r_\text{unstable}$ \\ [0.5ex] 
		\hline\hline
		2 & 22.304 & 46.526 \\ 
		\hline
		4 & 22.508 & 46.566  \\
		\hline
		14 & 22.991 & 46.649 \\
		\hline
		18 & 22.919 & 46.640 \\ [1ex] 
		\hline
	\end{tabular}
    \captionof{table}{Position in $r$ of the stable and unstable circular orbits for massless particles (light rings) for $Q=213$. }\label{tab:FS_light_radius}
\end{center}
To compare with the extremal RN solution, we recall the results in \cite{Pradhan:2011}, who identified a stable circular orbit at $r=M$ and an unstable circular orbit at $r=2M$. This lines up closely with the results in Table \ref{tab:FS_light_radius}, although as for massive particles, we observe our stable orbit lies at a radius slightly less than $M$.\\


\subsection{Particle Collisions}
\subsubsection{Massive Particles}

\begin{figure}[!h]
	\centering
	\begin{subfigure}{0.4\textwidth}
		\includegraphics[width=\textwidth]{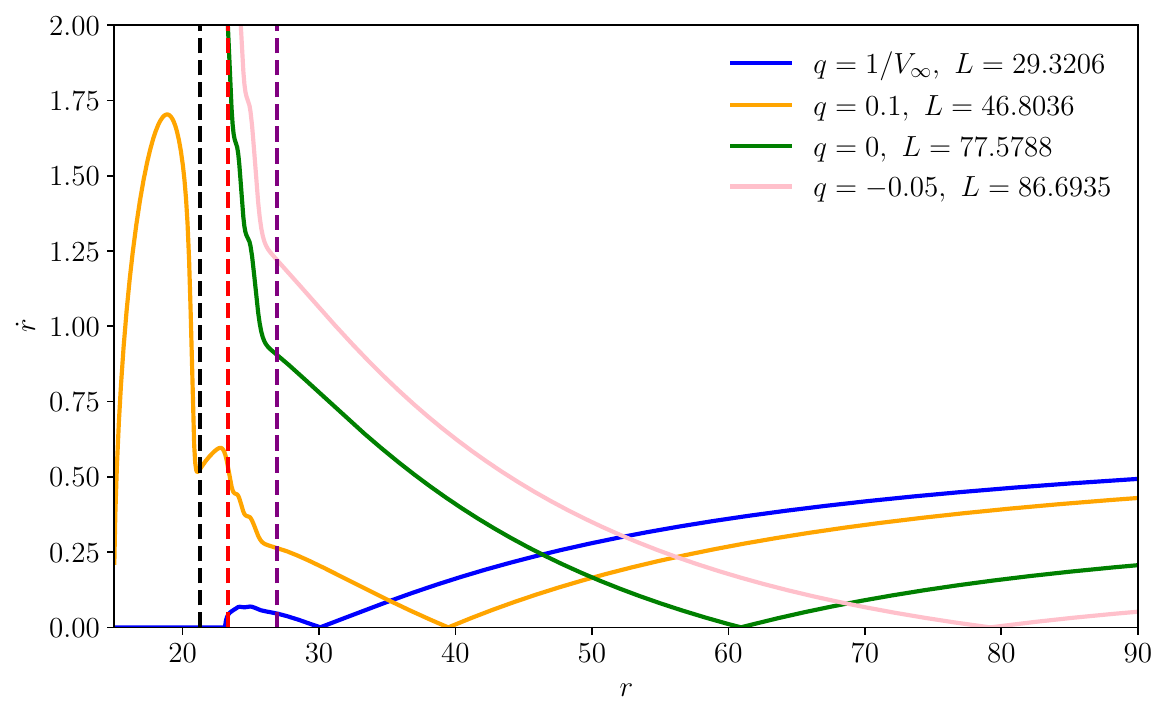}
	\end{subfigure}
    \begin{subfigure}{0.4\textwidth}
		\includegraphics[width=\textwidth]{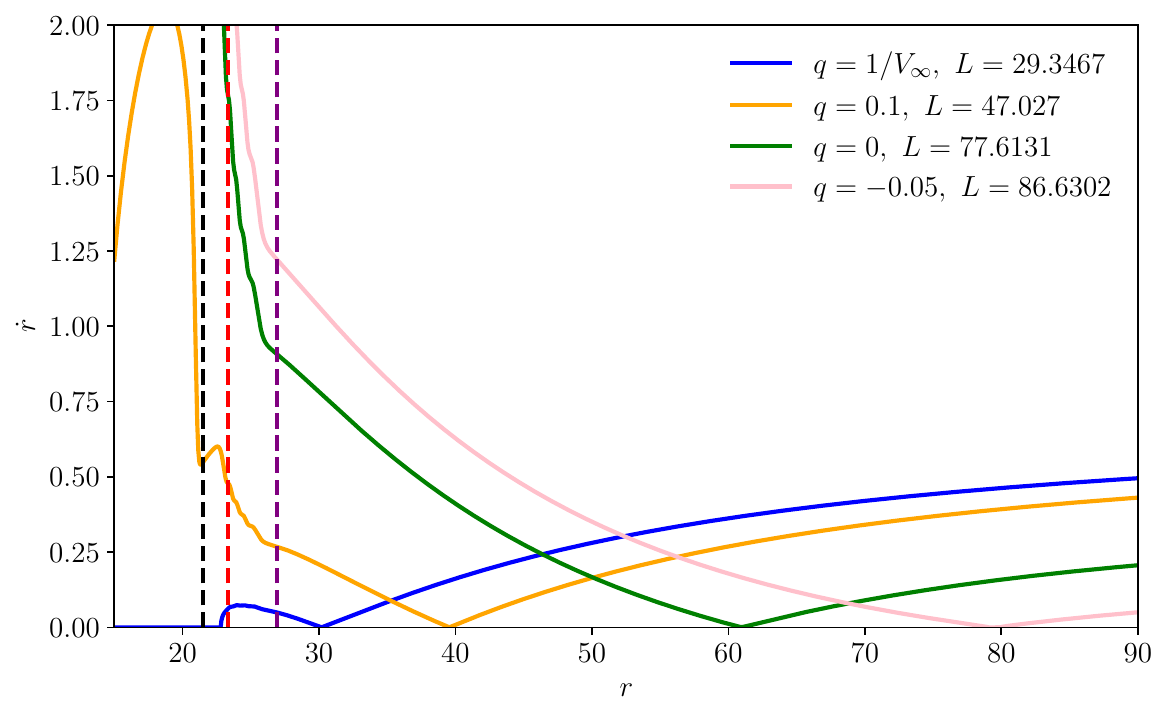}
	\end{subfigure}
    \caption{{\it Left}: We plot the radial velocity \(\dot{r}\) as function of the radial coordinate \(r\) for (un)charged particles moving in the frozen star space-time with \(Q=213\) and node \(k=2\), for several values of the angular momentum \(L\). The black and purple dashed lines indicate the inner and outer boundaries of the shell, respectively, while the red dashed line marks the radial position at which \(N(r)\rightarrow0\).  {\it Right}: Same as left, but for \(Q=213\) and \(k=4\). }
    \label{fig:massive_k4}
\end{figure}

\begin{figure}[!h]
	\centering
	\begin{subfigure}{0.4\textwidth}
		\includegraphics[width=\textwidth]{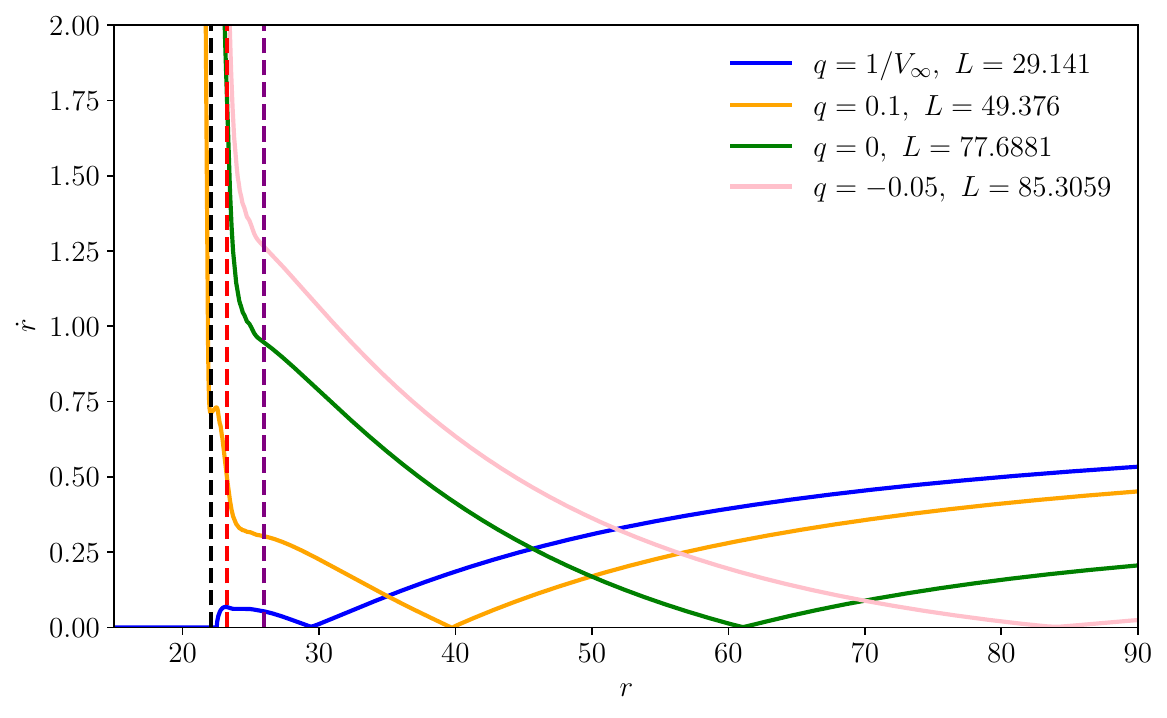}
	\end{subfigure}
    \begin{subfigure}{0.4\textwidth}
		\includegraphics[width=\textwidth]{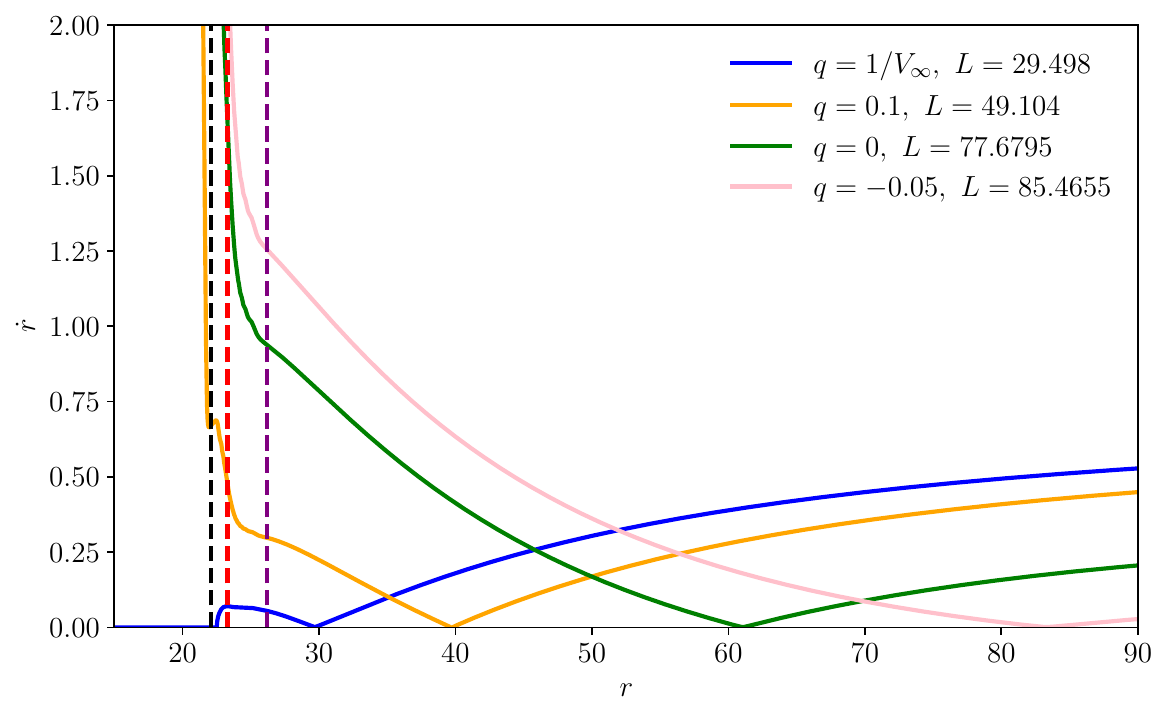}
	\end{subfigure}
    \caption{{\it Left}: We plot the radial velocity \(\dot{r}\) as function of the radial coordinate \(r\) for (un)charged particles moving in the frozen star space-time with \(Q=213\) and node \(k=14\), for several values of the angular momentum \(L\). The black and purple dashed lines indicate the inner and outer boundaries of the shell, respectively, while the red dashed line marks the radial position at which \(N(r)\rightarrow0\).  {\it Right}: Same as left, but for \(Q=213\) and \(k=18\).  }
    \label{fig:massive_k18}
\end{figure}

Within the shell region, we find that the conditions required for a collision to occur lead to either \(E<0\) or \(L^2<0\). Both possibilities are physically impossible, since a particle cannot possess negative energy in this context and the squared angular momentum is necessarily non-negative. The limiting case in which a collision can occur closest to the shell region corresponds to a particle with charge \(q=1/V_{\infty}\). This is shown for different values of $q$ in the space-time of the frozen boson star with $k=$ in Fig.~\ref{fig:massive_k4}
and for $k=18$ in Fig.~\ref{fig:massive_k18}. Clearly, the condition for a critical particle with $\dot{r}=0$ is only fulfilled far away from the shell, where the solution is already RN.

\begin{figure}[!h]
    \centering
    \includegraphics[width=0.6\textwidth]{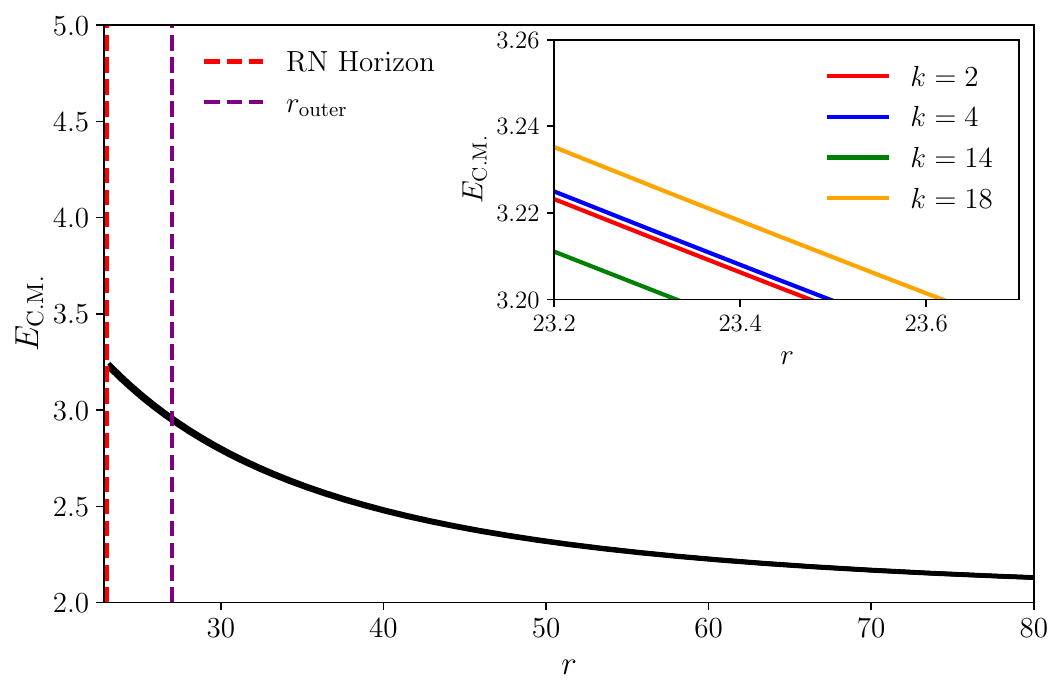}
    \caption{We show the center-of-mass energy as a function of \(r\) for collision between two particles with charge \(q=1/V_{\infty}.\)}
    \label{fig:ECM2}
\end{figure}

The question then is whether large values for the center of mass energy \(E_{\text{\tiny{C.M.}}}\) can be achieved. The value of this quantity in dependence of $r$ is shown in Fig.~\ref{fig:ECM2}. Here. the red dashed line indicates the location of the horizon of the corresponding RN solution, while the purple dashed line denotes the outer boundary of the shell. For \(k=18\), the configuration gives the largest value in the displayed region, while \(k=14\) gives the smallest and \(E_{\text{\tiny{C.M.}}}\) remains finite on the RN horizon
for all four $k$. This suggests that there is no singular behaviour in \(E_{\text{\tiny{C.M.}}}\) in this space-time.  

\begin{figure}[!h]
    \centering
    \includegraphics[width=0.6\textwidth]{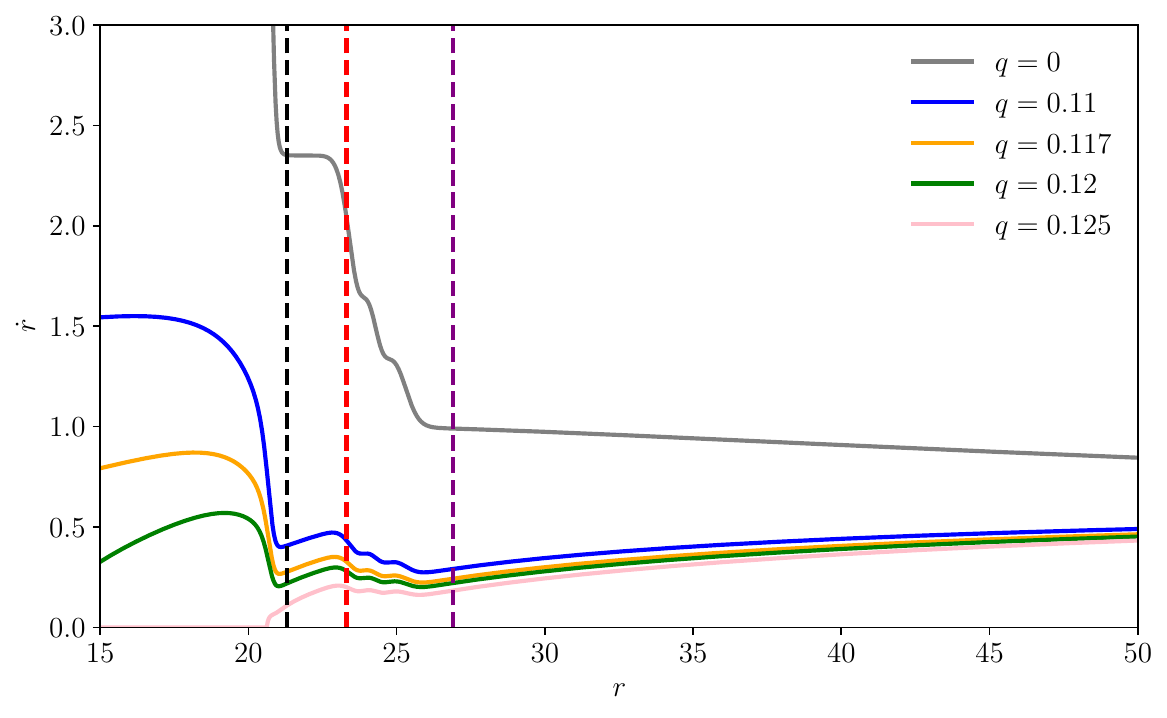}
    \caption{We plot the radial velocity, \(\dot{r}\), as a function of \(r\) for particles with zero angular momentum, \(L=0\) and varying electric charge \(q\) in the space-time with \(Q=213\) and node number \(k=14\).}
    \label{fig:dotrk14}
\end{figure}

An interesting feature of radial motion shown in Fig.~\ref{fig:dotrk14} is the distinct variation of \(\dot{r}\) within the shell region. In contrast to the relatively smooth behaviour observed outside the shell, the particles exhibit a non-monotonic radial motion as they propagate through the shell. This behaviour reflects the structure of the underlying space-time, since the radial velocity is determined by the combined effect of the metric functions \(N(r)\) and \(\sigma(r)\), as well as the electromagnetic potential \(V(r)\). The stronger radial variation of these background quantities within the shell consequently leads to a more pronounced modification of the particle trajectories. Thus, the features observed in \(\dot r\) inside the shell are a direct result of the non-trivial geometry and electromagnetic structure of the space-time in this region. Since the particles experience zero angular momentum, the differences between these trajectories are entirely associated with the electromagnetic interaction, rather than an angular-momentum barrier, and shows that how particles can reach the center only when the charge-dependent energy condition is satisfied.


\subsubsection{Massless Particles}
We now investigate the behaviour of the center-of-mass energy in the region where \(N(r)\) becomes very small, in particular at \(r=r_{\text{Min}}\), where \(N(r)\) reaches its minimum. We begin by considering a head-on-collision, with particle 1 ingoing and particle 2 outgoing. The corresponding signs of the radial motion are therefore \(\epsilon_1=-1\) and \(\epsilon_2=1\), giving \(\epsilon_1\epsilon_2=-1\). Substituting these values into the expression for the center-of-mass energy (\ref{CMEstaticMassless}), we obtain
\begin{equation}
\label{CMEstaticMassless-1}
\frac{E^2_{\text{\tiny{C.M.}}}}{2}=\frac{E_1 E_2}{N\sigma^2}-\frac{L_1L_2}{r^2}+\frac{1}{N}\sqrt{\frac{E_1^2}{\sigma^2}-\frac{NL^2_1}{r^2}}\sqrt{\frac{E^2_2}{\sigma^2}-\frac{NL^2_2}{r^2}} \ .
\end{equation}
We are interested in understanding the behaviour of (\ref{CMEstaticMassless-1}) when \(N\) is very small at \(r=r_{\text{Min}}\). In particular, we need to examine the terms proportional to \(1/N\) and determined whether they produce a divergence in \(E_{\text{\tiny{C.M.}}}\) as \(N\rightarrow0\). At the same time, it is important to check that the radial motion of both particles remains physically admissible in this limit, since a divergence in the formal expression for \(E_{\text{\tiny{C.M.}}}\) does not by itself guarantee that the collision can actually occur. The expansion of (\ref{CMEstaticMassless-1}) about \(r=r_{\text{Min}}\) gives
\begin{equation}
\label{CMEstaticMasslessExp}
\frac{E^2_{\text{\tiny{C.M.}}}}{2}\bigg|_{r=r_{\text{Min}}}=\frac{2E_1 E_2}{N\sigma^2}-\frac{1}{2r^2_{\text{Min}}}\bigg(\frac{E_2}{E_1}L^2_1+\frac{E_1}{E_2}L^2_2\bigg)-\frac{L_1L_2}{r^2_{\text{Min}}}+O(N) \ .
\end{equation}
Since \(N(r_{\text{Min}})\ll1\), the first term dominates over the terms that remain finite as \(N\rightarrow0\). Therefore, to leading order, we obtain
\[E^2_{\text{\tiny{C.M.}}}\big|_{r=r_{\text{Min}}}\sim\frac{4E_1 E_2}{N(r_{\text{Min}})\sigma^2(r_{\text{Min}})}\]
and consequently 
\[E_{\text{\tiny{C.M.}}}\big|_{r=r_{\text{Min}}}\sim\frac{2\sqrt{E_1 E_2}}{\sigma(r_{\text{Min}})\sqrt{N(r_{\text{Min}})}} \ .\]
Thus, as \(N(r_{\text{Min}})\) becomes increasingly small, the center-of-mass energy grows as \(N^{-1/2}\). This shows a head-on collision between an ingoing and an outgoing massless particle can produce a very large \(E_{\text{\tiny{C.M.}}}\) in the RN-like region of the frozen star. This conclusion follows from the leading \(1/N\) contribution, while the angular momentum dependent terms remain finite in the \(N\rightarrow0\) limit. Therefore, we do not require either particles to have vanishing angular momentum, provided that both particles are physically able to reach \(r=r_{\text{Min}}\) and their radial motion remains real at that point where the outgoing particle has \(\dot{r}_2>0\) and the ingoing particle has \(\dot{r}_1<0\). \\
\newline The conditions that describe an unstable circular orbit have been discussed earlier on, where \(r_c\) is the radius of the unstable circular orbit with its corresponding critical angular momentum. However, we should not impose the same three condition onto outgoing particle. If we want a genuinely outgoing particle at the collision point, then at the collision radius \(r_c\), we would need \(\dot{r}_2(r_c)>0\). So our two particles should generally not both satisfy the circular-orbit condition. If we are looking at particles traveling from infinity, then the outgoing particle will come from a turning point. So initially it travels inward, i.e. \(\dot{r}_2<0\) and reaches some radius \(r_t\) where \(\dot{r}_2(r_t)=0\). Since \(\dot{r}^2_2=-V_{\text{eff},2}(r)\), the turning point satisfies \(V_{\text{eff},2}(r_t)=0\) and we then require that it reaches the collision radius, i.e. \(r_t<r_c\). But unlike for a circular orbit, for an ordinary point we have \(V'_{\text{eff},2}(r_t)\neq0\). After reaching the turning point, the particle reverses direction, i.e. \(\dot{r}_2>0\) and begins to move outward and we require that 
\[\frac{E^2_2}{\sigma^2(r)}-\frac{N(r)L_2^2}{r^2}>0\] is satisfied
for \(r_t\leq r\leq r_c\). For a massless particle coming from infinity, the turning point is controlled by the competition between its energy and angular momentum, i.e. its impact parameter \(b=L_2/E_2\). Therefore, the radial trajectory depends on the impact parameter, not on \(E_2\) and \(L_2\) independently. It is useful to define 
\[R(r;E_2,L_2,q)\equiv \frac{E_2^2}{\sigma^2}-\frac{NL_2^2}{r^2}\]
so that \(\dot{r}^2=R(r)\). A turning point occurs when \(R(r_t)=0\), therefore \(|E_2|/\sigma(r_t)=|L_2|\sqrt{N(r_t)}/r_t\) is the condition we need to find the radius at which the second particle changes from ingoing to outgoing, or 
\[b^2=\frac{r_t^2}{N(r_t)\sigma^2(r_t)} \ .\]
Therefore, large value of \(L_2/E_2\) mean we have a stronger centrifugal barrier, so the particle can turn around. Small values of  \(L_2/E_2\) indicate that the particle travels further inward and increasing \(E_2\) while keeping \(L_2\) fixed decreases \(L_2/E_2\), making a turning point less likely. This suggests that we need the second particle to have a sufficiently large angular momentum relative to its energy.

\begin{figure}[!h]
    \centering
    \includegraphics[width=0.7\textwidth]{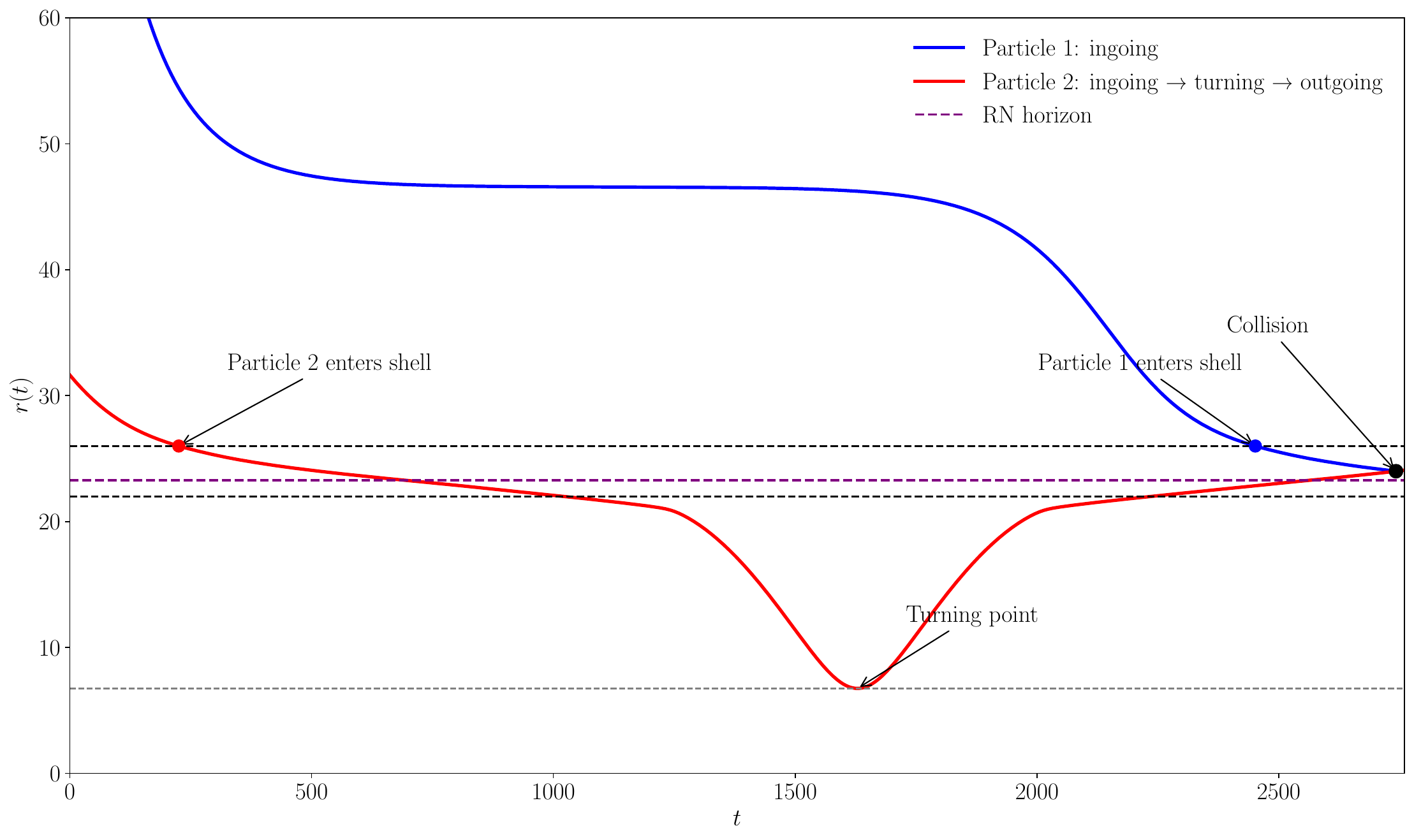}
    \caption{Collision of two massless test particles in the space-time of a frozen boson star}
    \label{massless}
\end{figure}

In Fig.~\ref{massless}, we shown the case for \(E_1 = E_2 = 1\), but \(L_1 = 93.1981\) and \(L_2 = 90\). The first particle has the angular momentum of a critical particle, corresponding to the unstable circular null orbit at approximately \(r = 46.5654\). The second particle has \(L_2 = 90\). This gives it an inner turning point at approximately \(r = 6.75154\). Therefore, particle 2 can initially move inward, reach this turning point, reverse its radial motion, and subsequently move outward again. This is what allows the two particles to approach the collision point from opposite directions. Particle 1 remains ingoing, while particle 2 becomes outgoing after its turning point. At the chosen collision radius, \(r_c = 24\), we therefore have particle 1 moving inward and particle 2 moving outward. This is particularly important for the center-of-mass energy calculation because the collision approaches the limit of RN horizon. The two particles are not simply passing through \(r = 24\) in the same direction; they are approaching the collision point from opposite radial directions. Not every choice of angular momentum produces a valid collision, even when the two energies are identical. Some choices of \(L\) produce a turning point outside the desired collision region, meaning that the particle cannot reach \(r = 24\). Other choices do not produce the required outgoing branch at the collision radius. In these cases, although one may still be able to substitute the parameters into the center-of-mass energy expression, there is no physically admissible pair of trajectories meeting at the required space-time point.




\clearpage\section{Conclusions}
We have studied the motion of test particles as well as their collision in the space-time of charged boson stars. We find that for small electric charge $Q$ of the boson star there is always one stable circular orbit for massive particles. This is also true for negatively charged  massive particles, while for positively charged particles there is a limit due to electromagnetic repulsion between the test particle and the boson star. While the space-time is globally regular, test particles can never reach $r=0$ for $L\neq 0$ due to the centrifugal barrier, but they can get arbitrarily close. They can go through the core of the boson star and also circle around the core. 
For massless particles we do not find circular orbits. This changes when increasing the charge $Q$ and reaching the frozen star limit of the boson stars. We find that circular orbits resemble those of RN, but that an additional stable orbit can exist inside the shell which interpolates between the de Sitter interior and the black hole exterior. The radius of this orbit is smaller than that of the event horizon of the corresponding RN solution with the same mass and charge. 

Typically, the collision point with $\dot{r}=0$ is far from the center of boson stars and
moves away with increasing $Q$. For frozen stars particles can have a collision point as close as possible to the shell for particle charge $q=1/V_{\infty}$ which is similar to the black hole case. We find that in these frozen star space-times, the center of mass energy of the collision can become arbitrarily large.

\vspace{2cm}

{\bf Acknowledgements} Katherine Horton is supported by the Engineering and Physical Sciences Research Council
under grant number EP/W524335/1.

\clearpage


\end{document}